\documentclass[numberedappendix,onecolumn]{emulateapj} 
\usepackage{rotating} 
\usepackage{amsmath,amssymb,bbold,mathrsfs}
\usepackage{subfigure}
\usepackage{bm}

\usepackage[usenames, dvipsnames]{color}
\usepackage{xcolor,cancel,soul}

\definecolor{g}{RGB}{0, 160, 0}
\setstcolor{g}

\begin{document}

\shorttitle{A novel Hanle-effect investigation of the small-scale magnetic activity of the quiet Sun}

\title{A novel investigation of the small-scale magnetic activity of \\the quiet Sun via the Hanle effect in the Sr {\sc i} 4607 \AA\ line}  

\author{T.\ del Pino Alem\'an$^{1}$, J.\ Trujillo Bueno$^{1,2,3}$\footnote{Affiliate scientist 
of the National Center for Atmospheric Research, Boulder, U.S.A.}, J. \v{S}t\v{e}p\'an$^{4}$, N. Shchukina$^{5}$}

\affil{$^1$Instituto de Astrof\'{\i}sica de Canarias, 38205, La Laguna, Tenerife, Spain}
\affil{$^2$Departamento de Astrof\'\i sica, Facultad de F\'\i sica, Universidad de La Laguna, Tenerife, Spain}
\affil{$^3$Consejo Superior de Investigaciones Cient\'{\i}ficas, Spain}
\affil{$^4$Astronomical Institute ASCR, Fri\v{c}ova 298, 251\,65 Ond\v{r}ejov, Czech Republic}
\affil{$^5$Main Astronomical Observatory, National Academy of Sciences, 27 Zabolotnogo Street, 03143, Kiev, Ukraine}

\begin{abstract}

One of the key research problems in stellar physics is to decipher the small-scale magnetic activity of the quiet solar atmosphere. Recent magneto-convection simulations that account for small-scale dynamo action have provided three-dimensional (3D) models of the solar photosphere characterized by a high degree of small-scale magnetic activity, similar to that found through theoretical interpretation of the scattering polarization observed in the Sr {\sc i} 4607 \AA\ line. Here we present the results of a novel investigation of the Hanle effect in this resonance line, based on 3D radiative transfer calculations in a high-resolution magneto-convection model having most of the convection zone magnetized close to the equipartition and a surface mean field strength ${\langle B \rangle}{\approx}170$ G. The Hanle effect produced by the model's magnetic field depolarizes the zero-field scattering polarization signals significantly, to the extent that the center-to-limb variation of the calculated spatially-averaged polarization amplitudes is compatible with the observations. The standard deviation of the horizontal fluctuations of the calculated scattering polarization signals is very sensitive to the model's magnetic field and we find that the predicted spatial variations are sufficiently sizable so as to be able to detect them, especially with the next generation of solar telescopes. We find that at all on-disk positions the theoretical scattering polarization signals are anti-correlated with the continuum intensity. To facilitate reaching new observational breakthroughs, we show how the theoretically predicted polarization signals and spatial variations are modified when deteriorating the signal-to-noise ratio and the spectral and spatial resolutions of the simulated observations.

\end{abstract}


\section{Introduction}\label{S-intro}

An important topic of research in astrophysics is the small-scale magnetic activity of the quiet regions of the solar atmosphere, which cover most of the solar surface at any given time during the solar cycle \citep[e.g., the reviews by][]{deWijnetal2009,SanchezMartinez2011,BMartinez2014}. Of great scientific interest is a precise determination of the mean field strength $\langle B \rangle$ of the quiet solar photosphere, including its variation with height. If the magnetic energy density carried by the small-scale magnetic fields of the solar photosphere is indeed as significant as indicated by an investigation based on the Hanle effect in atomic and molecular lines \citep{Trujilloetal2004}, such magnetism could perhaps be the main driver for heating the solar chromosphere and corona above quiet regions of the solar disk \citep{Trujilloetal2004,Amarietal2015,Rempel2017}.

The anisotropic radiation of the Sun's atmosphere induces atomic level polarization (i.e., population imbalances and quantum coherence between the atomic levels), which can produce the appearance of linear polarization in spectral lines (i.e., the so-called scattering line polarization) even without the need of a magnetic field. The Hanle effect is the magnetic-field-induced modification of the atomic level polarization and, therefore, of the scattering line polarization. \cite{Stenflo1982} suggested that this effect has diagnostic potential for inferring the presence of a tangled magnetic field at sub-resolution scales, because a magnetic field with unresolved mixed polarities tends to reduce the line scattering polarization amplitudes with respect to the zero magnetic field case. This can be shown by considering the idealized case of a one-dimensional (1D) stellar atmosphere model permeated by a micro-structured magnetic field (having random azimuth below the line's photon mean free path). The following approximate (Eddington-Barbier type) formula can be used to estimate the scattering polarization $Q/I$ amplitude at the center of a spectral line without lower-level polarization, such as that of Sr {\sc i} at 4607 \AA:

\begin{equation}
Q/I\,\approx\,{3\over{2\sqrt{2}}}(1-\mu^2){{{\cal H}^{(2)}}\over{1+{\delta}_u}}\,{{\cal A}}\, , \label{E-QIEddington}
\end{equation}
where ${\mu}={\rm cos}\,{\theta}$ (with $\theta$ the heliocentric angle), ${\delta}_u=D_u^{(2)}\,t_{\rm life}\,{\approx}\,D_u^{(2)}/A_{ul}$ is the upper-level rate of depolarizing elastic collisions with neutral hydrogen atoms in units of the Einstein $A_{ul}$ coefficient ($t_{\rm life}$ is the level's radiative lifetime), and ${{\cal A}}={\bar J}^2_0/\bar{J}^0_0$ is the degree of anisotropy of the spectral line radiation \citep[e.g.,][]{Trujillo2001}. ${\cal H}^{(2)}$ is the Hanle depolarization factor of a micro-structured magnetic field, whose value is unity for the zero field case and smaller than unity for the magnetized case \citep[e.g., see appendix A of][]{TrujilloManso1999}. The smallest value of ${\cal H}^{(2)}$ is reached for magnetic strengths $B>B_{\rm satur}$, where $B_{\rm satur}$ is the Hanle saturation field above which the scattering line polarization amplitude stops decreasing (e.g., $B_{\rm satur}\,{\approx}\,200$ G for the Sr {\sc i} 4607 \AA\ line). 

The main problem with the Hanle effect as a diagnostic tool of the quiet Sun magnetism is that it requires to compare the observed line scattering polarization amplitudes with those that the solar atmosphere would produce if all its physical properties were the same, but with no magnetic field. As reviewed by \cite{Trujilloetal2006} inferences based on the last scattering approximation \citep{Stenflo1982} and on some radiative transfer calculations in 1D semi-empirical models of the solar atmosphere \citep{Faurobertetal1995,Faurobertetal2001} have yielded artificially low values of the mean field strength (i.e., $\langle B \rangle\,{\sim}\,10$ G). 

To determine how much magnetic flux and energy reside at small (unresolved) scales, \cite{Trujilloetal2004} developed a technique based on comparisons of the scattering polarization amplitudes of the photospheric line of Sr {\sc i} at 4607 \AA\, observed in quiet regions of the solar disk at various distances from the limb with the linear polarization signals calculated in three-dimensional (3D) models of the quiet solar photosphere. The 3D model used by \cite{Trujilloetal2004} was based on the hydrodynamical solar surface convection simulations by \cite{Asplundetal2000}, which do not include magnetic fields. Since the scattering polarization observations considered by \cite{Trujilloetal2004} lacked spatial resolution, the theoretical Stokes profiles for each line-of sight were spatially-averaged so as to make a proper comparison with the observations. The observed $Q/I$ line-center amplitudes at each $\mu=\cos\theta$ position turned out to be significantly smaller than the calculated signals, by a factor ${\cal D}=(Q/I)/(Q/I)_{B=0}{\approx}0.4$. It is useful to note from Eq. \eqref{E-QIEddington} that ${\cal D}\,{\approx}\,{\cal H}^{(2)}$.

Because the 3D model used by \cite{Trujilloetal2004} is unmagnetized, in order to estimate $\langle B \rangle$ they made the following hypotheses on the quiet Sun magnetic field responsible for the Hanle depolarization: (a) it has an isotropic distribution of orientations at sub-resolution scales, and (b) the magnetic field strength is described by an exponential probability density function, ${\rm PDF}(B)={1\over{\langle B \rangle}}{\rm e}^{-B/{\langle B \rangle}}$. They concluded that an approximate {\em average} fit to the $Q/I$ line-center amplitudes observed between $\mu=0.6$ and $\mu=0.1$ is obtained with ${\langle B \rangle}{\approx}130$ G (i.e., $\sqrt{{\langle B^2 \rangle}}{\approx}180$ G), and that in the quiet solar photosphere ${\langle B \rangle}$ decreases with height. The height range in the solar photosphere that corresponds to ${0.1}\,{\le}\,{\mu}\,{\le}\,{0.6}$ for the Sr {\sc i} line is 200---400 km, approximately. \cite{Trujilloetal2004} pointed out that with a mean field strength of the order of 100 G the ensuing energy flux estimated using 1 ${\rm km}\,{\rm s}^{-1}$ for the convective velocity (thinking in rising magnetic loops) or the Alfv\'en speed (thinking in Alfv\'en waves generated by magnetic reconnection) turns out to be substantially larger than that required to balance the radiative energy losses from the solar chromosphere. An additional conclusion resulted from the constraints imposed by the Hanle effect in the Sr {\sc i} line (${\langle B \rangle}{\sim}100$ G) and in the C$_2$ lines of the Swan system (${\langle B \rangle}{\sim}10$ G), namely that the downward-moving intergranular plasma must be pervaded by relatively strong tangled magnetic fields at sub-resolution scales, with ${\langle B \rangle}\,{{>}}\,200$ G \citep[see][]{Trujilloetal2004,Trujilloetal2006}.

A few years later, \cite{VoglerSchussler2007} presented their 3D magneto-convection simulations of the quiet Sun photosphere, characterized by a topologically complex small-scale magnetic field with ${\langle B \rangle}\,{\approx}\,50$ G at the model's visible surface and ${\langle B \rangle}\,{\approx}\,15$ G at a height of about 300 km, resulting from dynamo amplification of a weak seed field. By solving the resonance line polarization transfer problem in this 3D photospheric model and contrasting the results with the observational data, \cite{ShchukinaTrujillo2011} concluded that the magnetic microactivity of the model proposed by \cite{VoglerSchussler2007} is significantly weaker than that of the real quiet Sun photosphere. They showed that the scattering polarization signals observed in the Sr {\sc i} 4607 \AA\ line can be explained after enhancing the magnetic strength of the 3D model by a scaling factor $f{\approx}10$. This is significantly larger than the scaling factor 3 needed by \cite{Danilovicetal2010} for explaining the histograms of the polarization signals produced by the Zeeman effect in the Fe {\sc i} lines at 6301.5 \AA\ and 6302.5 \AA. \cite{ShchukinaTrujillo2011} pointed out that such two different scaling factors do not imply a contradiction, because the Zeeman polarization signals of the Fe {\sc i} lines provide information on the low photosphere (heights $h{\approx}60$ km) while the Hanle signals of the Sr {\sc i} 4607 \AA\ line probe instead the middle solar photosphere (heights $h{\approx}300$ km). In any case, it is important to note that, due to non-linear feedback, conclusions inferred from a rescaled weak-field model are not necessarily identical to those obtained using the fully non-linear stronger field model.

Recently, \cite{Rempel2014} went a step further by performing improved radiative magneto-convection simulations, which show a significantly higher level of small-scale magnetic activity. Among the various numerical experiments performed by \cite{Rempel2014}, with spatial resolutions ranging from 2 till 32 km, of particular interest is a 3D model with a mean field strength of 170 G at the model's visible surface, which implies a subsurface root mean square (rms) field strength increasing with depth at the same rate as the equipartition field strength. This is the 3D model of the quiet solar photosphere we have chosen for doing the radiative transfer calculations of the scattering polarization in the Sr {\sc i} 4607 \AA\ line presented here. The mixed-polarity magnetic field of this 3D model resulted from a non-grey numerical experiment of magneto-convection with small-scale dynamo action, carried out using an open bottom boundary that allows for the presence of (small-scale) horizontal magnetic field in the upflow regions in order to mimic a deep magnetized convection zone, where ``small-scale" dynamo action takes place as well \citep[see][]{SteinNordlund2002,Stein2012}. It is important to note that a significantly lower magnetic field strength is found in 3D models that do not allow advection of magnetic flux through the bottom boundary \citep{Rempel2014}. Other 3D photospheric models resulting from magneto-convection experiments carried out without advecting any magnetic field through the bottom boundary can be seen in \cite{Khomenkoetal2017}, who showed that the Biermann battery term of the magnetic induction equation naturally provides the seed field that is then amplified via small-scale dynamo action till reaching a surface mean field strength ${\langle B \rangle}{\approx}110$ G. As mentioned above, the 3D photospheric model we have chosen for our investigation of the Hanle effect in the Sr {\sc i} 4607 \AA\ line is the most magnetized one among those discussed by \cite{Rempel2014}, which has a surface mean field strength ${\langle B \rangle}{\approx}170$ G. 

In addition to investigating whether the small-scale magnetic field of this 3D photospheric model is sufficient for explaining the scattering polarization observations of the Sr {\sc i} 4607 \AA\ line, we take the opportunity to study the spatial variations of the calculated linear polarization signals, as well as the possibility of observing them with the present and the next generation of solar telescopes. 


\begin{figure}[htp]
\centering 
\includegraphics[width=.5\textwidth]{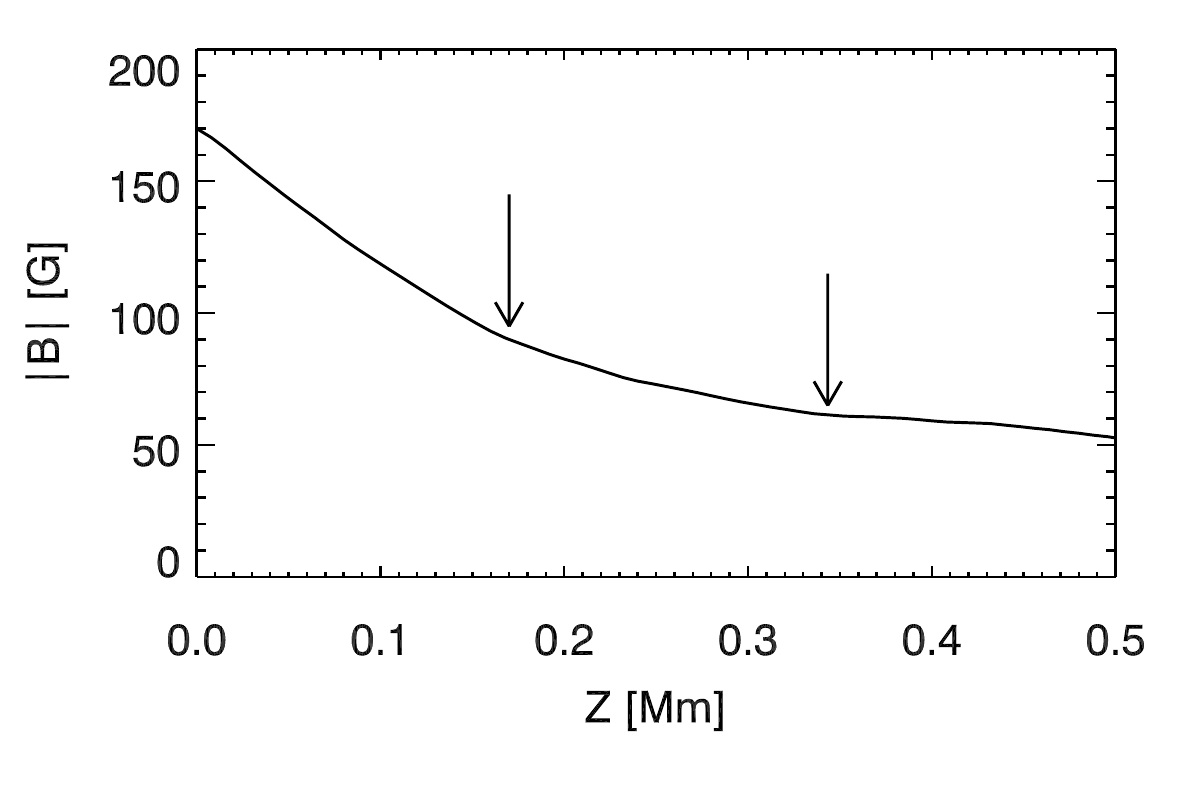}\hspace*{-2em}
\includegraphics[width=.5\textwidth]{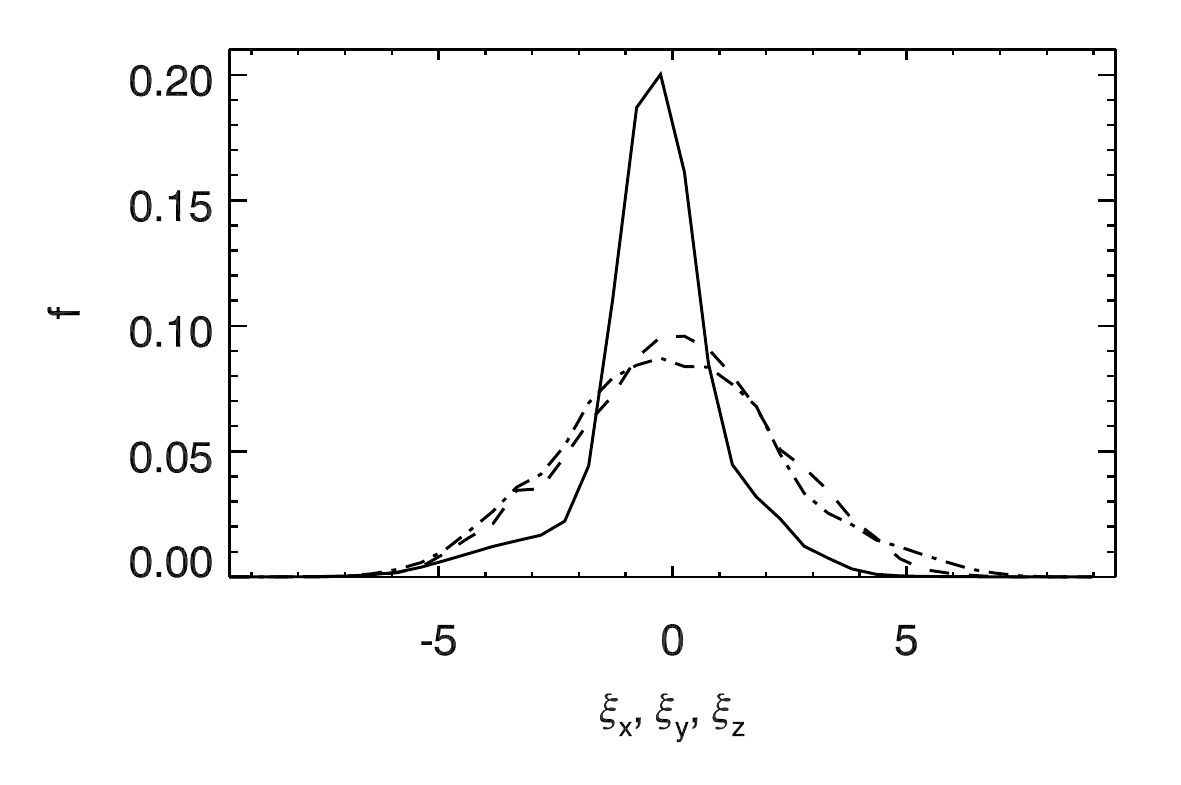}
\caption{Left panel: variation with height of the magnetic field strength, averaged on each horizontal plane of the 3D model atmosphere. The left (right) arrow indicates the approximate mean height where the optical depth at the center of the Sr {\sc i} 4607 \AA\ line is $\tau = 1$ ($\tau = 0.1$) for a line of sight with $\mu=1$. Right panel: histograms of the horizontal (dashed and dashed-dotted curves) and vertical (solid curve) components of the model's plasma velocity, normalized to the line's Doppler width, on the corrugated surface where the line-center optical depth along the disk center line of sight is unity.}
\label{F-modBandV} 
\end{figure}

\section{The Physical Problem}\label{Sproblem}

\begin{figure}[htp]
\centering 
\includegraphics[width=1.0\textwidth]{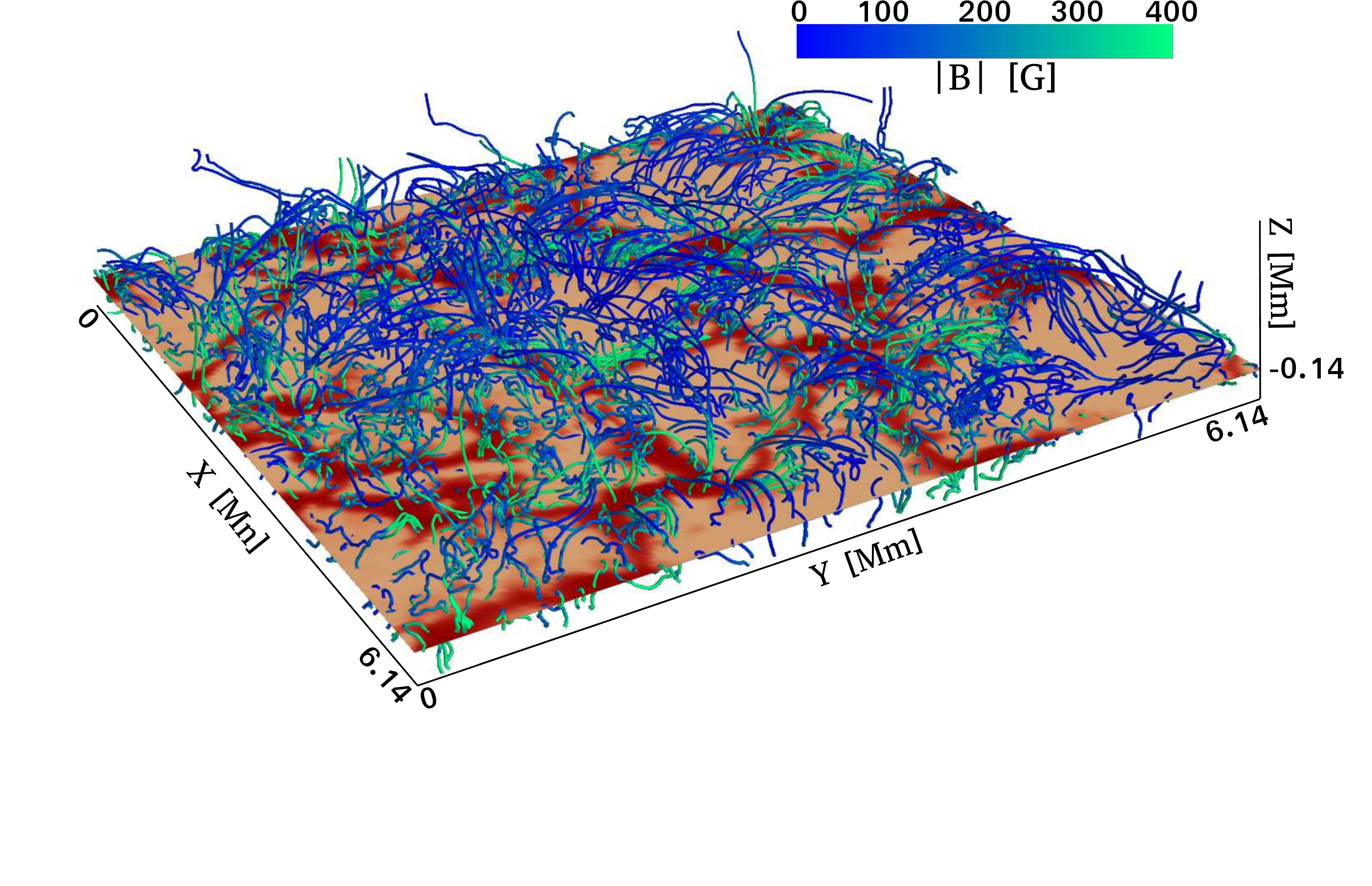}
\vspace{-70pt}
\caption{Visualization of the magnetic field lines in Rempel's (2014) 3D model of the quiet solar photosphere, which has a mean field strength of 170 G at the model's visible surface. The red-color plane shows the temperature in a horizontal slice at a height of $140$ km below the model's visible surface, with the darker (lighter) red colors representing regions with lower (higher) temperature.}
\vspace{20pt}
\label{F-Blines} 
\end{figure}

\begin{figure}[htp]
\centering 
\includegraphics[width=.6\textwidth]{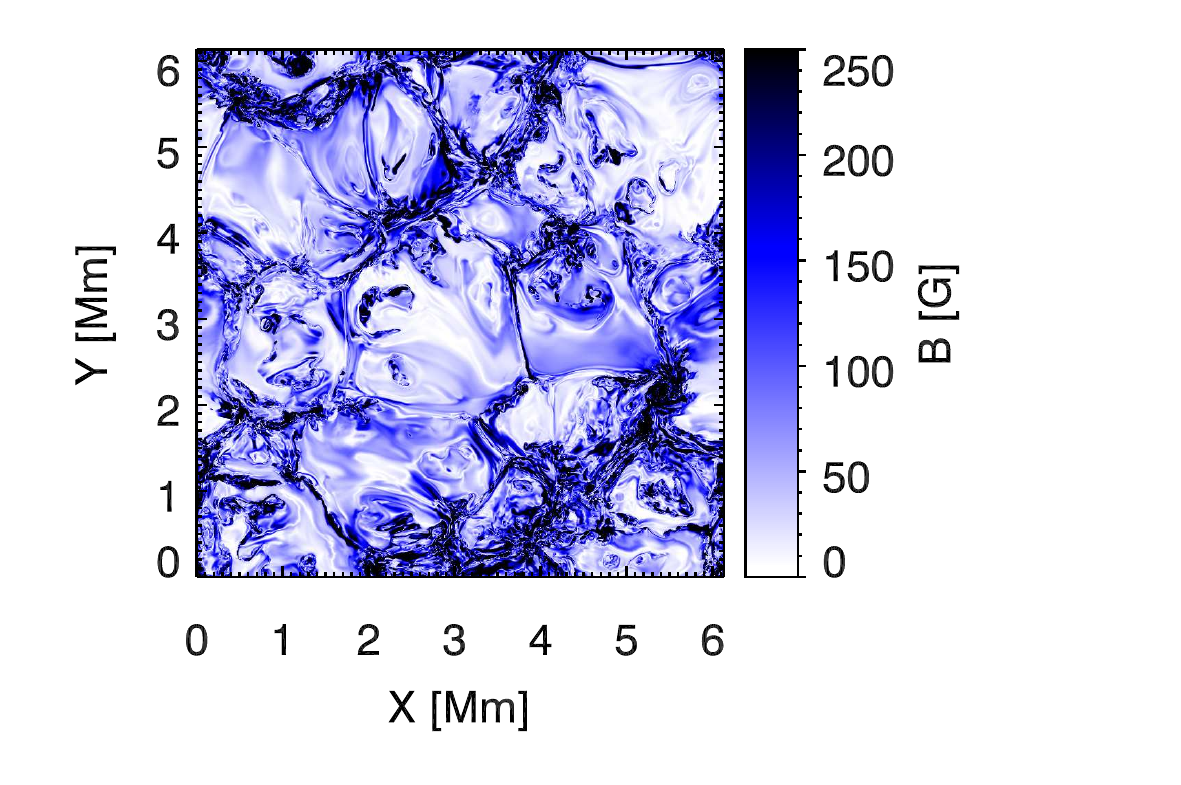}\hspace*{-6em}
\includegraphics[width=.6\textwidth]{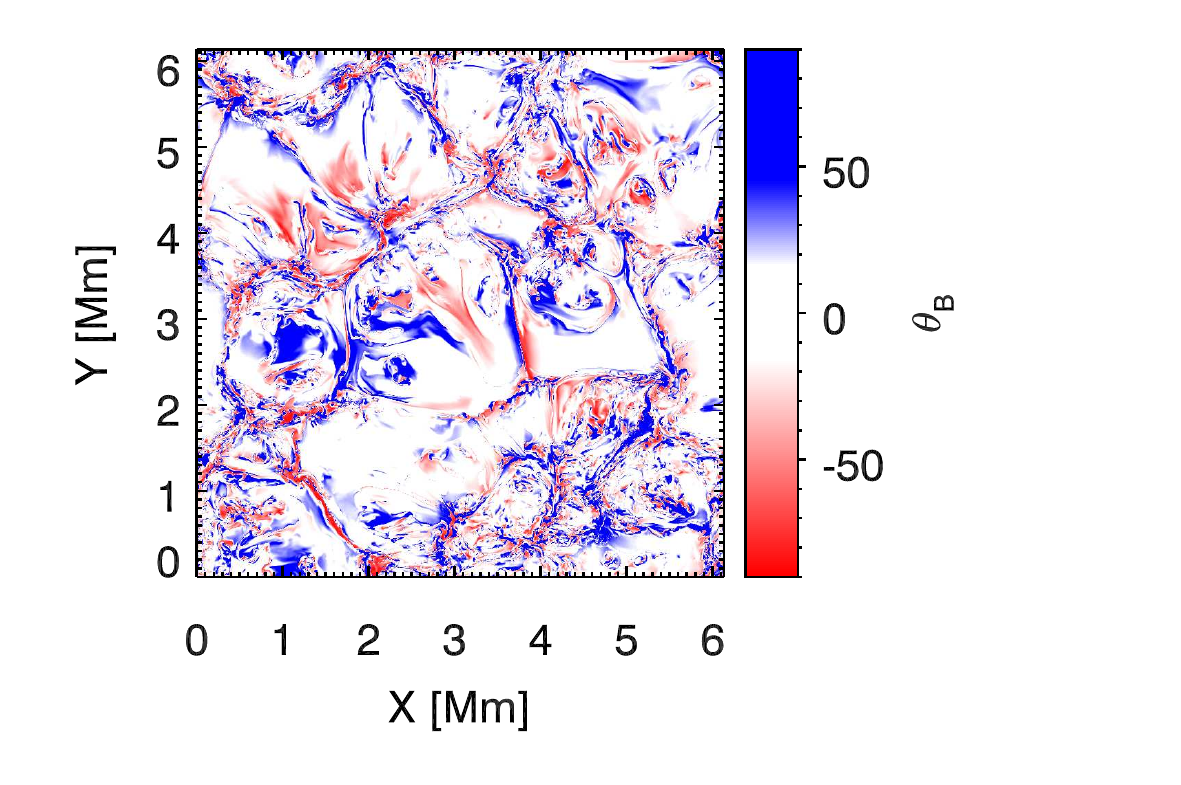}\\
\includegraphics[width=.6\textwidth]{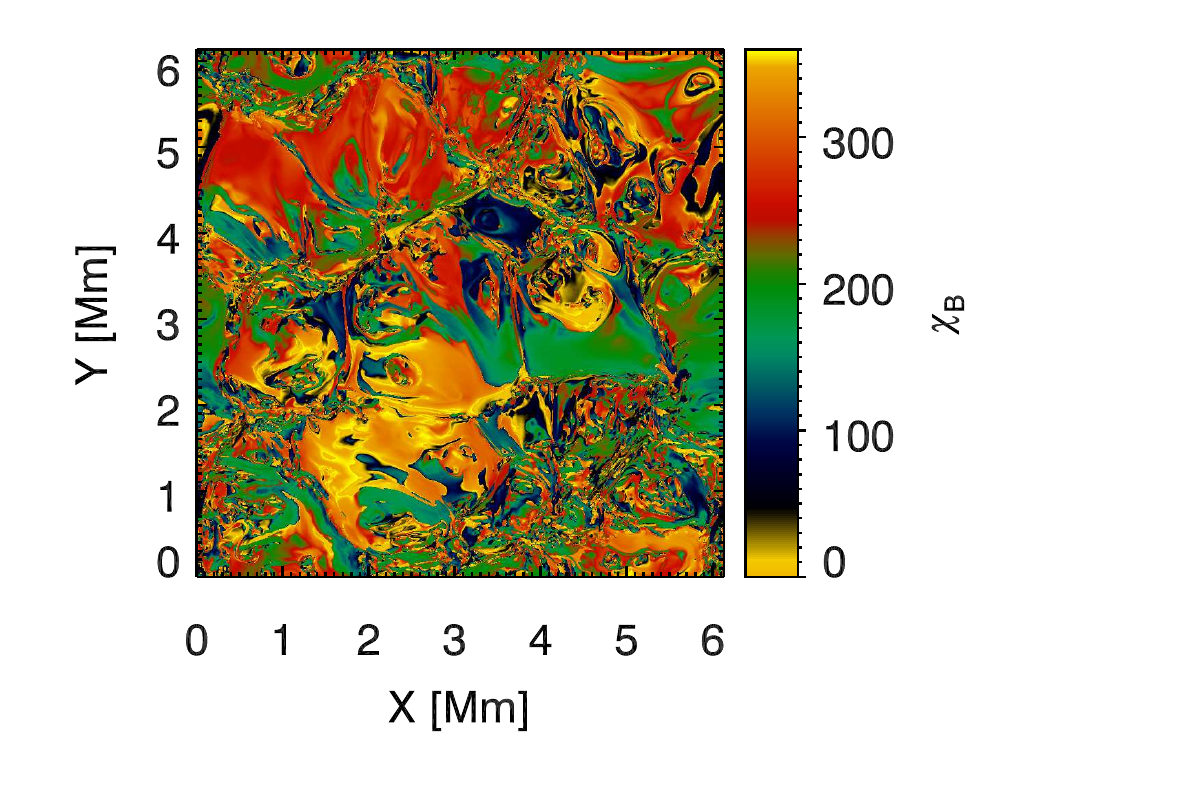}\hspace*{-6em}
\includegraphics[width=.6\textwidth]{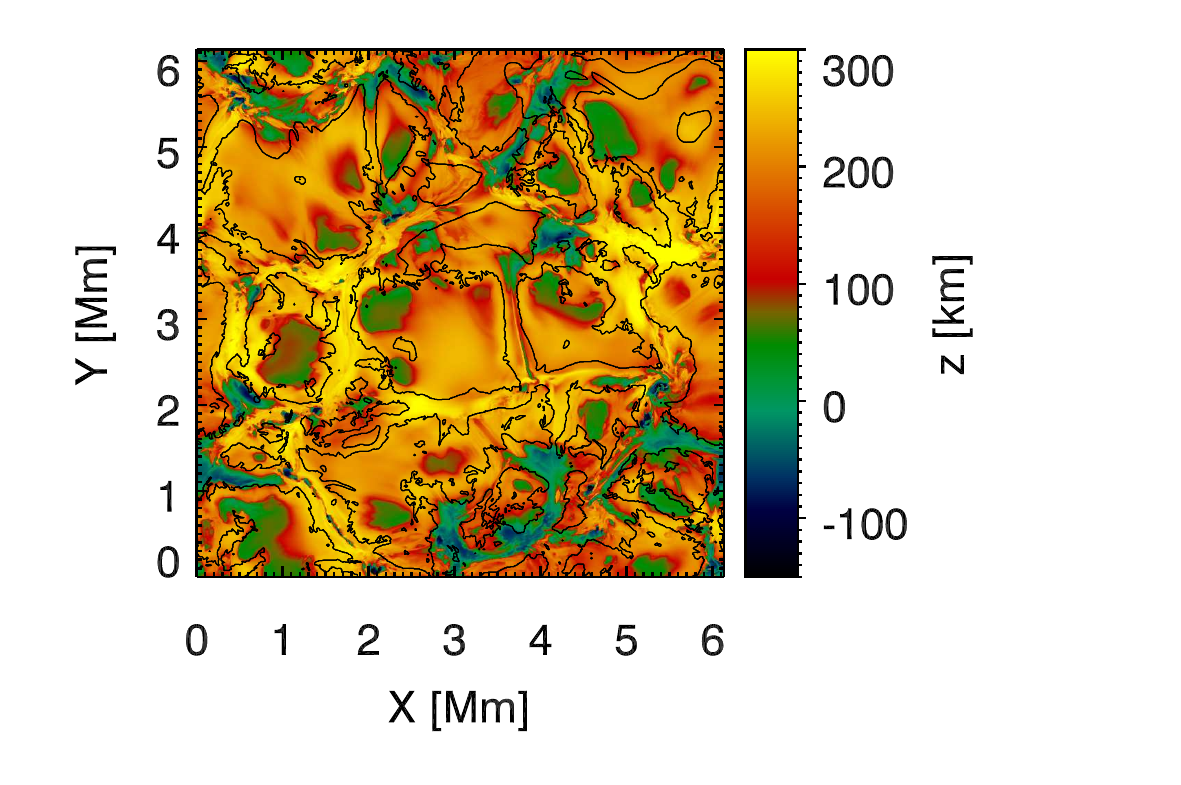}\\
\caption{Magnetic field strength (top-left panel), magnetic field inclination (top-right panel), magnetic field azimuth (bottom-left panel), and geometric height (bottom-right) at the heights in the 3D model where the line-center optical depth is unity along the disk center line of sight. The areas inside (outside) the contours in the bottom-right panel correspond to the upflowing (downflowing) regions of the plasma flow at the heights where the continuum optical depth is unity along the disk center line of sight.}
\label{F-BZtau1} 
\end{figure}

The 3D model of the quiet solar photosphere used in this investigation is a snapshot taken from one of the magneto-convection numerical experiments of \cite{Rempel2014}, which is characterized by the variation with height of the mean field strength shown in the left panel of Fig. \ref{F-modBandV} (with ${\langle B \rangle}\,{\approx}\,170$ G at the model's visible surface) and by the dynamical activity quantified in the right panel of Fig. \ref{F-modBandV}. This model, which resulted from a magneto-convection simulation with non-grey radiative transfer, shows a subsurface rms field strength increasing with depth at the same rate as the equipartition field strength, which implies that most of the convection zone is magnetized close to the equipartition. \cite{Rempel2014} considers this solution an upper limit for the quiet Sun field strength. Figure \ref{F-Blines} visualizes the model's magnetic field lines; note the topological complexity of this small-scale magnetism, and that the stronger fields (green lines) are predominantly associated with both lower photospheric layers and inter-granular lanes (red color areas). Fig. \ref{F-BZtau1} shows the spatial variation of the strength $B$, inclination $\theta_B$ and azimuth $\chi_B$ of the model's magnetic field at the heights (shown in the lower right panel) where the line-center optical depth is unity along the $\mu=1$ (disk center) line of sight. Note in the top left panel of this figure that ${\langle B \rangle}{\sim}10$ G in the (upflowing) cell centers, while ${\langle B \rangle}{>}200$ G in the (downflowing) inter-granular lanes, in agreement with \cite{Trujilloetal2004,Trujilloetal2006}.

The original grid of the 3D snapshot has $768\times 768\times 384$ points, with a regular spacing of $8$ km in the three dimensions. For our calculations, we have cut the MHD model in the vertical direction in order to include only the region of the atmosphere that is relevant for the formation of the Sr {\sc i} 4607\AA\ line. The atmospheric model we use in our calculations has $768\times 768\times 137$ grid points, with the vertical axis going from $360$ km below to $736$ km above the average height where the optical depth of the continuum at 4607\AA\ is unity, and the same $8$ km grid resolution.

\begin{figure}[htp]
\centering 
\includegraphics[width=.4\textwidth]{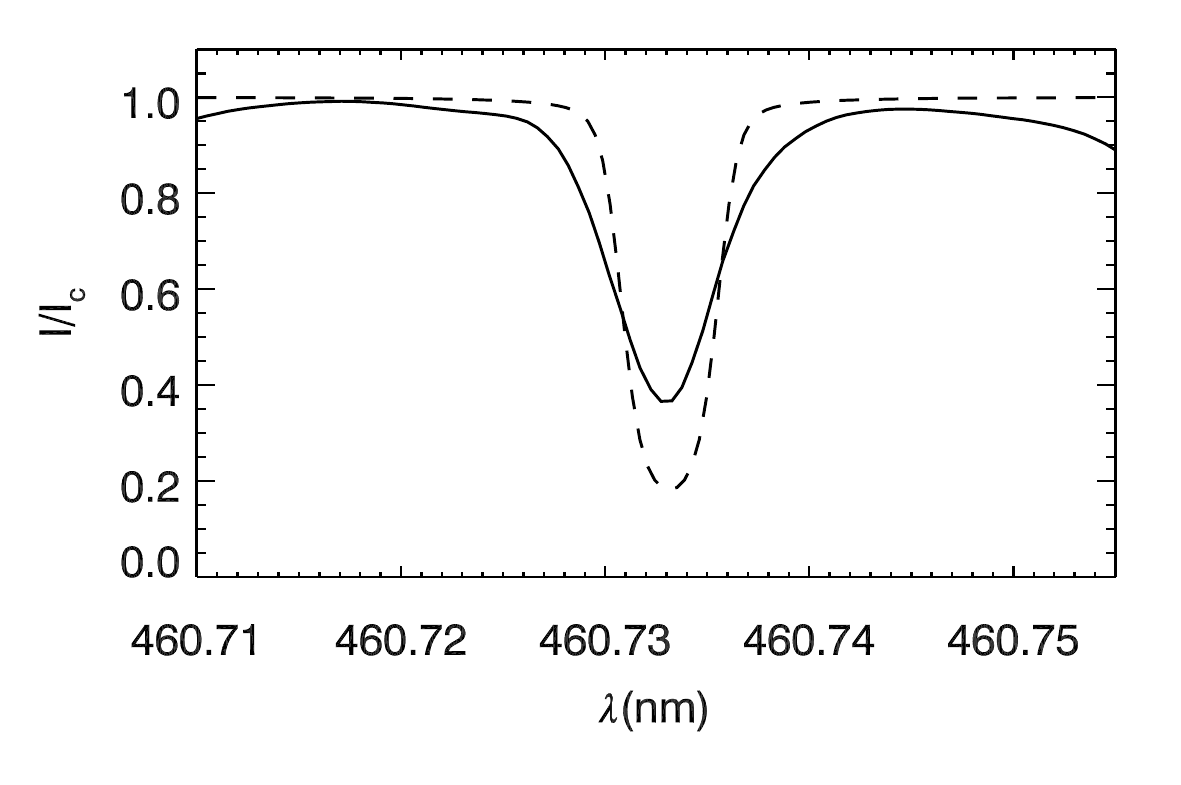}
\includegraphics[width=.4\textwidth]{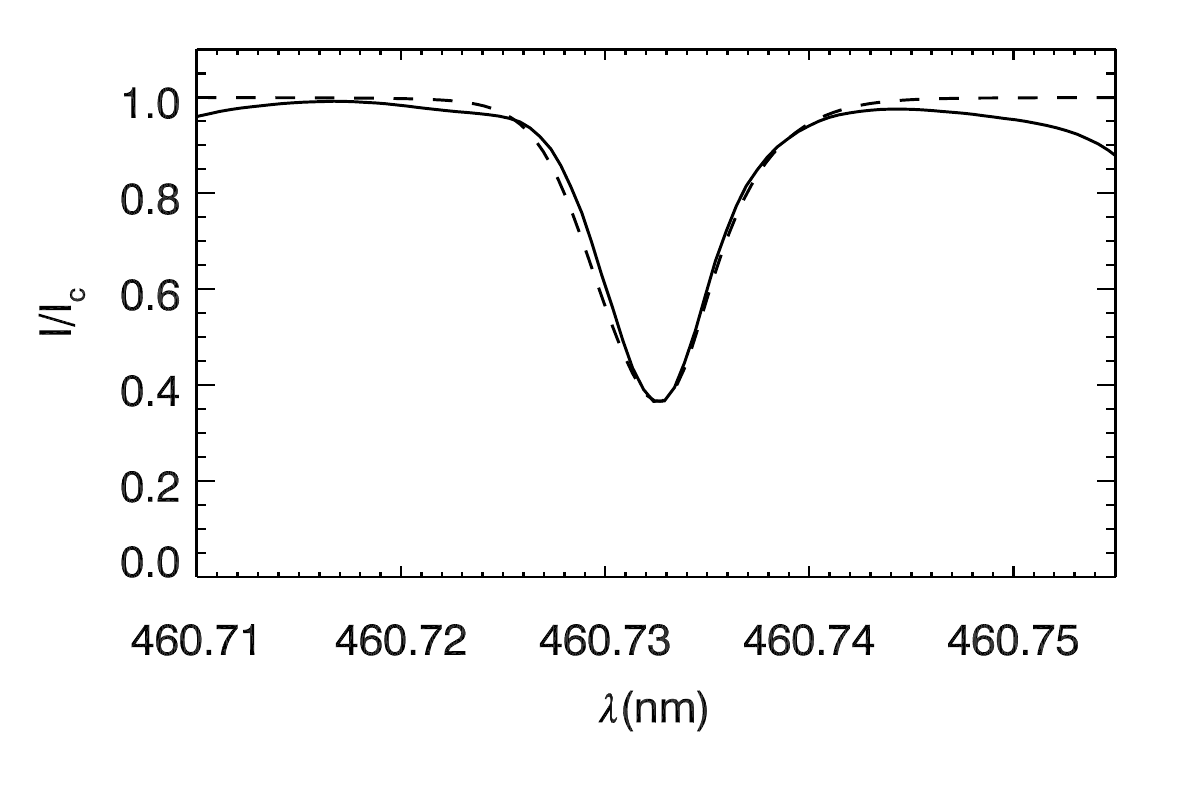}
\caption{Observed intensity profile (solid curve) and calculated emergent intensity profiles averaged over the field of view (dashed curves), normalized to the continuum intensity, for the disk center line of sight. The macroscopic plasma velocities are neglected in the left panel and taken into account in the right panel.}
\label{F-IFTS} 
\end{figure}

\begin{figure}[htp]
\centering 
\includegraphics[width=.55\textwidth]{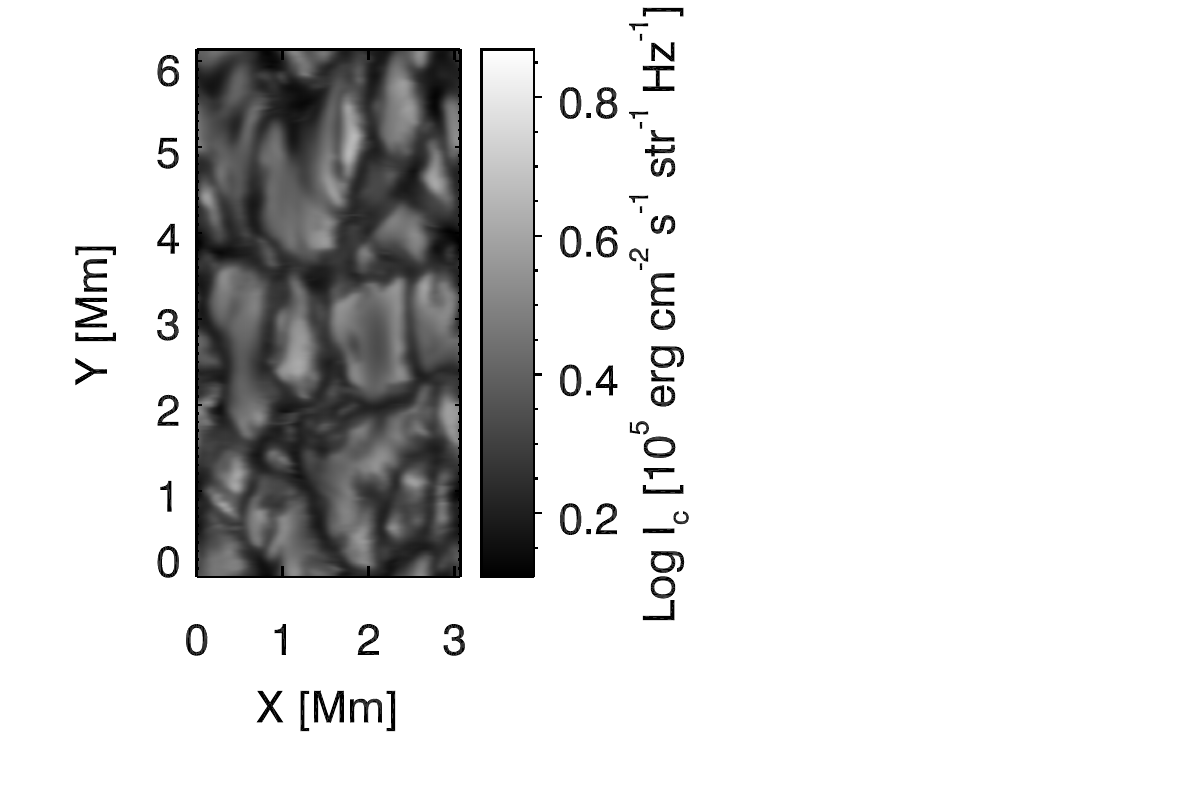}\hspace*{-10em}
\includegraphics[width=.55\textwidth]{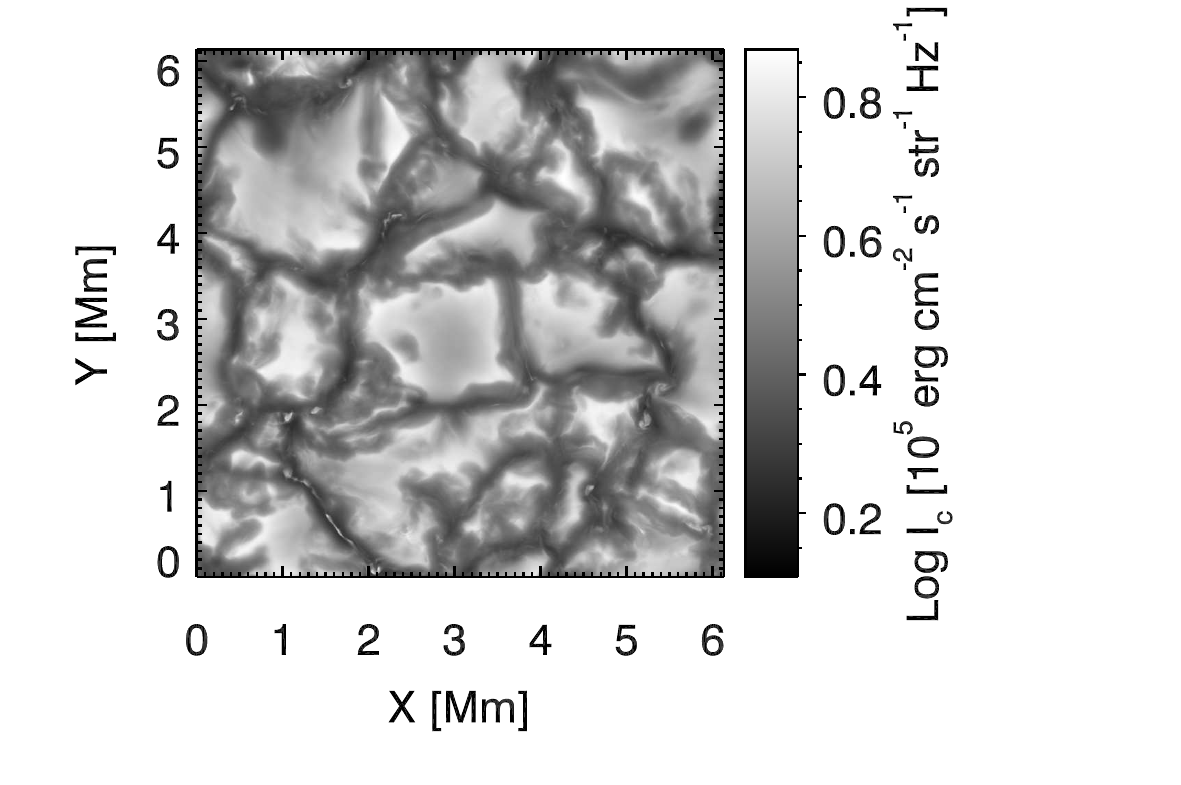}\\
\includegraphics[width=.55\textwidth]{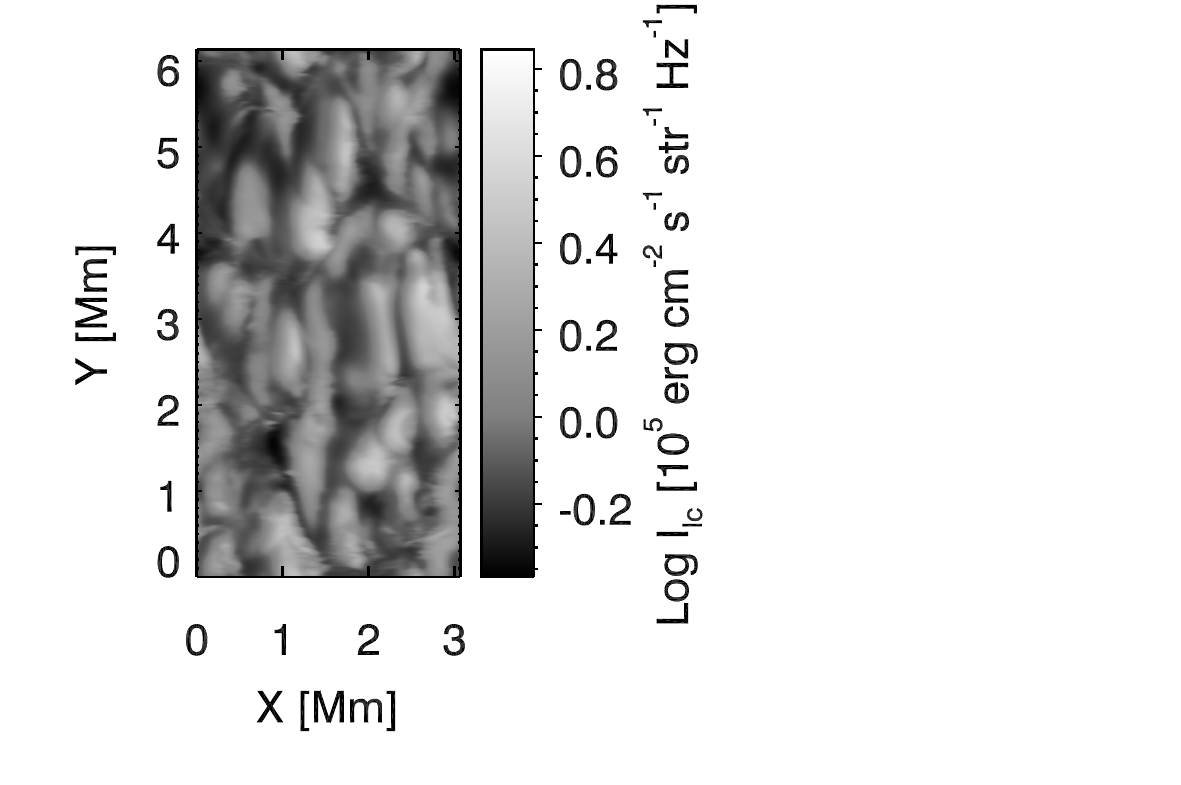}\hspace*{-10em}
\includegraphics[width=.55\textwidth]{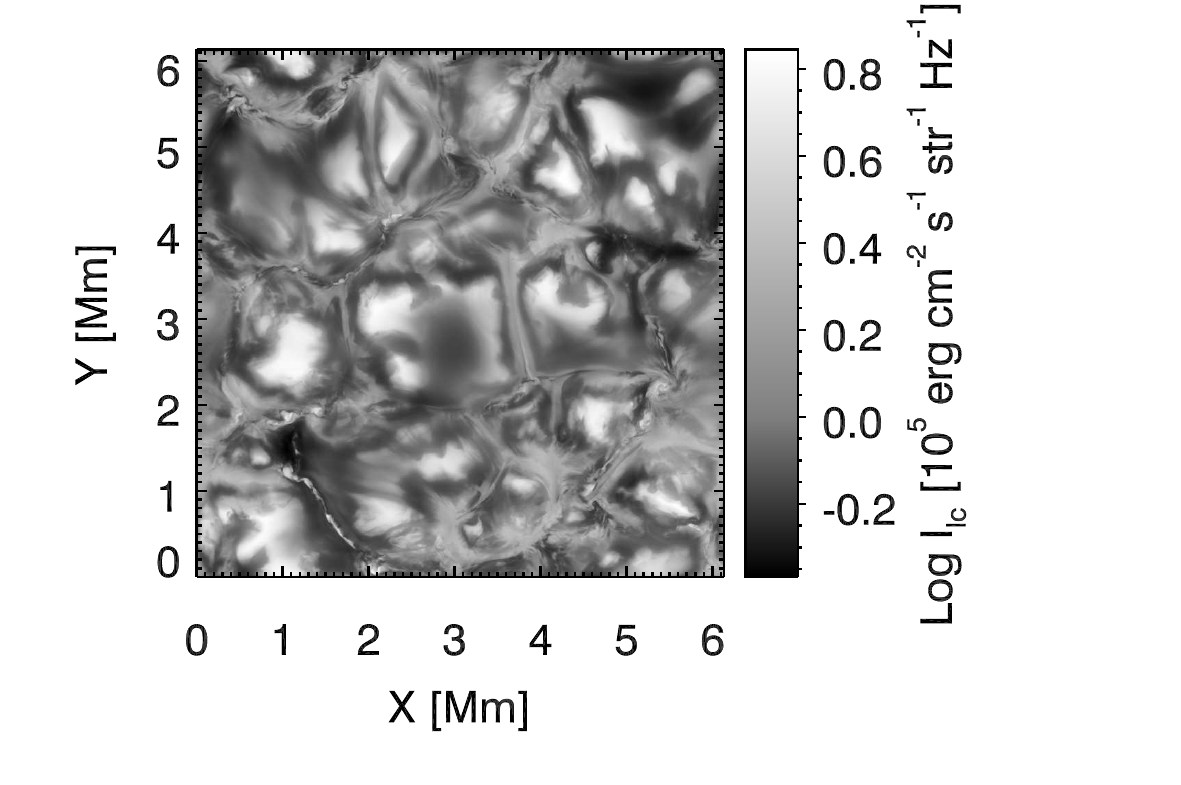}
\caption{ The spatial variation of the calculated continuum intensities (upper panels) and of the line-center intensities (lower panels) for the $\mu=0.5$ (left panels) and $\mu=1$ (right panels) lines of sight.}
\label{F-continuum} 
\end{figure}

The solid curves of Figure \ref{F-IFTS} show the intensity profile of the Sr {\sc i} 4607 \AA\ line observed in quiet regions at the solar disk center without spatial resolution \citep[see the atlas of][]{BWallaceetal1998}. The dashed curve is the theoretical Stokes $I(\lambda)$ profile we have obtained after spatially averaging the intensity profiles calculated at each point of the model's upper boundary for a line of sight with $\mu=1$. In the left panel the dashed curve has been obtained ignoring the model's macroscopic velocities at each iterative step needed to obtain the self-consistent solution. In the right panel the impact of the Doppler shifts caused by such velocities have been taken into account, and we consider the excellent agreement with the observed profile as an indication that the thermodynamical and dynamical structure of the 3D model is sufficiently realistic to allow for a reliable determination of the scattering line polarization corresponding to the zero-field reference case. Figure \ref{F-continuum} shows the spatial variation of the calculated intensity at the line center of the Sr {\sc i} 4607 \AA\ line (bottom panels) and at the nearby continuum (upper panels)\footnote{Note that in each figure of this paper visualizing spatial variations, we take into account the projection effects by means of which the off-disk-center images appear contracted by a factor $\mu$ along the line of sight direction, i.e., along the direction perpendicular to the nearest limb.}.

The calculations of the emergent Stokes profiles have been carried out with the radiative transfer code PORTA \citep[see][]{StepanTrujillo2013}, which solves the non-LTE multilevel problem of the generation and transfer of polarized radiation in 3D cartesian models of stellar atmospheres taking fully into account the effects of horizontal radiative transfer and the Doppler shifts caused by the model's macroscopic velocities. Therefore, we have taken into account the breaking of the axial symmetry of the incident radiation field at each point within the medium caused by (1) the model's horizontal thermal and density inhomogeneties, (2) the spatial gradients of the non-radial components of the model's macroscopic velocity and (3) the Hanle effect of the model's magnetic field. 

The calculation with PORTA required to solve first the problem of the strontium ionization balance in order to obtain the number density of strontium atoms in the lower and upper levels of the Sr {\sc i} 4607 \AA\ line at each spatial point of the 3D model. To this end, we solved the unpolarized radiation transfer problem using a model atom with 15 Sr {\sc i} levels and the ground level of Sr {\sc ii}, with energies taken from the NIST atomic spectra database (\citealt{NIST}). The Sr {\sc i} model contains most of the levels with principal quantum numbers $4 \le n \le 7$ and azimuthal quantum numbers  $l \le 3$. Our atomic model takes into account 12 bound-bound radiative transitions. Their oscillator strengths are taken from the NIST database when available\footnote{https://www.nist.gov/pml/atomic-spectra-database}, otherwise they are calculated using the Coulomb approximation (\citealt{BSobelmanetal1995}). The photoionization cross sections for the $n$s, $n$p and $n$d levels have been calculated using the quantum defect method (\citealt{Peach1967}), while for the 4f level we applied the hydrogen-like approximation (\citealt{BLang1974}). The bound-bound inelastic collisional rates with electrons were calculated using the approximation of \cite{VanRegemorter1962} for radiatively electric dipole allowed transitions, and with the approximation of \cite{BelyVanRegemorter1970} if the transition is forbidden. The collisional ionization rates with electrons were calculated following the formulae by \cite{BCox2000}.

The model atom used to compute with PORTA the Stokes profiles of the Sr {\sc i} line at 4607 \AA\ contains the Sr {\sc i} ground level $^1{\rm S}_0$ and the line's upper level $^1{\rm P}_1$, whose angular momentum values are $J_\ell=0$ and $J_u=1$, respectively. Given that we are dealing with a resonance line, the two-level atom approximation is excellent (we have checked this via 1D radiative transfer calculations applying the multilevel code described in \cite{delPinoTrujillo2017}). Of the four stable isotopes of strontium, one has hyperfine structure with a relative abundance of only 7\%. In our 3D radiative transfer investigations the calculation of the emergent Stokes profiles has been carried out assuming that 100\% of the strontium atoms has nuclear spin $I=0$, which we expect to be a suitable approximation. Therefore, in our radiative transfer modeling the only level of the Sr {\sc i} 4607 \AA\ line that can be polarized is the upper level whose radiative lifetime is $t_{\rm life}{\approx}1/A_{ul}$, with $A_{ul}\,{\approx}\,2\times10^8 \, {\rm s}^{-1}$ the transition's Einstein coefficient for spontaneous emission. For the abundance of strontium in the solar atmosphere we have taken ${\rm A}_{\rm Sr}=2.9$ \citep[i.e., the photospheric value given by][]{AndersGrevesse1989}.

\begin{figure}[htp]
\centering 
\includegraphics[width=.6\textwidth]{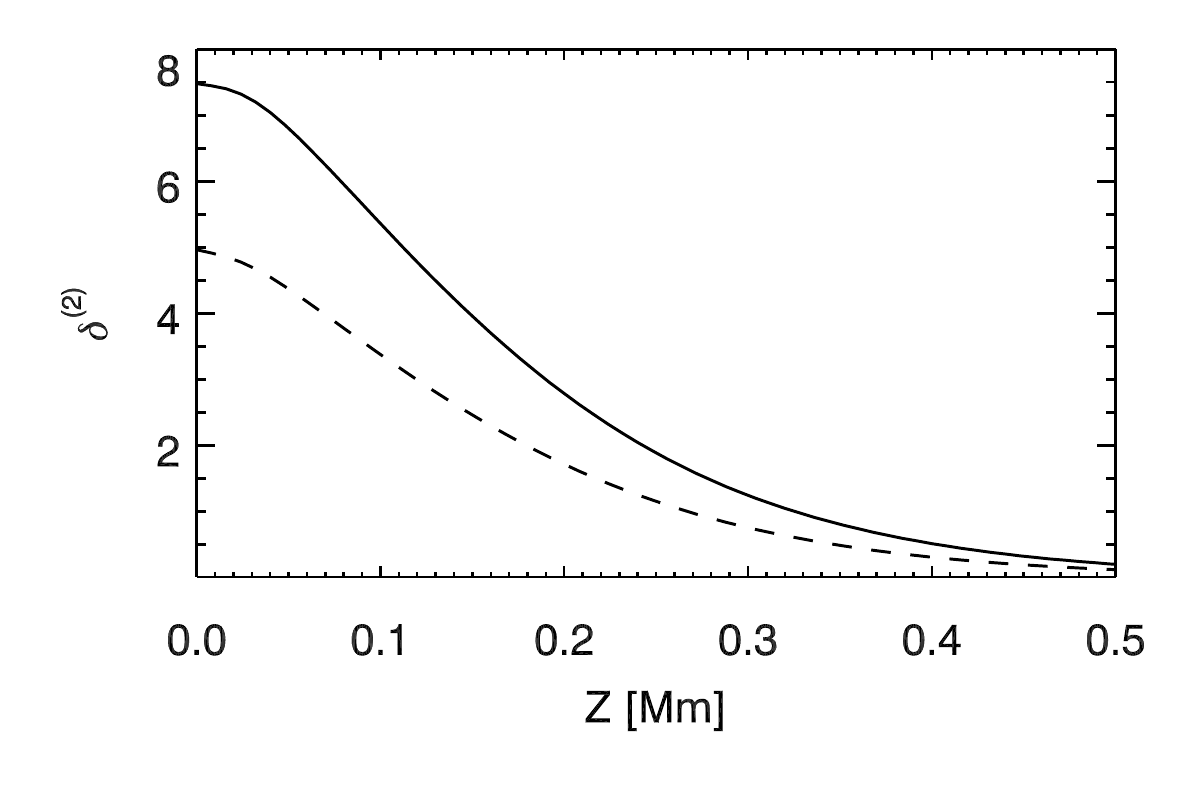}
\caption{At each height of the 3D model the figure shows the ratio between the horizontally-averaged rate of depolarizing elastic collisions with neutral hydrogen atoms and the line's Einstein coefficient for spontaneous emission. The solid curve corresponds to the elastic collisional rates given by \cite{Faurobertetal1995}, while the dashed curve to those given by \cite{Mansoetal2014}. Note that $\delta^{2}$ decreases with height and that it is of the order of unity between 200 and 400 km, which corresponds to the height range where the line-center optical depth of the Sr {\sc i} 4607 \AA\ line is unity for lines of sight with ${0.1}\,{\le}\,{\mu}\,{\le}\,{0.6}$.}
\label{F-delta} 
\end{figure}

If depolarization by elastic collisions with neutral hydrogen atoms were negligible for the Sr {\sc i} 4607 \AA\ line, then the critical magnetic field for the onset of the Hanle effect would be

\begin{equation}
B_{\rm H}\,=\,1.137{\times}10^{-7}A_{ul}/g_{u}\,=\,23 {\rm G}, \label{E-BH}
\end{equation}
where $g_u=1$ is the Land\'e factor of the line's upper level. However, the center of the Sr {\sc i} line under consideration originates in the bulk of the solar photosphere, approximately between 200 and 400 km above the corrugated surface of continuum optical depth unity for the disk center line of sight. In such a height-range the neutral hydrogen number density varies between $10^{17}$ ${\rm cm}^{-3}$ and $10^{16}$ ${\rm cm}^{-3}$, approximately, and ${\delta}^{(2)}_u{\approx}D_u^{(2)}/A_{ul}$ (i.e., the upper-level rate of elastic collisions in units of the Einstein $A_{ul}$ coefficient) is of order unity (see Fig. \ref{F-delta}). Therefore, since the effect of collisional quenching is significant for the Sr {\sc i} resonance line, the critical magnetic field for the onset of the Hanle effect in this line is actually larger, as indicated by the following approximate expression \citep{Trujillo2003}

\begin{equation}
B_{\rm c}\,{\approx}\,(1+{\delta}^{(2)}_u)B_{\rm H}. \label{E-BHdelta}
\end{equation}

The $D_u^{(2)}$ collisional rate is proportional to the neutral hydrogen number density, which increases approximately exponentially with depth in the solar atmosphere (\citealt{LambTerHaar1971}; see Eq. 7.108 of \citealt{BLandiLandolfi2004}). In this investigation we have calculated the Stokes profiles of the Sr {\sc i} 4607 \AA\ line using for $D_u^{(2)}$ the expression given by \cite{Faurobertetal1995}, which is the same one used in all our previous 3D radiative transfer investigations. As pointed out by \cite{ShchukinaTrujillo2011} the ensuing elastic collisional rates coincide with those obtained applying the semi-classical theory of \cite{AnsteeOmara1995}. Significantly smaller elastic collisional rates are obtained using the expressions for $D_u^{(2)}$ given by \cite{Mansoetal2014}, which results from ab-initio quantum mechanical calculations (see dashed curve of Fig. \ref{F-delta}). Such lower elastic collisional rates are similar to those provided by \cite{Kerkeni2002}, as well as to the rates obtained applying equation (7.108) of \cite{BLandiLandolfi2004}. 
In Section \ref{S-magnetization} we show the scattering polarization amplitudes that result from the $D_u^{(2)}$ collisional rates of \cite{Faurobertetal1995} and \cite{Mansoetal2014}, as well as their comparison with the available observations.

As mentioned above, the two-level atom model is suitable for calculating the Stokes profiles of the Sr {\sc i} resonance line. Moreover, doing the calculations in the Hanle regime (i.e., neglecting the impact of the Zeeman effect on the line profile) and neglecting the effects of partial frequency redistribution are suitable approximations for calculating the line-center scattering polarization amplitudes of the Sr {\sc i} 4607 \AA\ line \citep{Alsinaetal2017}. Given that the Zeeman effect can be neglected in this investigation, the circular polarization is zero and, consequently, we do not have atomic level orientation (i.e., both the $J^1_Q$ components of the radiation field tensor and the $\rho^1_Q$ multipolar components of the atomic density matrix are zero). Therefore, since $J_u=1$ the number of multipolar components of the atomic density-matrix needed to specify the atomic excitation of the upper level at each point within the model atmosphere is six: $\rho^0_0(u)$, $\rho^2_0(u)$, ${\rm Re}[{\rho}^2_1(u)]$, ${\rm Im}[{\rho}^2_1(u)]$, ${\rm Re}[{\rho}^2_2(u)]$ and ${\rm Im}[{\rho}^2_2(u)]$. To obtain at each spatial grid point the self-consistent values of such atomic density-matrix elements, PORTA solves jointly the radiative transfer equation for the Stokes vector and the following statistical equilibrium equations \citep[c.f.,][]{MansoTrujillo2011}\footnote{There is a typing error in equations (9) of that 2011 paper, since the expressions must be with $+{\Gamma}$ instead of $-{\Gamma}$.}:

\begin{equation}
S^0_0\,=\,(1-\epsilon){\bar{J}}_0^0\,+\,{\epsilon}\,B_{\nu}, \label{E-S00}
\end{equation}

\begin{eqnarray}
     \label{E-SKQ}
  \left( \begin{array}{l} 
      {S^2_0} \\\\ 
      {{\rm Re}[{S}^2_1}] \\\\
      {{\rm Im}[{S}^2_1}] \\\\ 
      {{\rm Re}[{S}^2_2}] \\\\
      {{\rm Im}[{S}^2_2}]
    \end{array} \right) = 
    w^{(2)}_{J_uJ_l} \, {{(1-\epsilon)} \over { 1+{\delta}^{(2)}_u(1-\epsilon) }} 
  \left( \begin{array}{r}
      {\bar J}^2_0 \\\\
      {\rm Re}[{\bar J}{}^{2}_{1}] \\\\
      -{\rm Im}[{\bar J}{}^{2}_{1}] \\\\
      {\rm Re}[{\bar J}{}^{2}_{2}] \\\\
      -{\rm Im}[{\bar J}{}^{2}_{2}]
    \end{array} \right)\,
    +{\Gamma_u}\, {{(1-\epsilon)} \over { 1+{\delta}^{(2)}_u(1-\epsilon) }}\,
    \left( \begin{array}{cccccc}
        0 & M_{12} & M_{13} & 0 & 0 \\\\ 
        M_{21} & 0 & M_{23} & M_{24} & M_{25} \\\\
        M_{31} & M_{32} & 0 & M_{34} & M_{35} \\\\
        0 & M_{42} & M_{43} & 0 & M_{45} \\\\
        0 & M_{52} & M_{53} & M_{54} & 0 
      \end{array} \right) 
    \left( \begin{array}{l}
        {S^2_0} \\\\
        {{\rm Re}[{S}^2_1}] \\\\
        {{\rm Im}[{S}^2_1}] \\\\
        {{\rm Re}[{S}^2_2}] \\\\
        {{\rm Im}[{S}^2_2}]
     \end{array} \right)\,, 
\end{eqnarray}
where $w^{(2)}_{J_uJ_\ell}=1$ for the Sr {\sc i} 4607 \AA\ line and ${S_Q^K}\,=\,{\frac{2h{\nu}^3}{c^2}}{\frac{2{J}_\ell+1}{\sqrt{2{J}_u+1}}}{\rho}_Q^K$, with ${\rho}^K_Q$ the above-mentioned multipolar components of the upper-level density matrix normalized to the overall population of the transition's lower level (with $Q=0,\dots,K$). In these equations ${\Gamma_u}=8.79\times10^6g_{u}B/A_{u\ell}$ (with $B$ in gauss and $A_{u\ell}$ in ${\rm s}^{-1}$) and $\epsilon=C_{u\ell}/(A_{u\ell}+C_{u\ell})$ is the collisional destruction probability due to inelastic collisions with electrons (which is much smaller than unity in the region of formation of the Sr {\sc i} resonance line), with $C_{u\ell}$ the de-excitation rate of inelastic collisions. As seen in the Appendix, the $M_{ij}$-coefficients of the magnetic kernel ${\bf M}$ depend on the inclination ($\theta_B$) of the magnetic field vector with respect to the local vertical Z-axis and on its azimuth ($\chi_B$). These equations for the ${S_Q^K}$ multipolar components have a clear physical interpretation: in the absence of magnetic fields (i.e., ${\Gamma_u}=0$) each $K=2$ multipolar component of the upper-level density matrix is proportional to the corresponding multipolar component of the radiation field tensor ${\bar J}{}^{2}_{Q}$ (see Eqs. \eqref{E-JKQ} in the appendix); in the presence of a magnetic field the magnetic kernel ${\bf M}$ couples locally the $K=2$ components among them (the Hanle effect). One of the key points investigated in this work is the Hanle effect caused by the magnetic field of Rempel's (2014) 3D model of the solar photosphere. 

Once the self-consistent values of such ${S_Q^K}$ quantities have been computed iteratively at each spatial grid point we can obtain the line source function components using the following equations \citep{MansoTrujillo2011}:

\begin{eqnarray}
  \label{E-SIline}
S^{{\rm line}}_I&=\,S^0_0 + w^{(2)}_{J_uJ_l}\Big{\{}
  \frac{1}{2\sqrt{2}} (3 \mu^2-1)S^2_0 - \sqrt{3} \mu \sqrt{1-\mu^2} 
  ({\rm Re}[{S}^2_1]\cos \chi  - {\rm Im}[{S}^2_1]\sin \chi ) \\ \nonumber
  &\quad+ \frac{\sqrt{3}}{2} (1-\mu^2) ({\rm Re}[{S}^2_2]\cos 2\chi
  - {\rm Im}[{S}^2_2]\sin 2\chi) \Big{\}},
\end{eqnarray} 

\begin{eqnarray}
  \label{E-SQline}
S^{{\rm line}}_Q&=\, w^{(2)}_{J_uJ_l}\Big{\{} \frac{3}{2\sqrt{2}}(\mu^2-1) S^2_0 -
  \sqrt{3}  \mu \sqrt{1-\mu^2} (
  {\rm Re}[{S}^2_1]\cos \chi - {\rm Im}[{S}^2_1]\sin \chi ) \\ \nonumber
  &\quad- \frac{\sqrt{3}}{2} (1+\mu^2) ({\rm Re}[{S}^2_2]\cos 2\chi
  - {\rm Im}[{S}^2_2]\sin 2\chi)\Big{\}}, 
\end{eqnarray}
and 

\begin{eqnarray}
  \label{E-SUline}
S^{{\rm line}}_U&=\, w^{(2)}_{J_uJ_l}\sqrt{3} \Big{\{} \sqrt{1-\mu^2} (
  {\rm Re}[{S}^2_1]\sin \chi +  {\rm Im}[{S}^2_1]\cos \chi) \\ \nonumber
  &\quad+\mu ({\rm Re}[{S}^2_2]\sin 2\chi + {\rm Im}[{S}^2_2]\cos 2\chi)\Big{\}},
\end{eqnarray} 
where $\theta={\rm arccos}({\mu})$ and $\chi$ are the inclination with respect to the solar local vertical and azimuth of the ray, respectively, and $w^{(2)}_{J_uJ_l}=1$ for the Sr {\sc i} 4607 \AA\ line. We point out that in these equations the reference direction for Stokes $Q$ is in the plane formed by the ray's propagation direction and the vertical Z-axis.

Finally, we obtain the source-function components $S_I=r_{\nu}S^{\rm line}_I+(1-r_{\nu})S^{\rm c}_I$, $S_Q=r_{\nu}S^{\rm line}_Q$ and $S_U=r_{\nu}S^{\rm line}_U$ and solve the radiative transfer equations $dX/d{\tau}=X-S_X$ in order to obtain the Stokes profiles $X=I,Q,$ and $U$ of the emergent spectral line radiation.\footnote{The transfer equations for Stokes $I$, $Q$ and $U$ are decoupled because (a) the lower level of the Sr {\sc i} line is unpolarized ($J_l=0$) and (b) the Zeeman effect is not considered in this investigation.} Note that $r_{\nu}={\kappa_l}{\phi_{\nu}}/({\kappa_l}{\phi_{\nu}}+{\kappa}_c)$, with ${\kappa_l}$ the line-integrated opacity, ${\kappa}_c$ the continuum opacity, and $\phi_{\nu}$ the normalized Voigt profile that includes the wavelength shift caused by the Doppler effect. $S^{\rm c}_I = \epsilon_{c}/\kappa_c$ is the source function of the continuum, with $\epsilon_{c}$ the emissivity in the continuum. In our 3D radiative transfer calculations of the scattering polarization in the Sr {\sc i} 4607 \AA\ line we do not include the polarization of the continuum radiation caused by Rayleigh and Thomson scattering \citep[see][]{TrujilloShchukina2009}. This is a suitable approximation because at 4607 \AA\ the continuum polarization amplitude is much smaller than that of the line itself.

\section{The pumping radiation and the induced atomic level polarization}\label{S-JKQrhoKQ}

The radiation field tensors defined in the Appendix characterize the symmetry properties of the radiation field at each spatial point of the 3D model atmosphere. While ${\bar J}^0_0$ is the familiar mean intensity and ${\bar J}^2_0$ quantifies the radiation anisotropy (i.e., whether the illumination of the atomic system is predominantly vertical or horizontal), the other four tensors account for the breaking of the axial symmetry (with respect to the local vertical Z-axis) of the incident radiation at each point within the medium. In a plane-parallel, horizontally homogeneous and static model atmosphere the only way to break such an axial symmetry is through the presence of an inclined magnetic field. The same is true in a 1D dynamical model atmosphere when only the vertical component of the plasma's macroscopic velocity is considered (e.g., \citealt{Carlinetal2013}). Obviously, in the real solar atmosphere (and in the 3D model considered in this paper) the thermal and density horizontal inhomogeneities of the plasma and the spatial gradients of the non-radial components of the macroscopic velocities break the axial symmetry of the pumping radiation field without the need of the model's magnetic field \citep[see][and references therein]{StepanTrujillo2016}. Equation \eqref{E-SKQ} demonstrates the need for calculating correctly all such radiation field tensors. 

The so-called 1.5D approximation is not suitable for investigating the scattering line polarization in a 3D model. This can be easily understood by noting that in the zero-field reference case the only non-zero radiation field tensors would be ${\bar J}^0_0$ and ${\bar J}^2_0$, when only the vertical component of the macroscopic velocity is considered. Instead, in the full 3D case the above-mentioned six radiation field tensors are non-zero even if the non-radial velocity components are disregarded. In our 3D radiative transfer investigations with PORTA, we have taken into account all such symmetry breaking causes, including the non-radial velocity components. 

\begin{figure}[htp]
\centering 
\includegraphics[width=.36\textwidth]{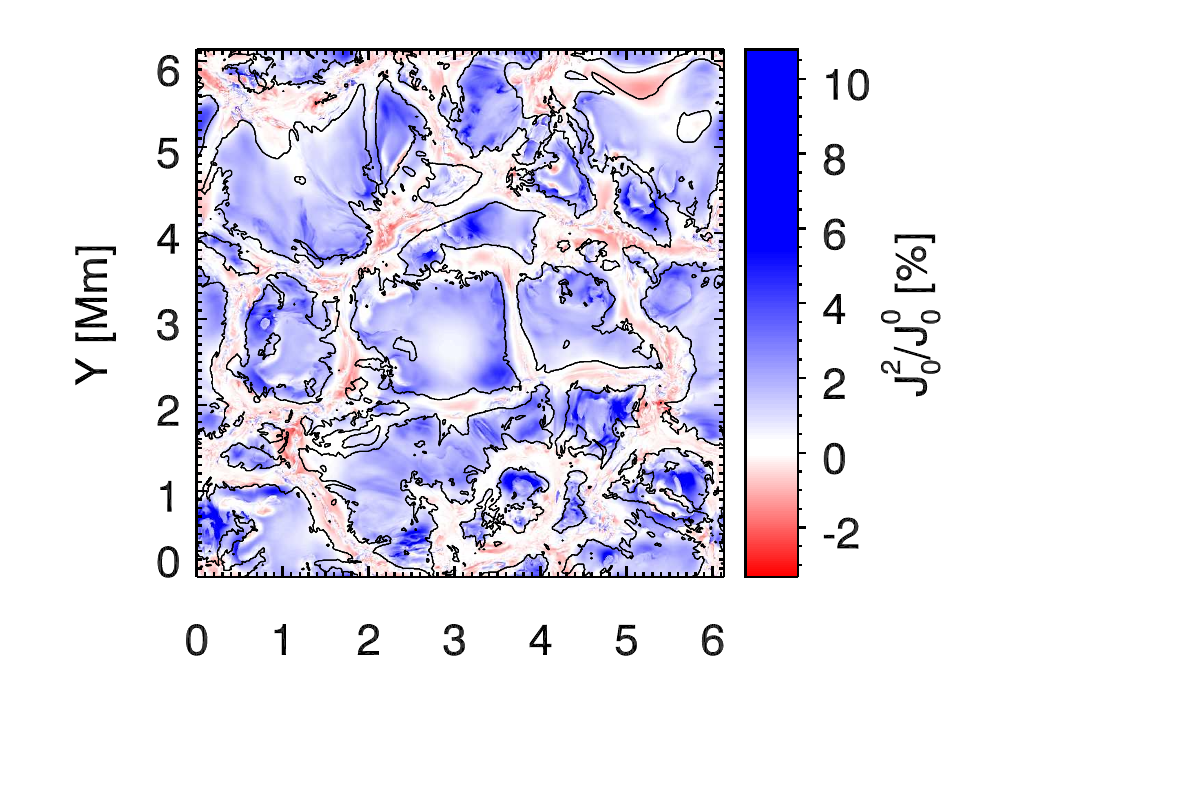}\hspace*{-6em}
\includegraphics[width=.36\textwidth]{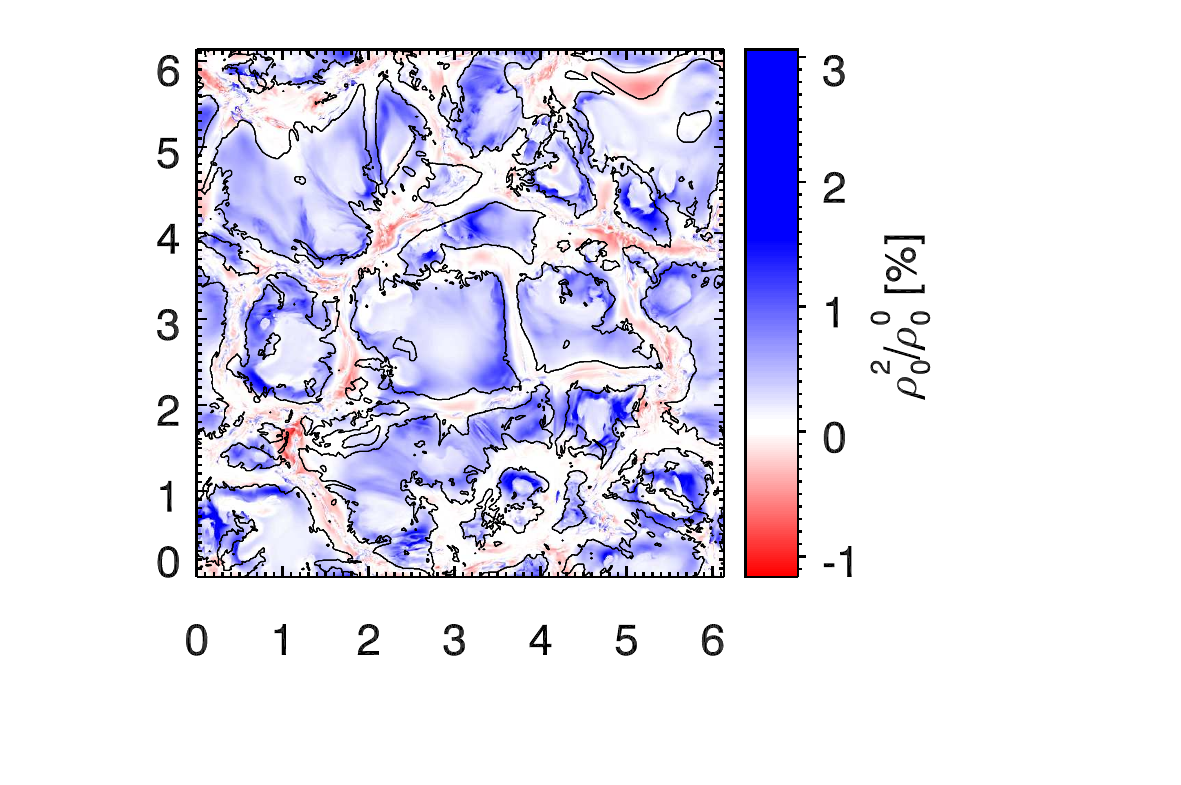}\hspace*{-6em}
\includegraphics[width=.36\textwidth]{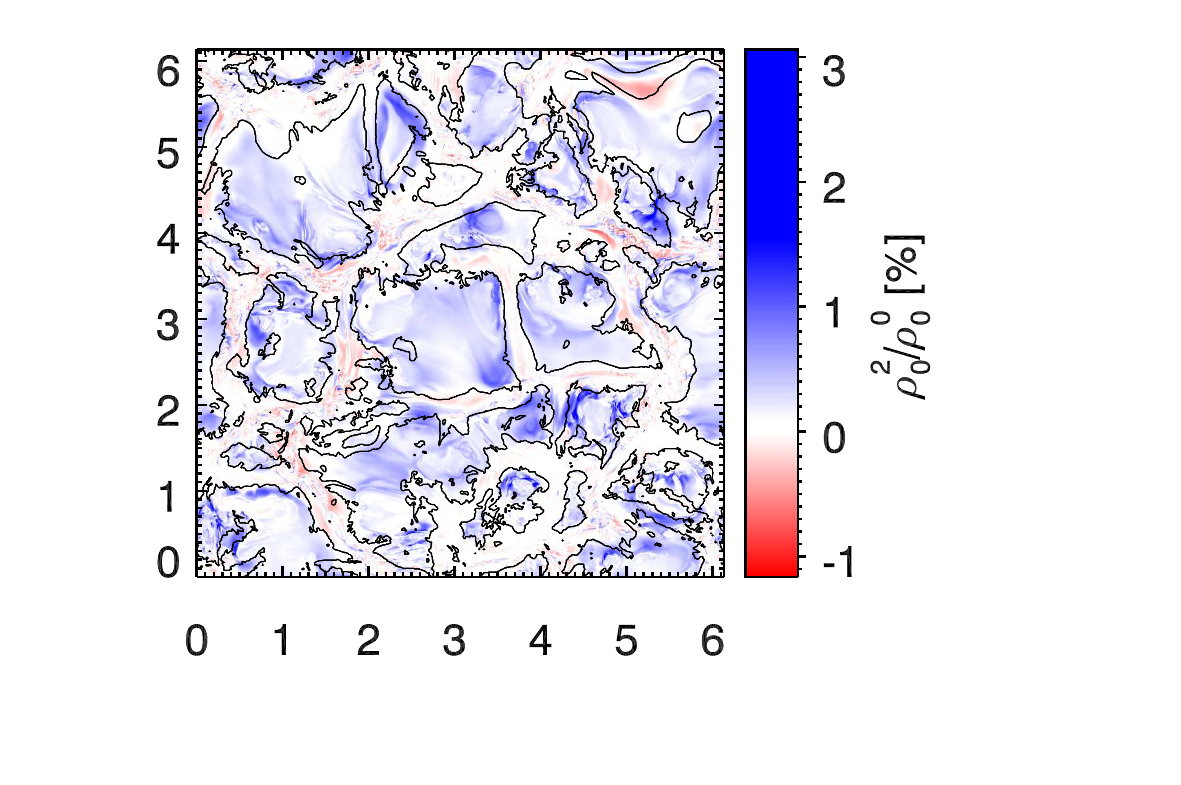}\\
\includegraphics[width=.36\textwidth]{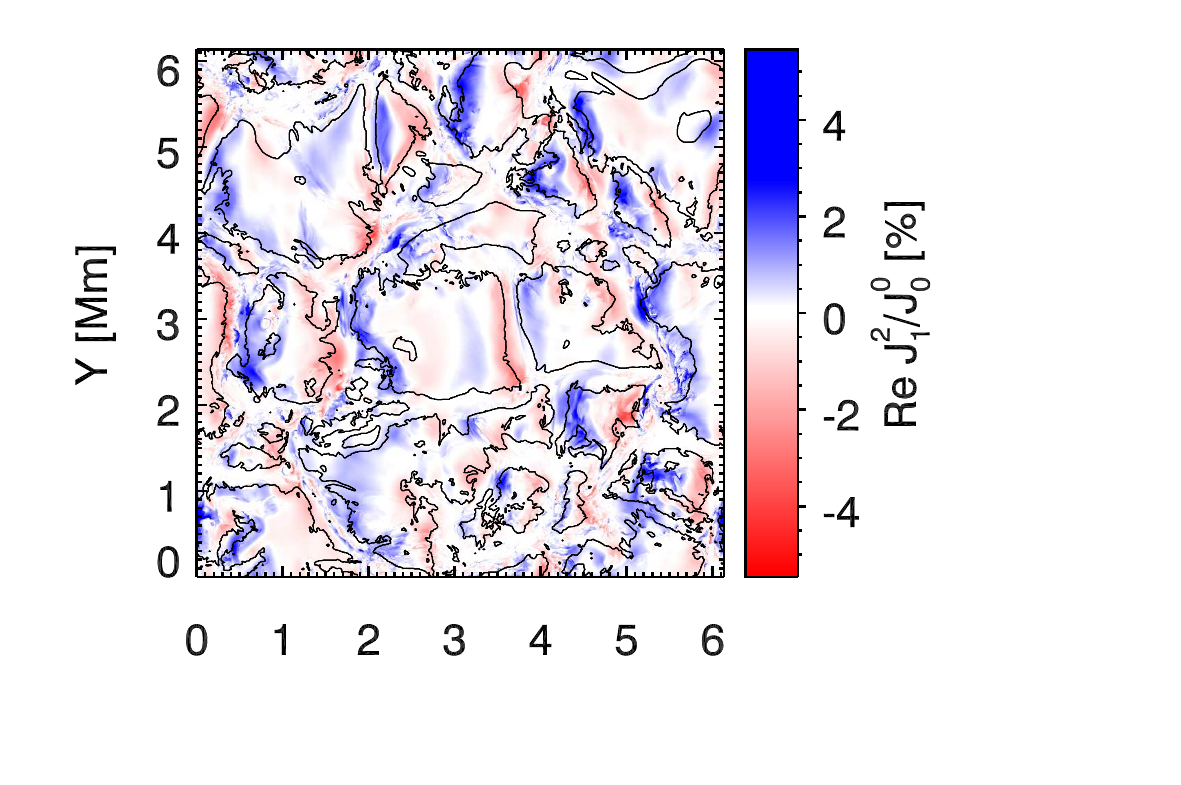}\hspace*{-6em}
\includegraphics[width=.36\textwidth]{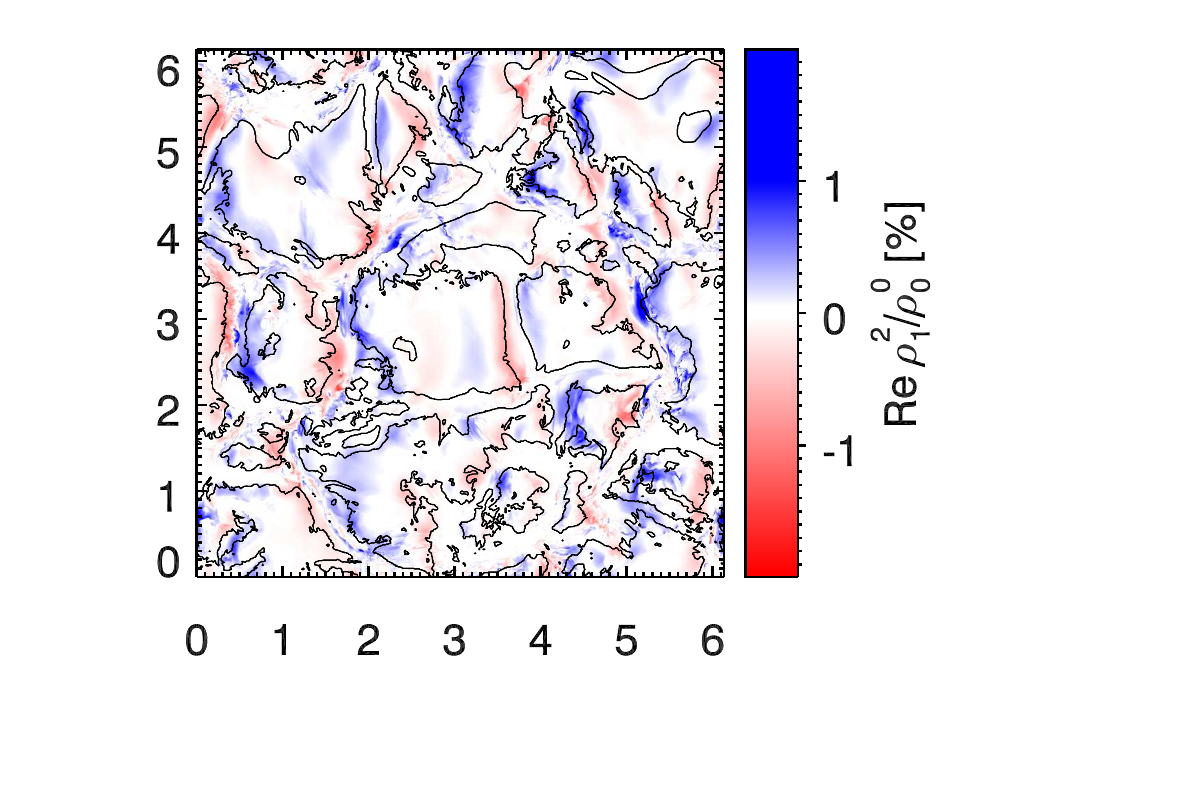}\hspace*{-6em}
\includegraphics[width=.36\textwidth]{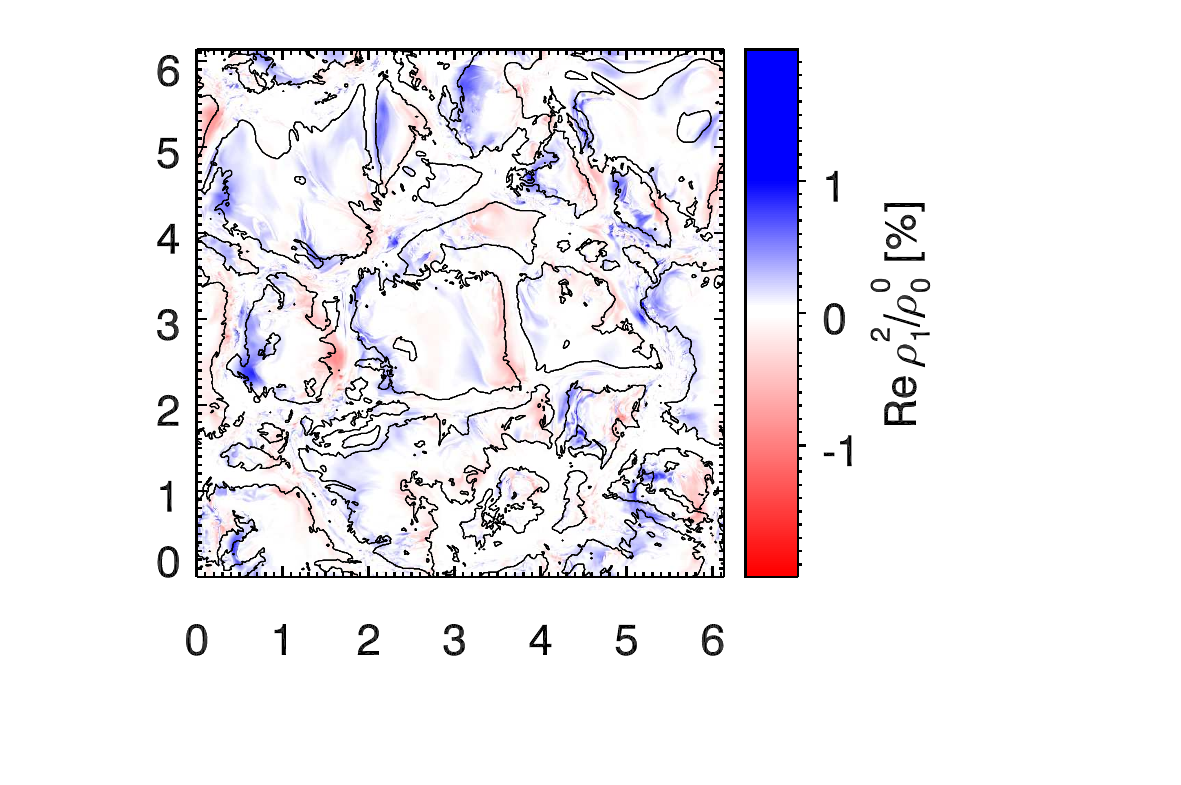}\\
\includegraphics[width=.36\textwidth]{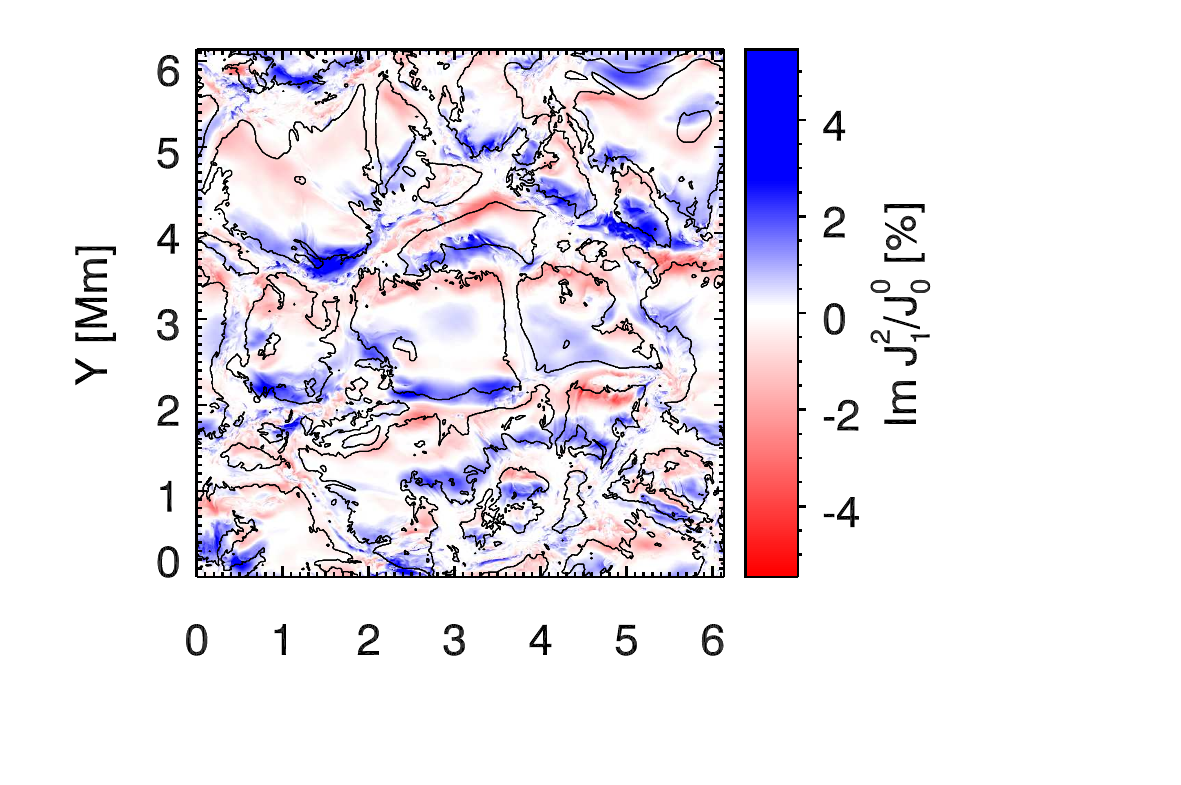}\hspace*{-6em}
\includegraphics[width=.36\textwidth]{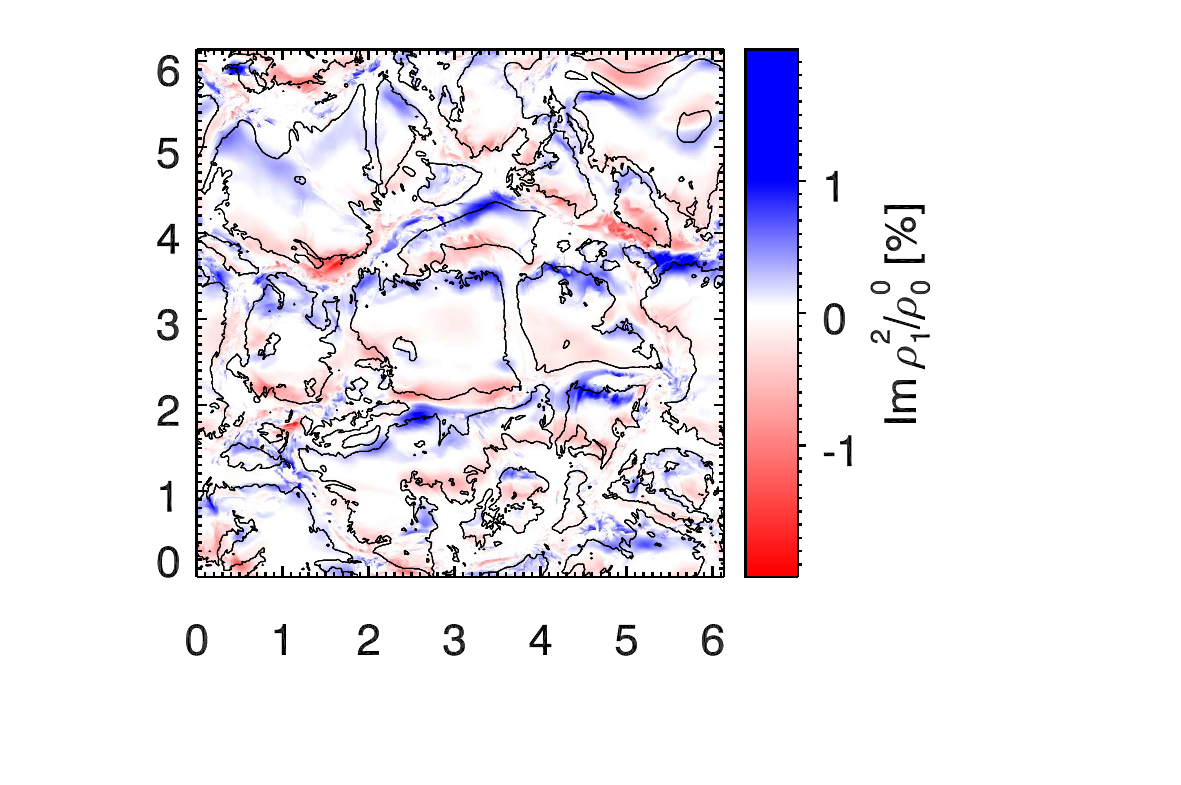}\hspace*{-6em}
\includegraphics[width=.36\textwidth]{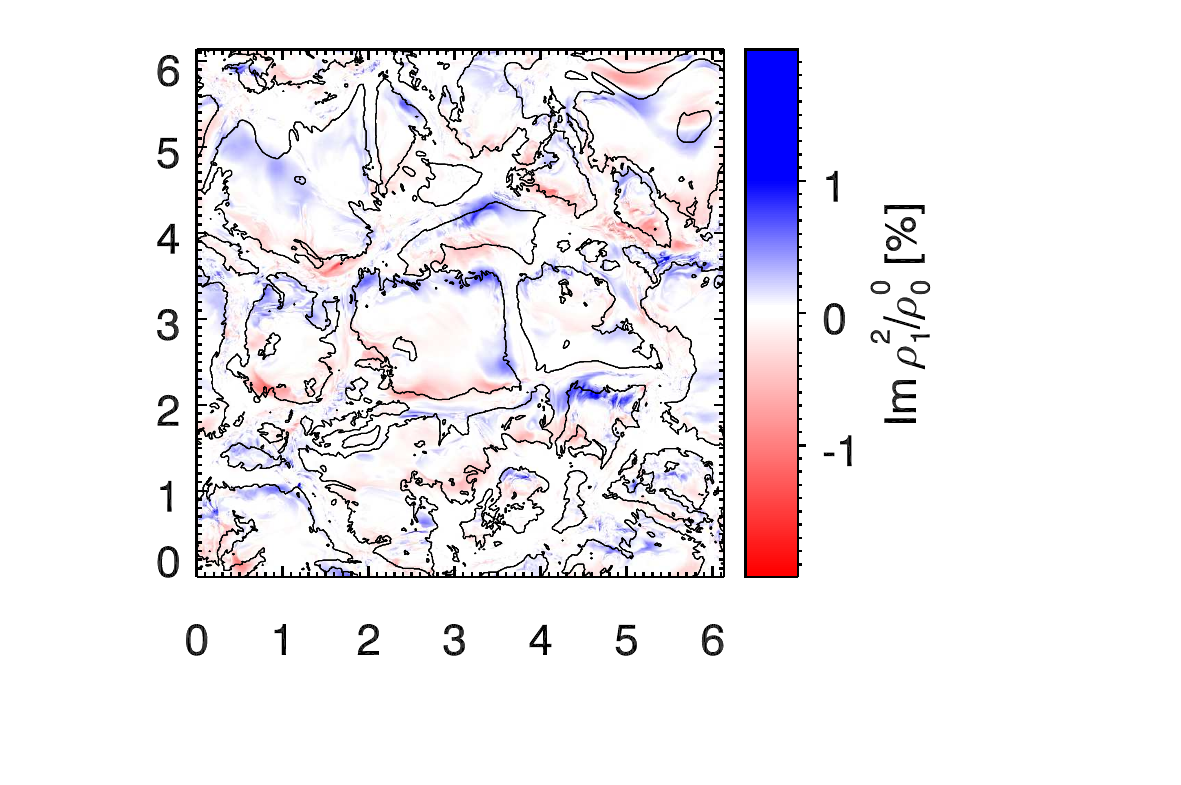}\\
\includegraphics[width=.36\textwidth]{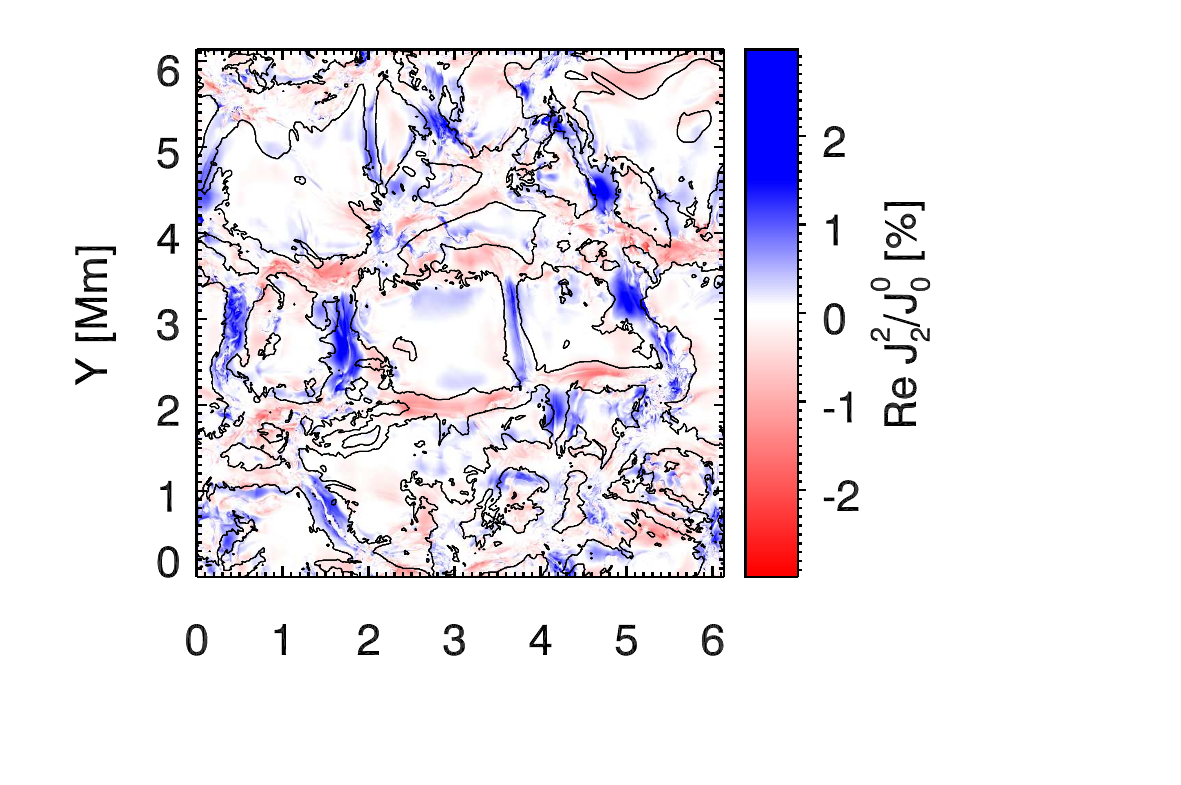}\hspace*{-6em}
\includegraphics[width=.36\textwidth]{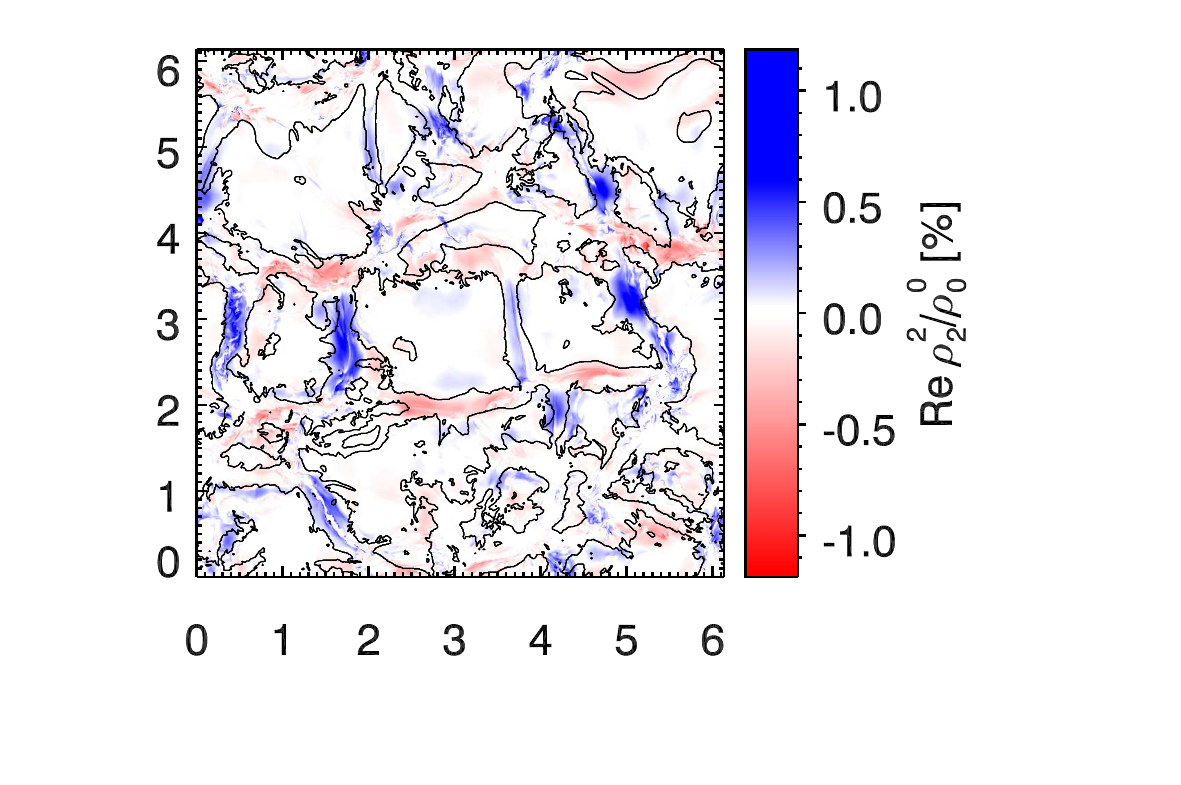}\hspace*{-6em}
\includegraphics[width=.36\textwidth]{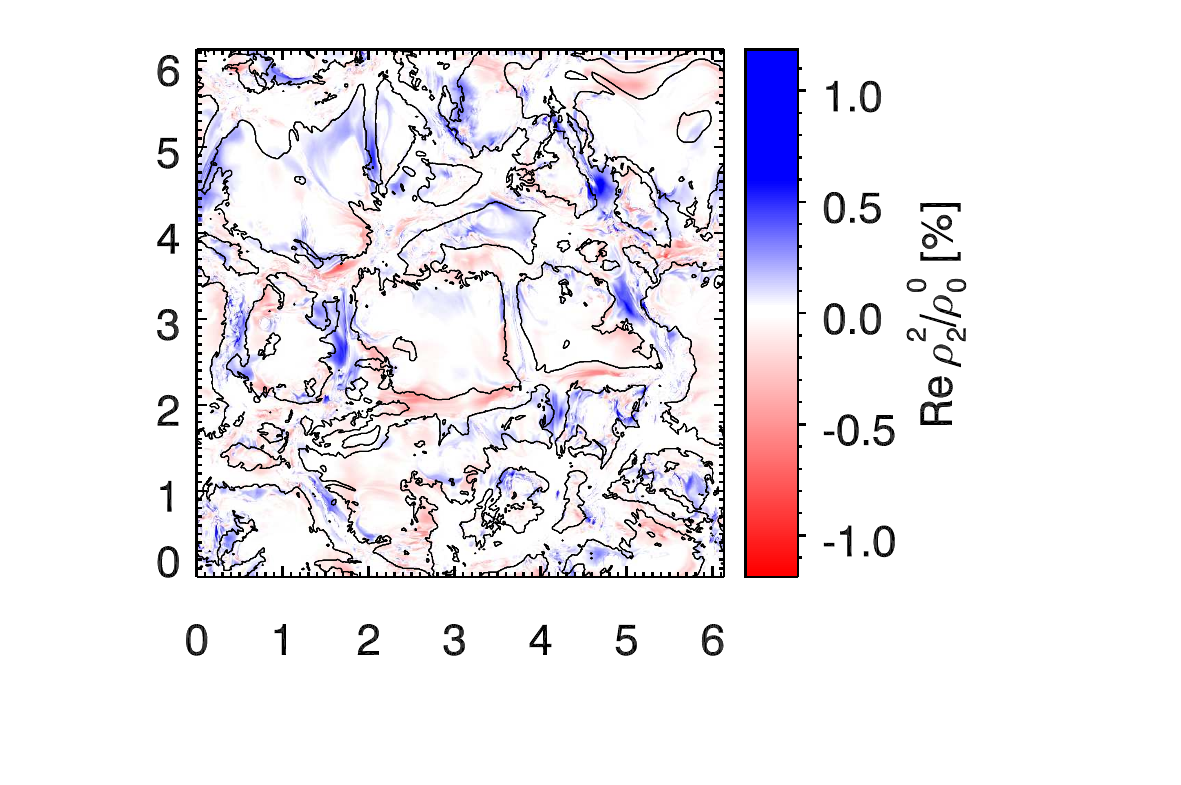}\\
\includegraphics[width=.36\textwidth]{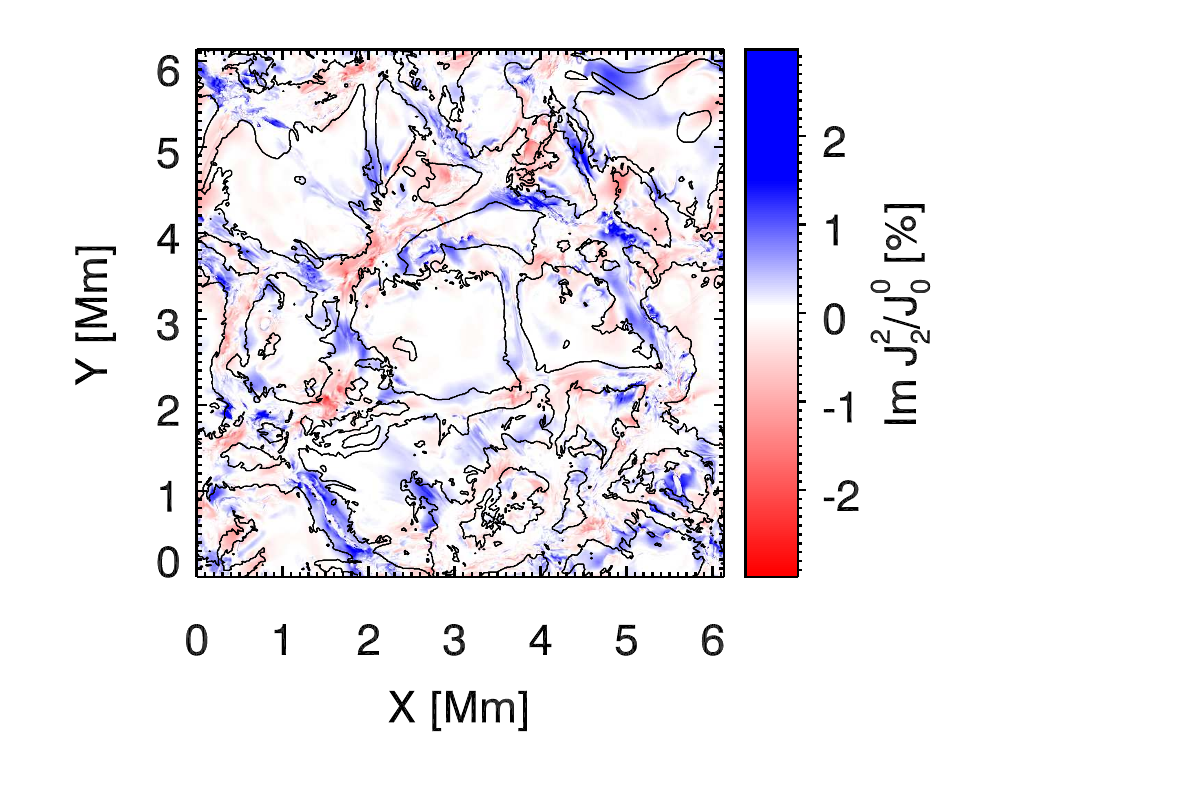}\hspace*{-6em}
\includegraphics[width=.36\textwidth]{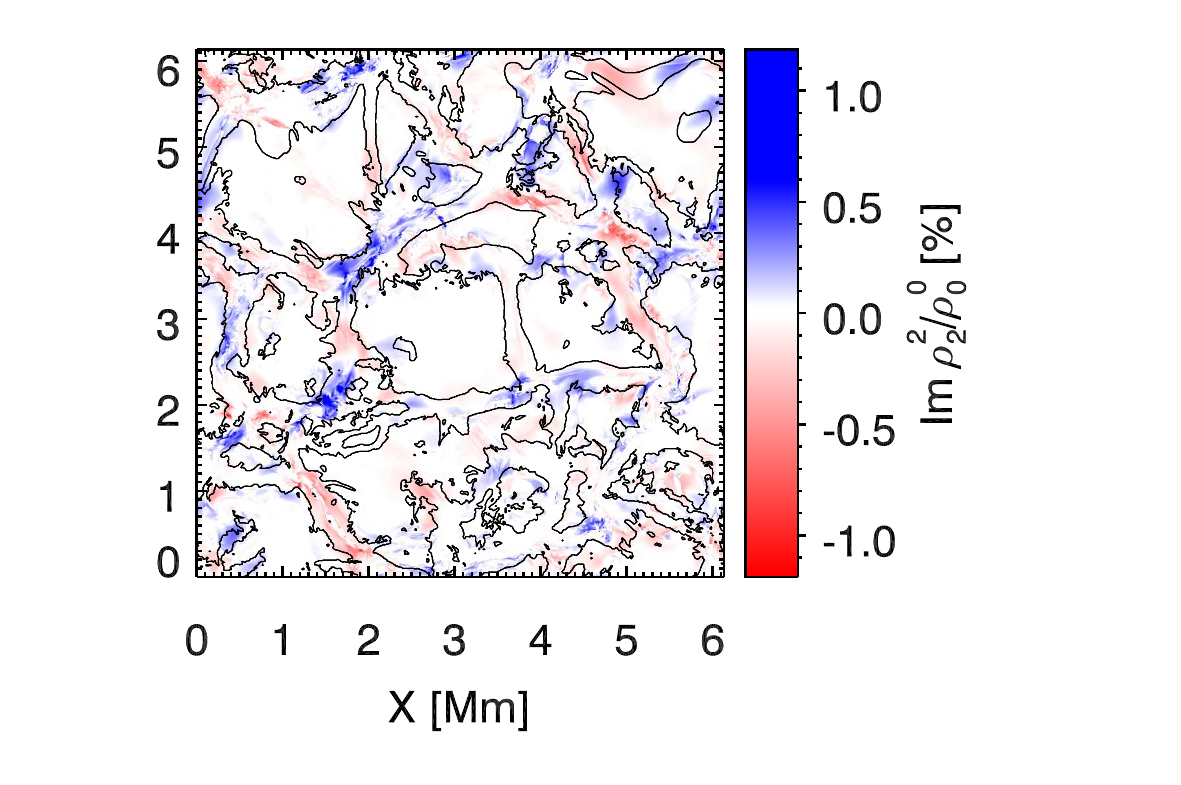}\hspace*{-6em}
\includegraphics[width=.36\textwidth]{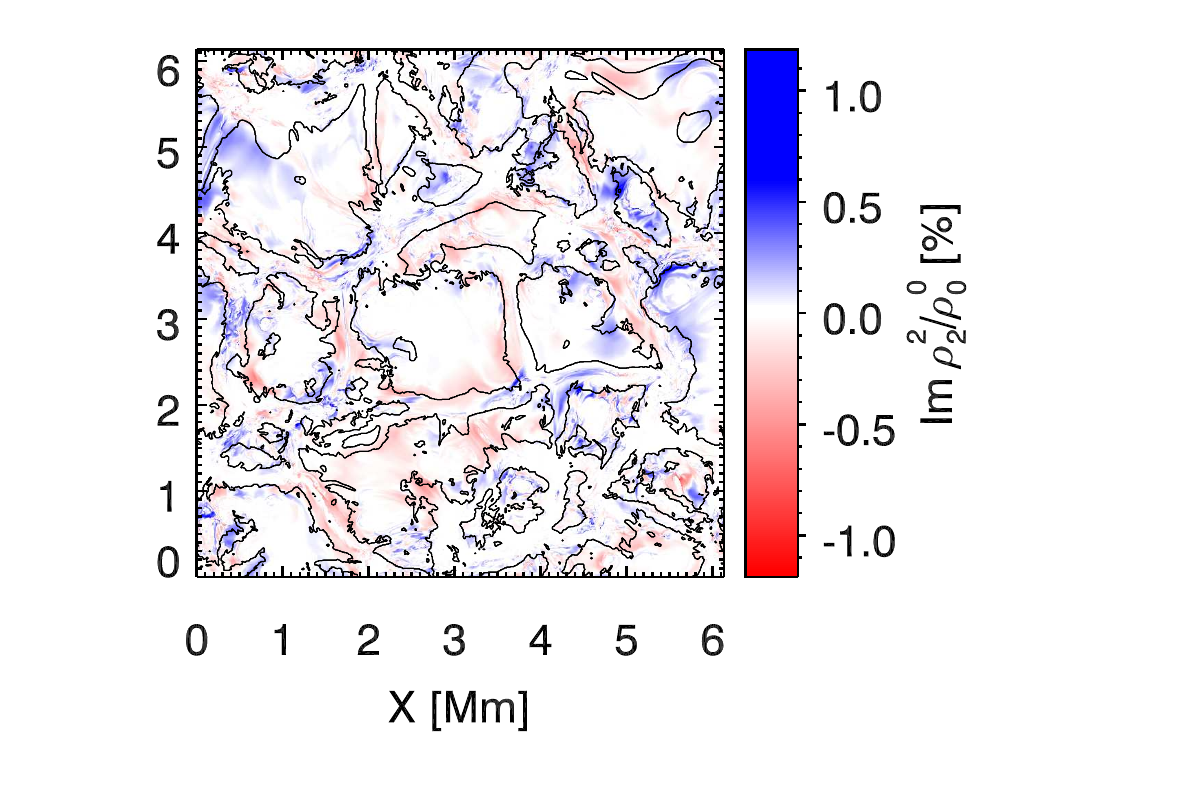}
\caption{Radiation field tensors normalized to the mean intensity $\bar{J}^0_0$ (left column) and density matrix elements of the upper level $u$ of the Sr {\sc i} 4607\AA\ line normalized to the population $\rho^0_0(u)$ component (middle and right columns), for the different multipolar components. The density matrix elements in the middle column are computed ignoring the model's magnetic field, while the ones in the right column take into account the Hanle effect due to the model's magnetic field. 
The contours delineate the upflowing regions at the model's visible surface. Note that the maximum and minimum values of the color table corresponding to the $K=2$, $Q=2$ panels are about a factor two smaller than those corresponding to the $K=2$, $Q=1$ panels.}
\label{F-JKQrhoKQ} 
\end{figure}

Figure \ref{F-JKQrhoKQ} shows the spatial variation of $\bar{J}^K_Q/\bar{J}^0_0$ and of the upper level's ${\rho}^K_Q/{\rho}^0_0$ at the heights tracing the corrugated surface where the line-center optical depth is unity for the disk-center line of sight. The left panels show the radiation field tensors, and we point out that virtually the same results are found when taking into account or neglecting the Hanle effect of the model's magnetic field. Obviously, the radiation field anisotropy ($\bar{J}^2_0/\bar{J}^0_0$) is practically magnetically insensitive, because at each point within the 3D model $Q\ll I$ and Stokes $I$ is virtually insensitive to the Hanle effect. Note that, although $\bar{J}^2_0/\bar{J}^0_0$ is more sizable inside the granular regions, the largest values are found around the granular-intergranular borders. Interestingly, in Rempel's (2014) 3D model of the solar photosphere the magnetic field has little effect on the breaking of the axial symmetry (with respect to the local vertical Z direction) of the spectral line radiation at each spatial grid point, which is instead dominated by Stokes $I$. Therefore, in this model the impact of the Hanle effect on the Stokes $Q$ and $U$ parameters only manifests through the coupling among the $S^K_Q$ quantities due to the magnetic kernel of Eq. \eqref{E-SKQ}. The impact of the Hanle effect on the $\rho^2_Q$ components is seen by comparing the right column of Fig. \ref{F-JKQrhoKQ} with the central column (zero-field reference case). Note that the ${\rho}_1^2$ and ${\rho}_2^2$ multipolar components are sizable mainly at the granular-intergranular borders and in the integranular lanes, and that in these regions the Hanle effect is particularly significant. In this 3D photospheric model the Hanle effect mainly depolarizes, but we point out that in a non-negligible fraction of its spatial points it increases the scattering  line polarization.

\section{The fully resolved case}\label{Sresolve}

With the self-consistently calculated values of the multipolar components of the atomic density matrix it is straightforward to obtain the $S_I$, $S_Q$ and $S_U$ source function components (see Section 2) and to solve the corresponding transfer equation to compute the $I(\lambda)$, $Q(\lambda)$ and $U(\lambda)$ profiles of the emergent radiation. Here we show the spatial variation of the Stokes signals calculated at the spatial resolution of the 3D model, considering three lines of sight characterized by $\chi=0^{\circ}$, and by $\mu=0.1$ (close to the limb observation), $\mu=0.5$ and $\mu=1$ (disk center observation). We consider the $Q/I$ and $U/I$ line-center amplitudes, as well as the $P$ amplitudes, where $P=\sqrt{Q^2+U^2}/I$. Finally, we take the opportunity to illustrate the important impact of the dynamical state of the 3D model on such linear polarization signals, highlighting the importance of the symmetry breaking caused by the non-radial velocity components.

\subsection{The $Q/I$, $U/I$ and $P/I$ signals}

\begin{figure}[htp]
\centering 
\includegraphics[width=.4\textwidth]{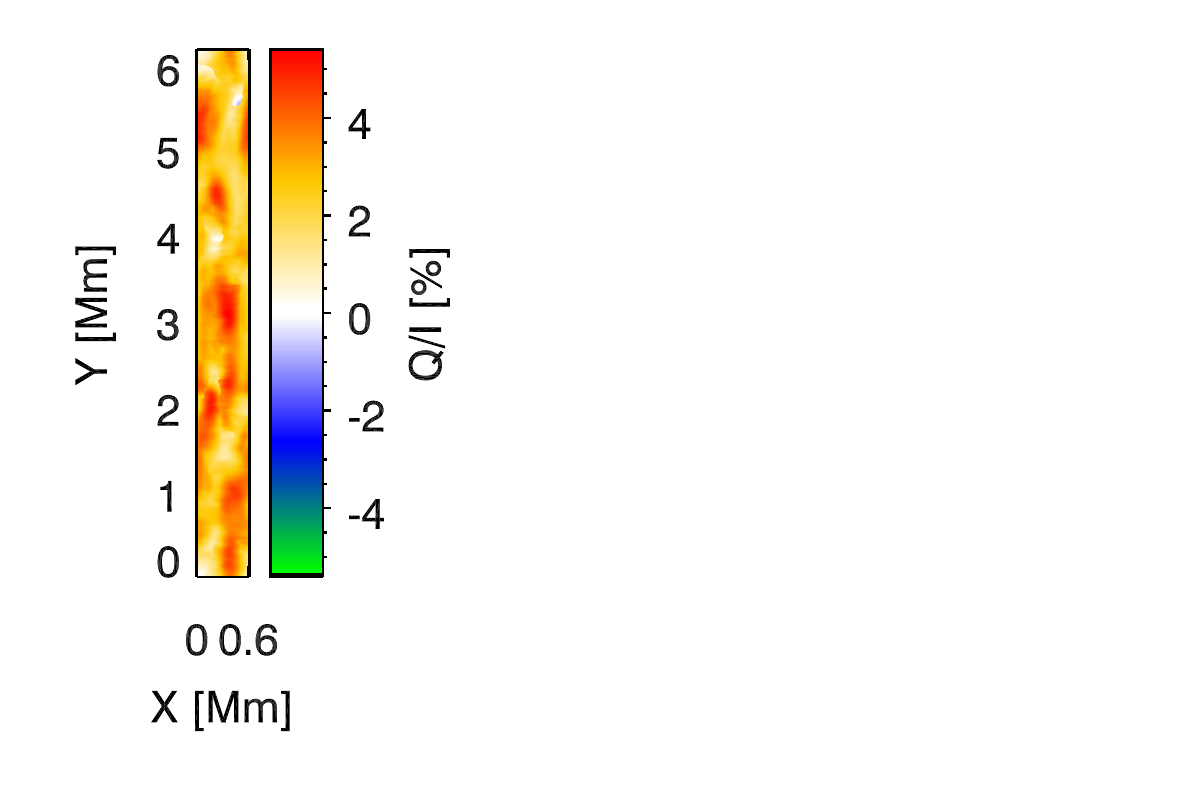}\hspace*{-14em}
\includegraphics[width=.4\textwidth]{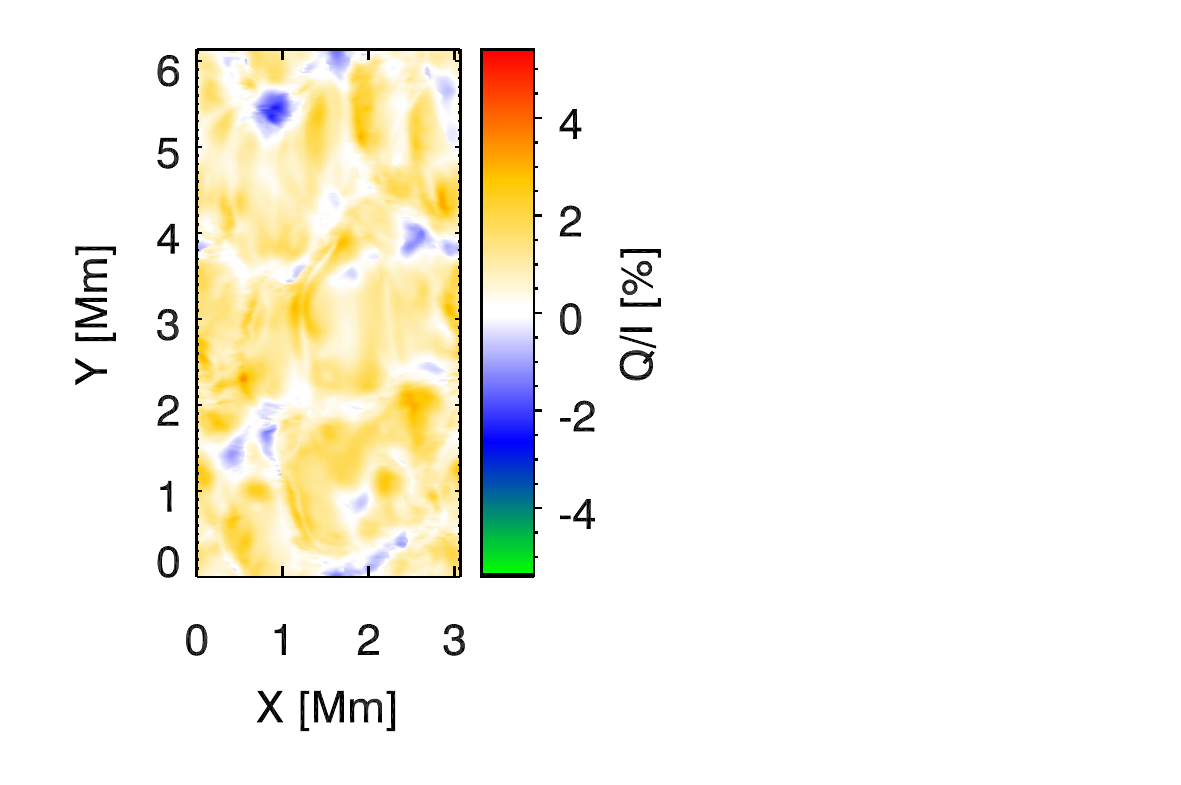}\hspace*{-10em}
\includegraphics[width=.4\textwidth]{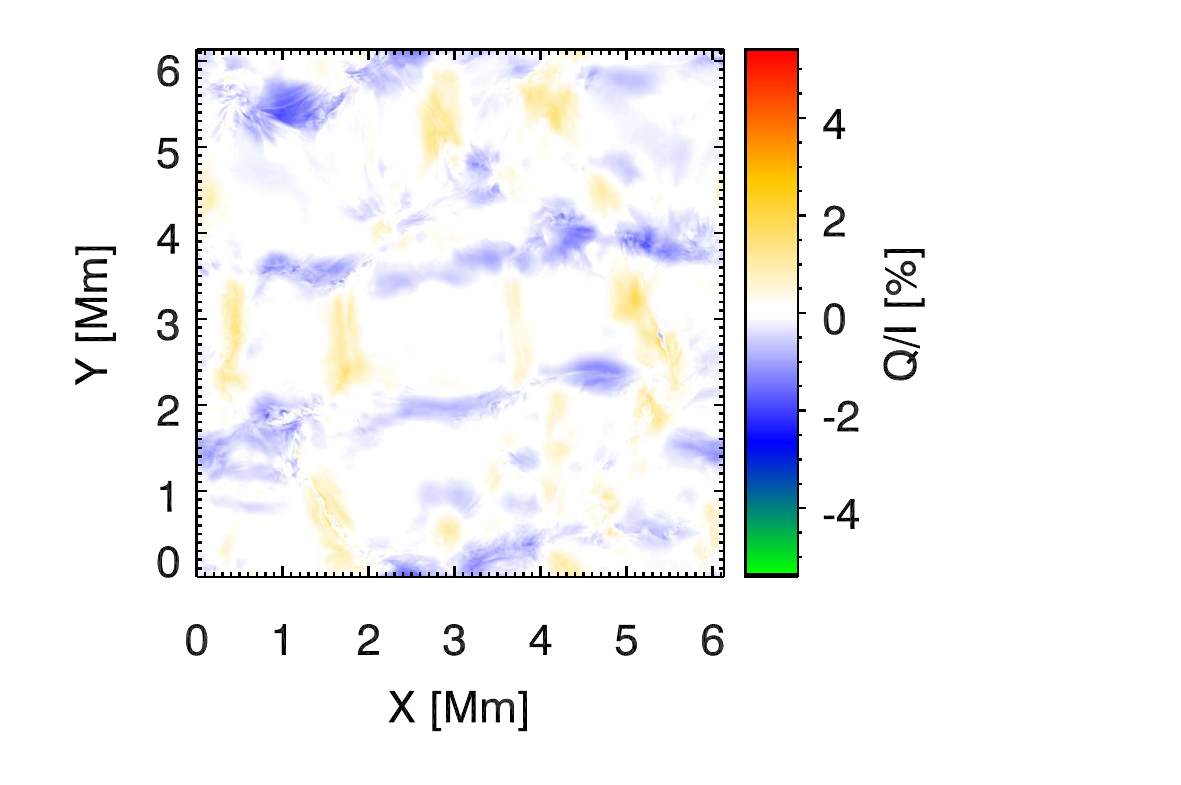}\\
\includegraphics[width=.4\textwidth]{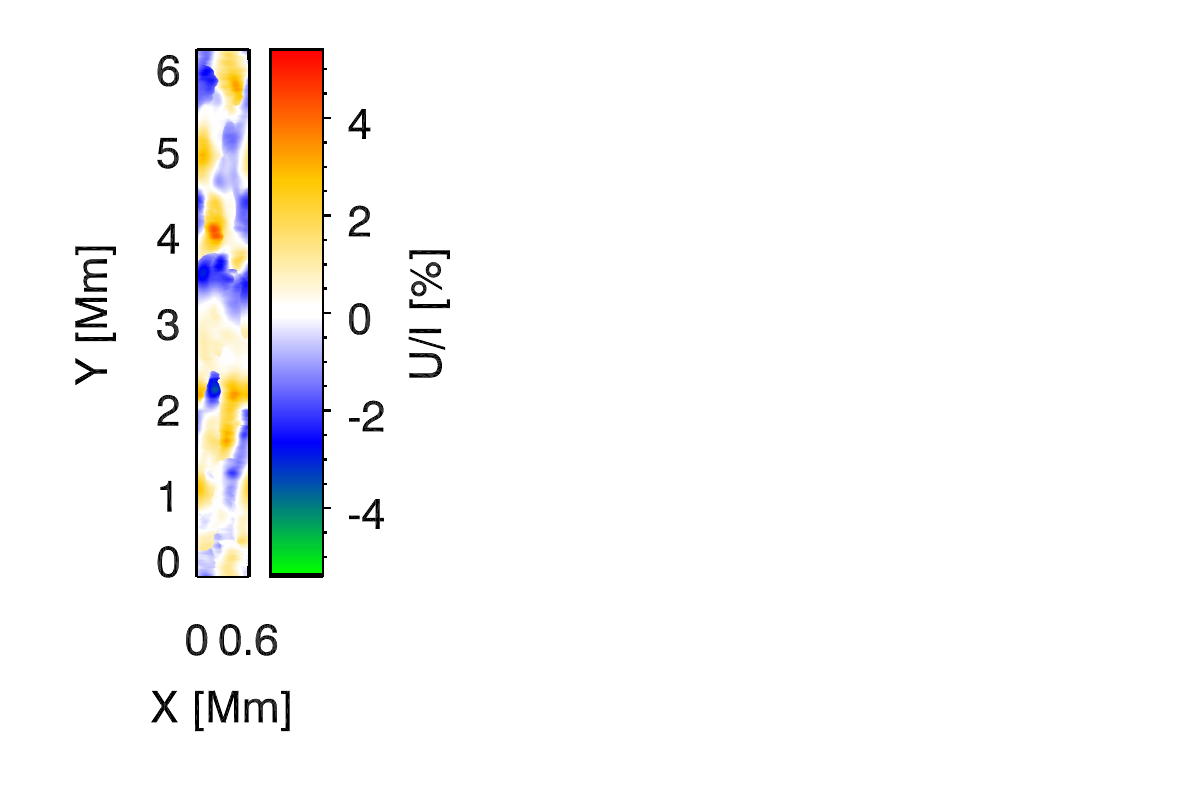}\hspace*{-14em}
\includegraphics[width=.4\textwidth]{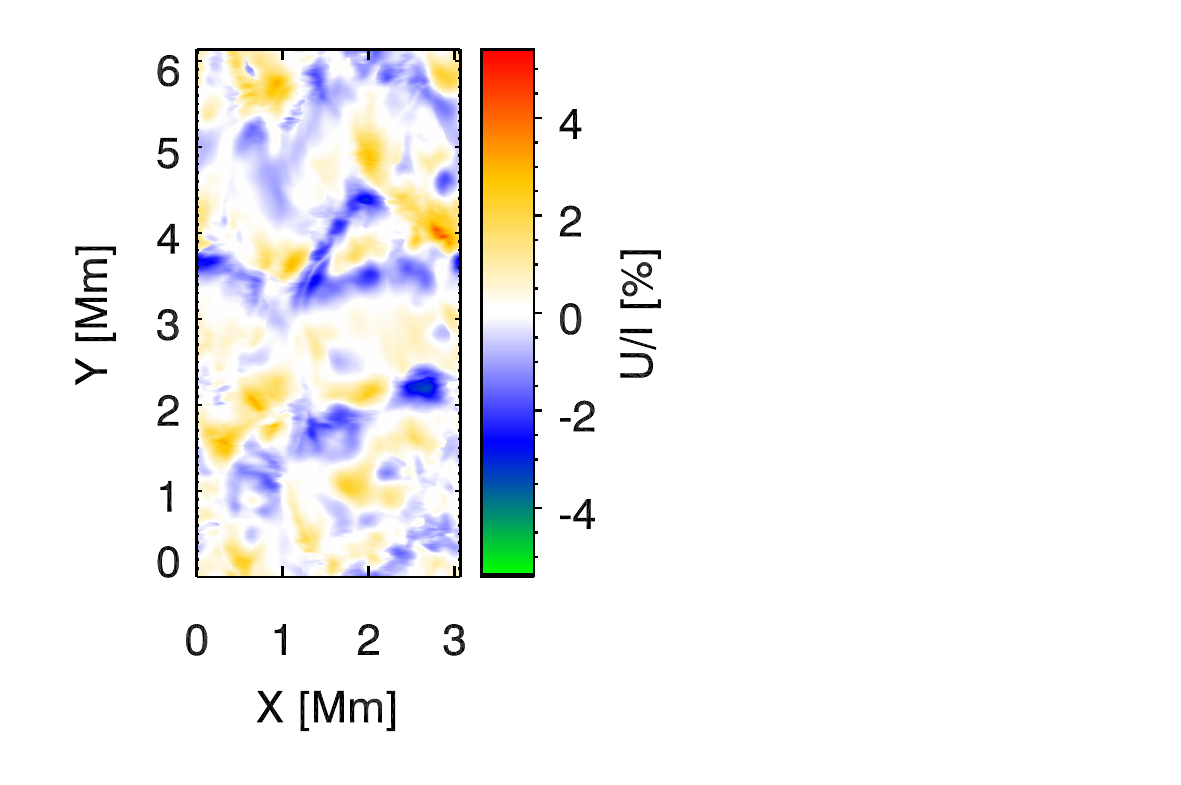}\hspace*{-10em}
\includegraphics[width=.4\textwidth]{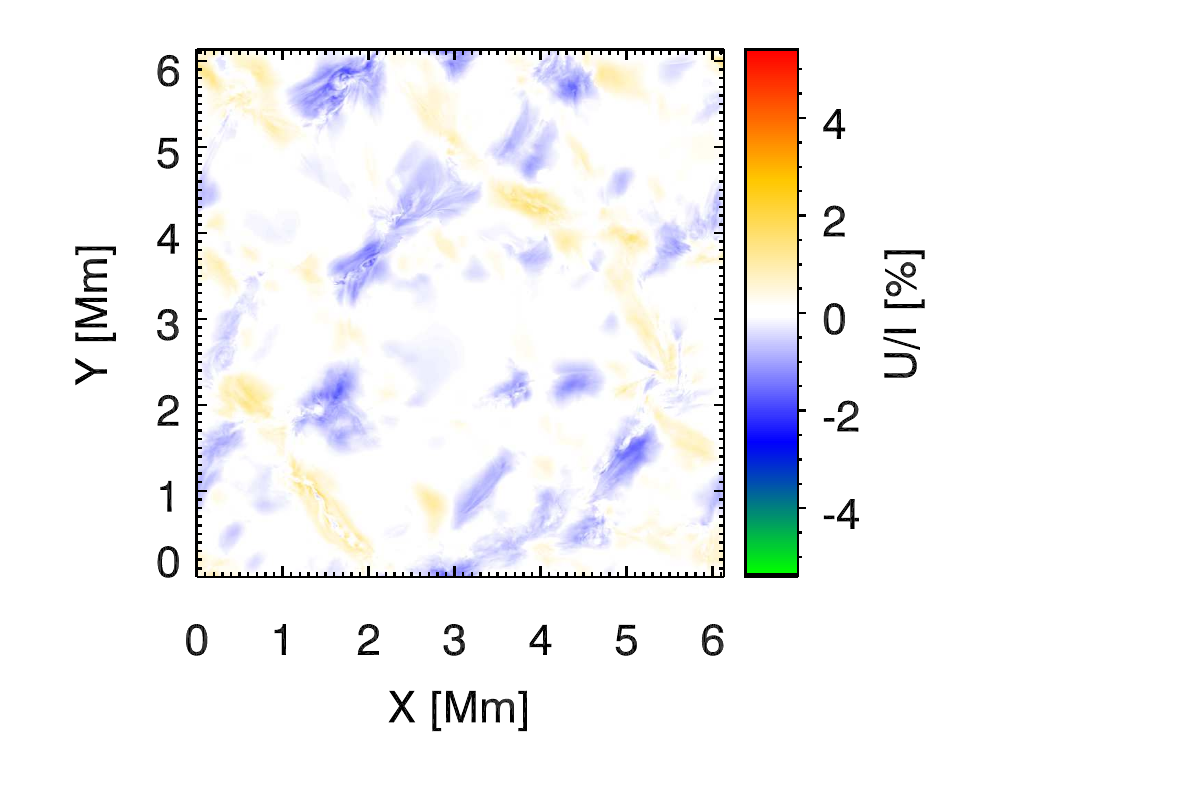}
\caption{Fractional linear polarization $Q/I$ (upper panels) and $U/I$ (lower panels) at the wavelength where the fractional total linear polarization $P$ is maximum at the spatial pixel under consideration, for lines of sight with $\mu=0.1$ (left column), $\mu=0.5$ (middle column), and $\mu=1$ (right column), neglecting the Hanle effect due to the magnetic field of the 3D atmospheric model.}
\label{F-QU} 
\end{figure}

Figure \ref{F-QU} shows the $Q/I$ (upper panels) and $U/I$ (lower panels) amplitudes neglecting the Hanle effect produced by the magnetic field of the 3D model atmosphere. Consider first the disk-center $Q/I$ and $U/I$ signals of the right panels of Fig. \ref{F-QU}, where both signs (positive and negative) are equally likely. They are due to the breaking of the axial symmetry of the incident radiation field at each point of the model, as can be easily understood by particularizing Eqs. \eqref{E-SIline} and \eqref{E-SQline} to the $\mu=1$ (with $\chi=0^{\circ}$) line of sight and by noting that $S^{{\rm line}}_Q{=}\sqrt{3}{\rm Re}[{S}^2_2]$ and $S^{{\rm line}}_U{=}-\sqrt{3}{\rm Im}[{S}^2_2]$, and that these  quantities are governed by Eqs. \eqref{E-SKQ}. In particular, notice that for the zero-field reference case 

\begin{equation}
{Q\over{I}}(\mu=1,\chi=0^{\circ})\,{\approx}\,{\sqrt{3}\over{(1+\delta^{(2)}_u)}}\,{{\rm Re}[{\bar J}^2_2]\over{\bar J}^0_0}\,, \label{E-QImu1}
\end{equation}
\begin{equation}
{U\over{I}}(\mu=1,\chi=0^{\circ})\,{\approx}\,{\sqrt{3}\over{(1+\delta^{(2)}_u)}}\,{{\rm Im}[{\bar J}^2_2]\over{\bar J}^0_0}\,, \label{E-UImu1}
\end{equation}
at the height in the model atmosphere where the line-center optical depth is unity along the line of sight. These approximate formulae explain why the patterns of the forward-scattering $Q/I$ and $U/I$ signals are similar to those seen in the ${{\rm Re}[{\bar J}^2_2]/{\bar J}^0_0}$ and ${{\rm Im}[{\bar J}^2_2]/{\bar J}^0_0}$ panels of Fig. \ref{F-JKQrhoKQ}, with positive and negative values across the field of view.

As seen in Fig. \ref{F-QU}, the $Q/I$ amplitudes increase when going from the disk center case (right panels) to the close to the limb case (left panels), and the closer to the limb the more predominantly positive they are.  This can be easily understood by noting from Eqs. \eqref{E-SQline} and \eqref{E-SKQ} that ${\bar J}^2_0$ makes an increasing contribution to $S^{{\rm line}}_Q$ for $\mu{<}1$,  and that at the heights in the model atmosphere where the line-center optical depth is unity for each line of sight under consideration ${\bar J}^2_0/{\bar J}^0_0$ is predominantly positive. The fact that for all line of sights positive and negative signs are equally likely for the $U/I$ signals can be easily understood by noting that the ${\bar J}^2_0$ tensor does not appear in the $S^{{\rm line}}_U$ expression (see Eq. \eqref{E-SUline}).

\begin{figure}[htp]
\centering 
\includegraphics[width=.4\textwidth]{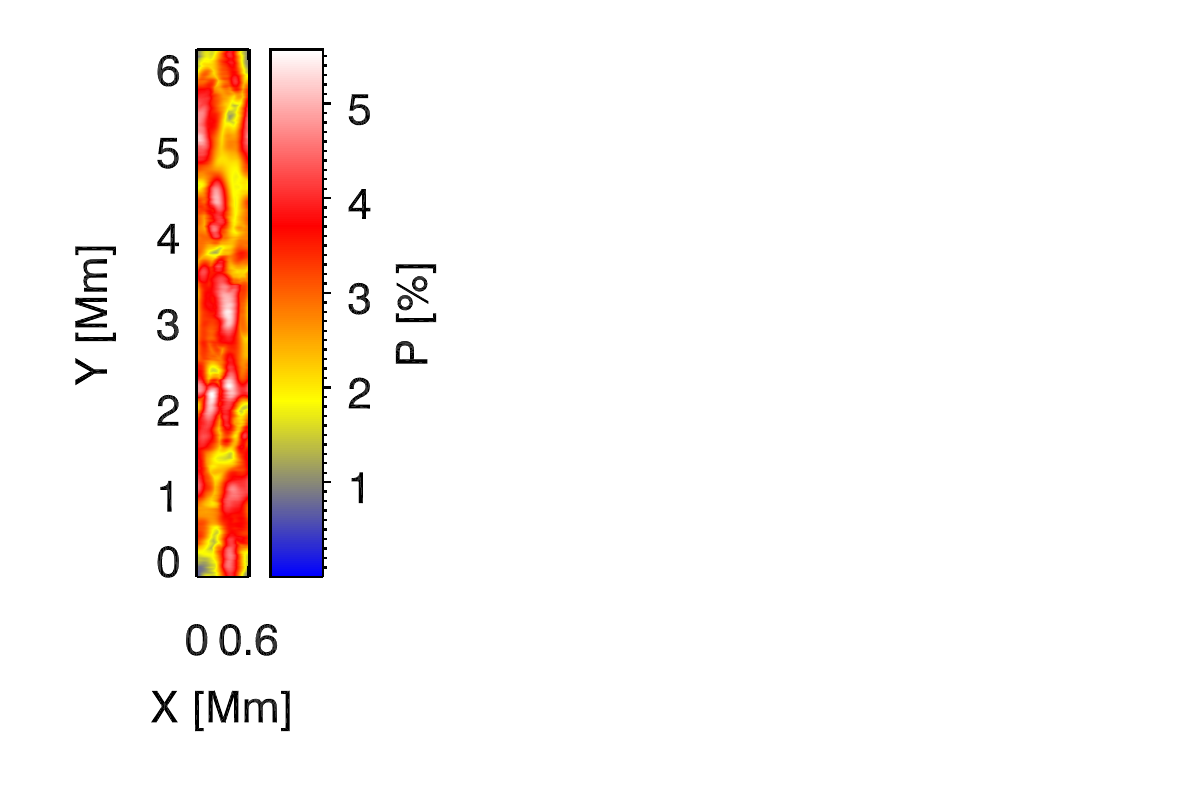}\hspace*{-14em}
\includegraphics[width=.4\textwidth]{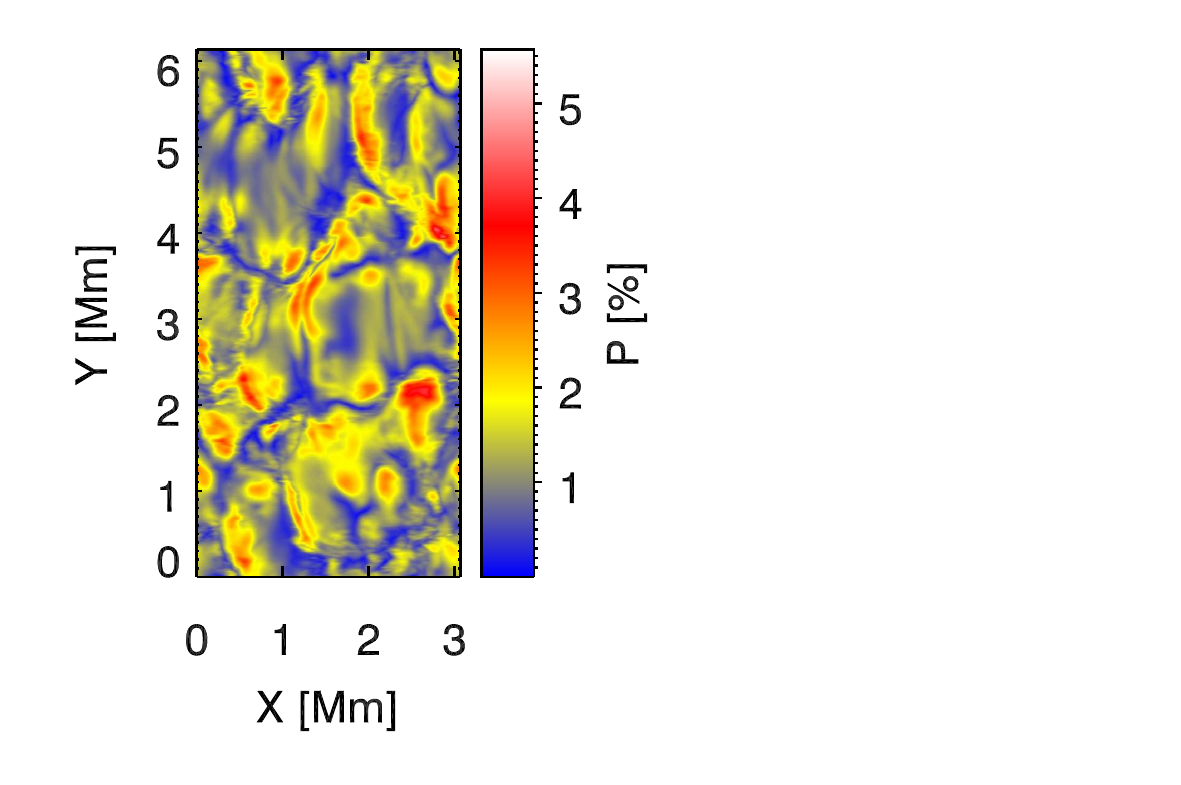}\hspace*{-10em}
\includegraphics[width=.4\textwidth]{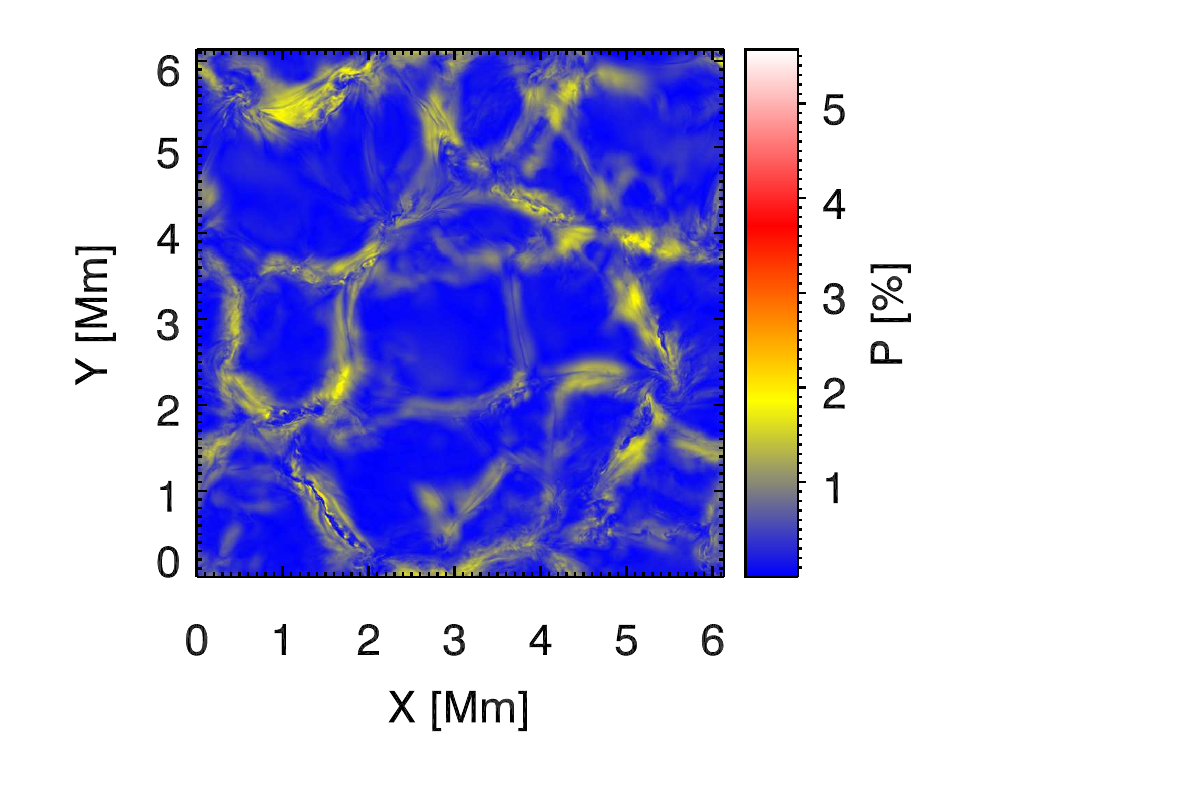}\\
\includegraphics[width=.4\textwidth]{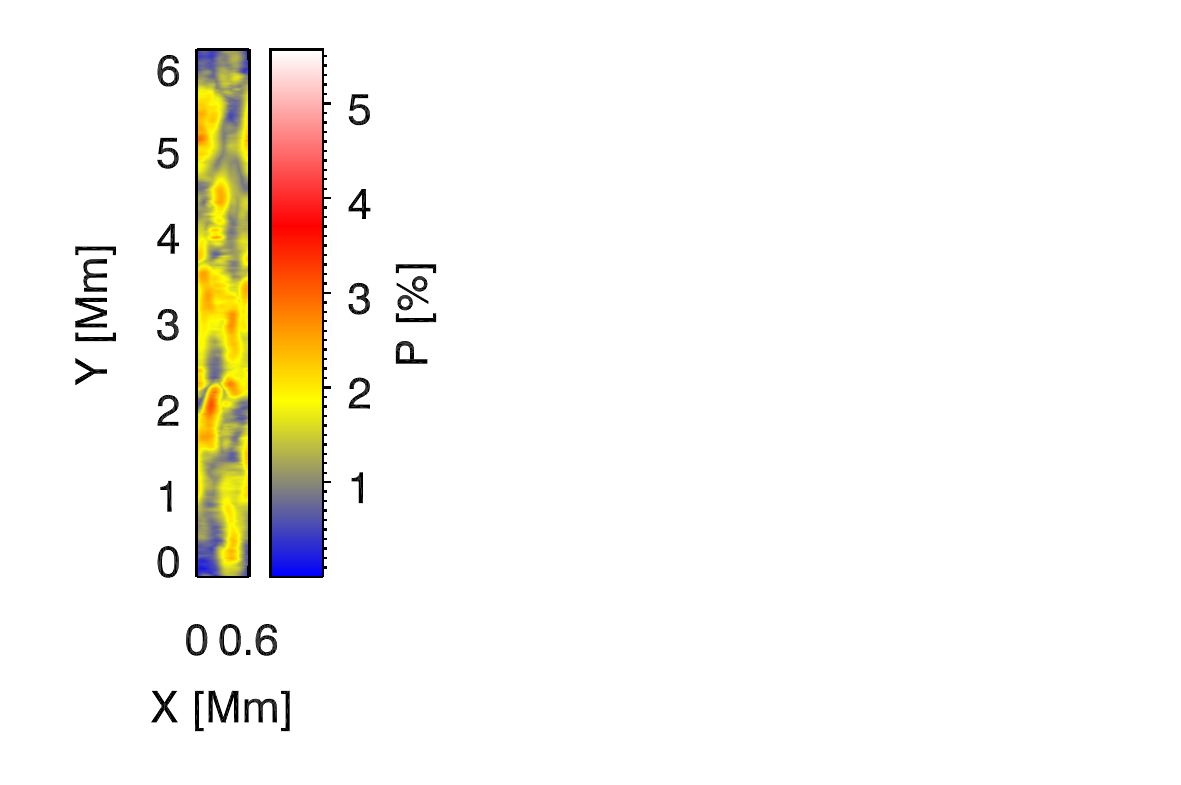}\hspace*{-14em}
\includegraphics[width=.4\textwidth]{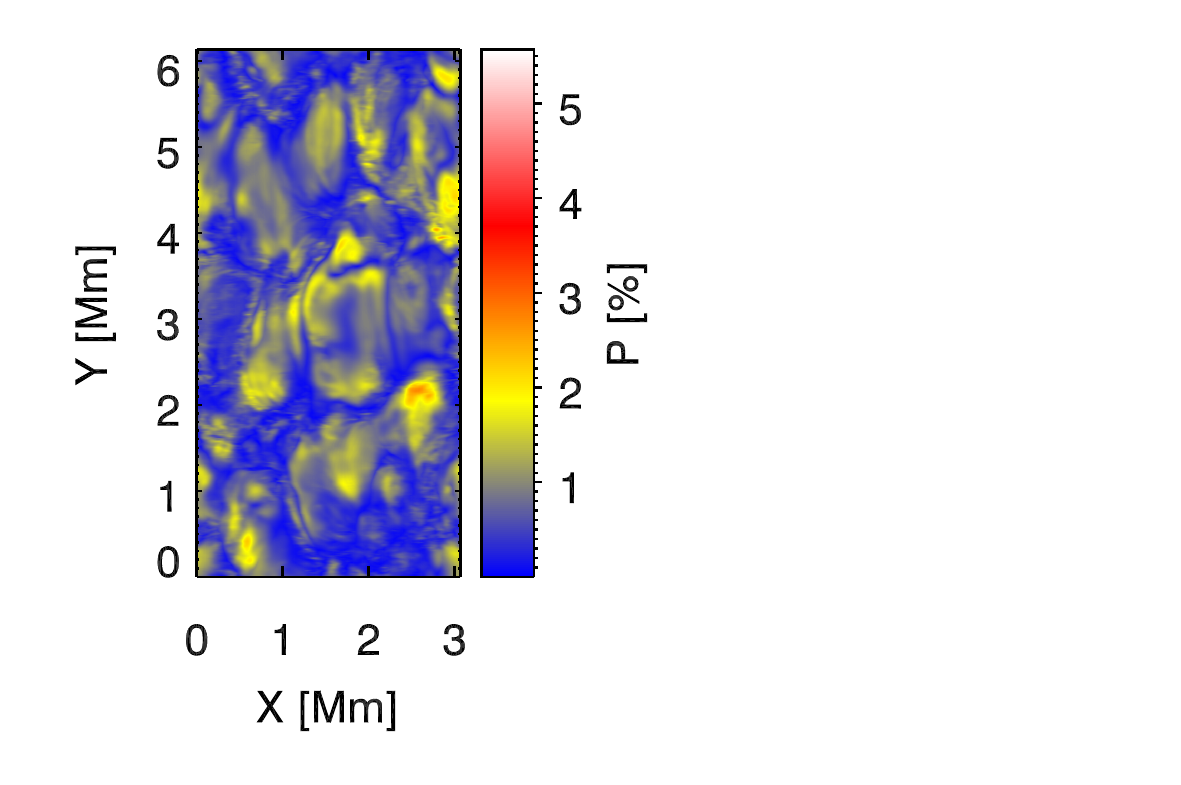}\hspace*{-10em}
\includegraphics[width=.4\textwidth]{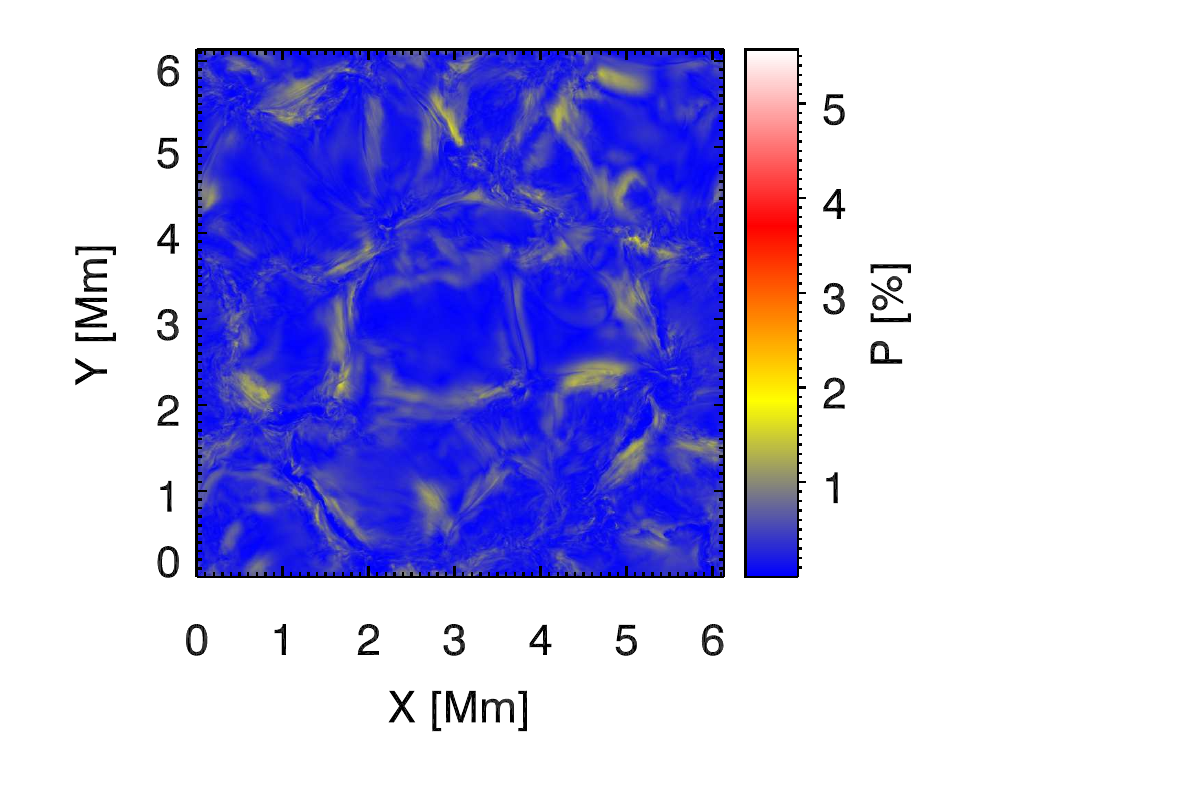}
\caption{Fractional total linear polarization $P$ at the wavelength where $P$ is maximum at the spatial pixel under consideration, for lines of sight with $\mu=0.1$ (left column), $\mu=0.5$ (middle column), and $\mu=1$ (right column), neglecting (top row) and taking into account (bottom row) the Hanle effect due to the magnetic field of the 3D atmospheric model. }
\label{F-P} 
\end{figure} 

The amplitude of the fractional total linear polarization $P$ of the Sr {\sc i} 4607 \AA\ line is shown in Fig. \ref{F-P}, neglecting (upper panels) and taking into account (lower panels) the Hanle effect caused by the model's magnetic field. A comparison between the upper and lower panels immediately shows that the Hanle effect mainly produces depolarization for all lines of sight. It is important to note that at the high spatial resolution (8 km) of Rempel's (2014) 3D model the calculated linear polarization signals are very significant, going from about 5\% for close to the limb observations ($\mu=0.1$) to about 1\% at the solar disk center ($\mu=1$). In particular, note that in the forward-scattering geometry of the observation at disk center the linear polarization signals are significant mainly in the inter-granular regions of the solar granulation pattern and at the borders between granules and inter-granules, where the breaking of the axial symmetry of the pumping radiation field is particularly important (see the left panels of Fig. 7).

\begin{figure}[htbp]
\centering
\includegraphics[width=.33\textwidth]{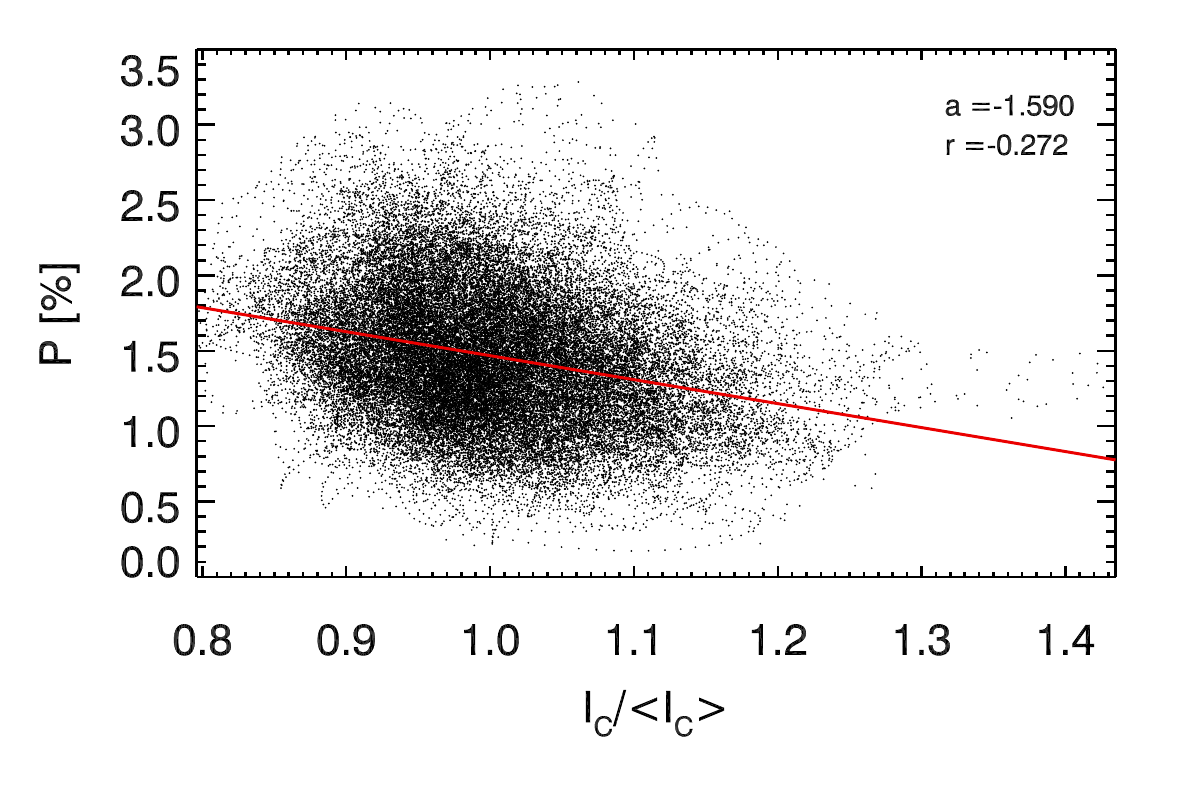}
\includegraphics[width=.33\textwidth]{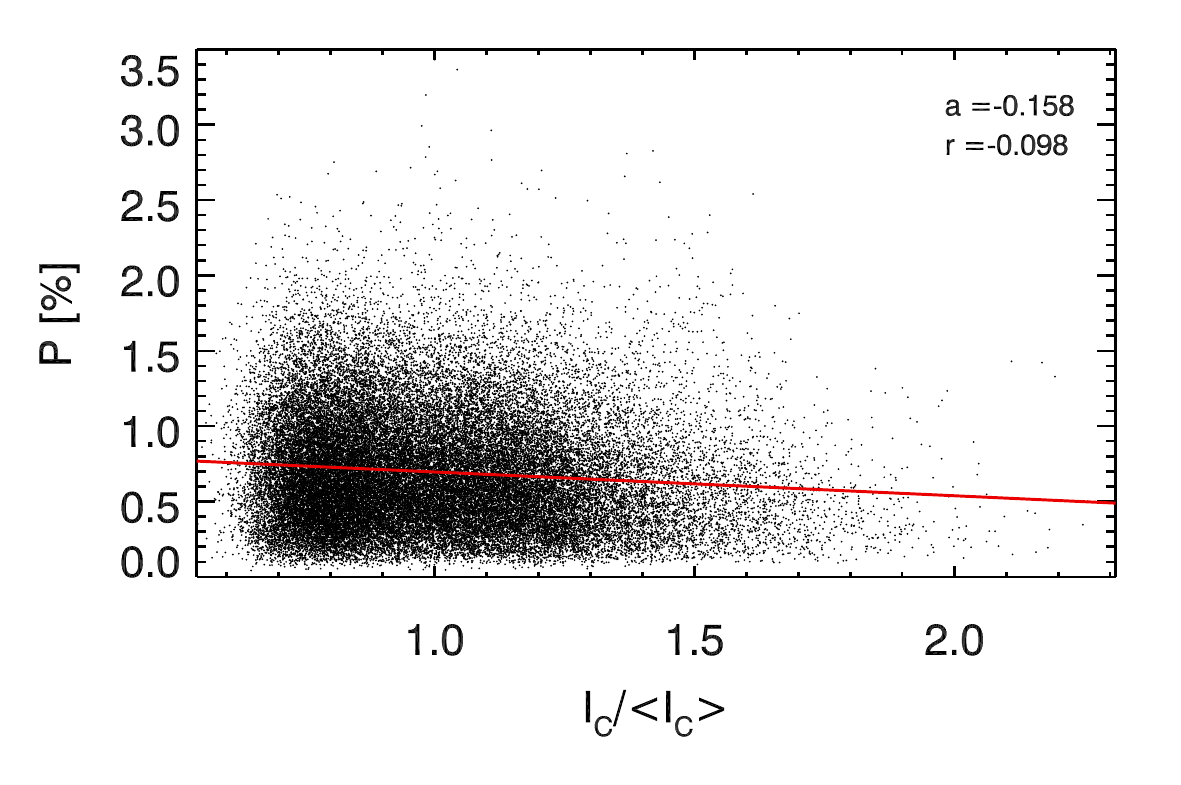}
\includegraphics[width=.33\textwidth]{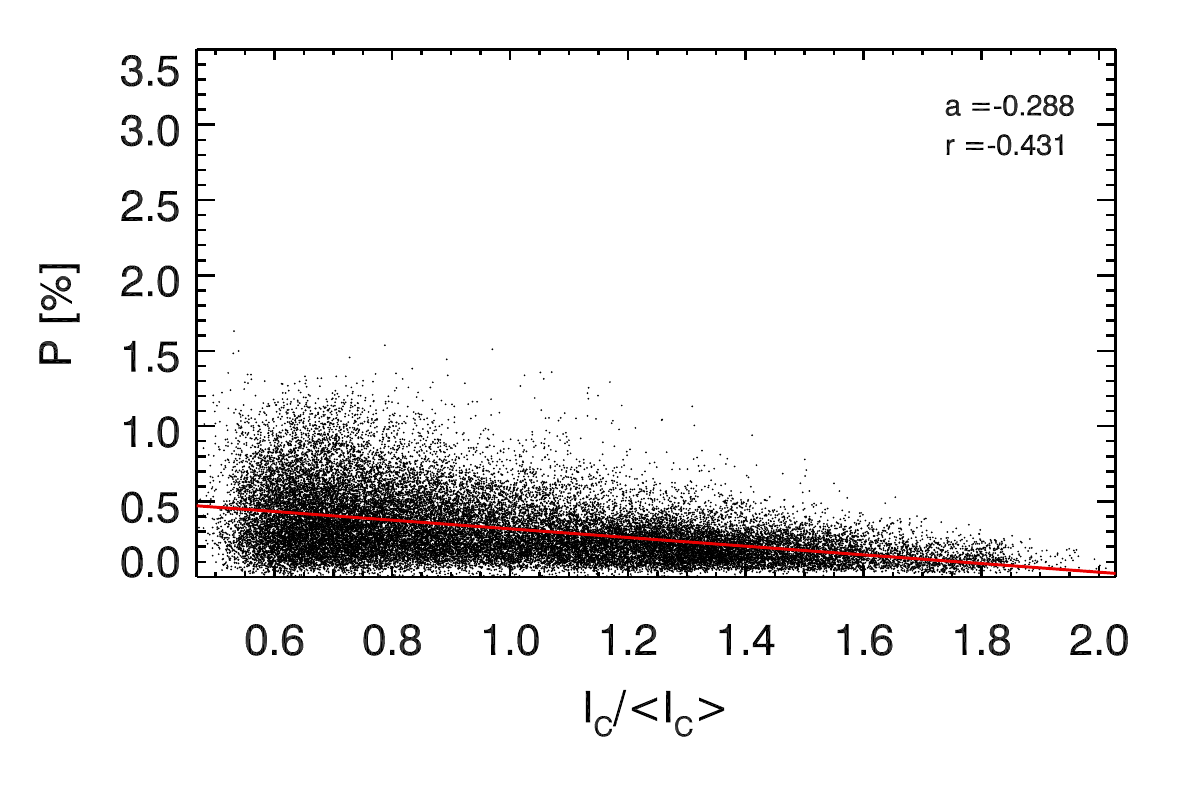}
\caption{Theoretical scatter plots of the total fractional scattering polarization amplitudes $P$ (see lower panels of Fig. \ref{F-P}) against the continuum intensity, calculated all over the field of view for line of sights with $\mu=0.1$ (left panel), $\mu=0.5$ (central panel) and $\mu=1$ (right panel). The $a$ and $r$ values shown in each panel correspond to the slope of the (red lines) least squares fits to a linear function, and the Pearson correlation coefficient, respectively. The peculiar dotted curves that can be seen in the left panel are due to the correlation between nearby points in the atmospheric model due to the large-scale organization of the atmospheric structure in the snapshot, more clearly seen in this panel because of the geometrical shrink produced by the inclined line of sight (i.e., the distance between the simulation nodes in the plane normal to the line of sight is ten times smaller in one of the axis than in the perpendicular one, in the plane of the sky).}
\label{F-scatter-P}
\end{figure}

Figure \ref{F-scatter-P} shows a scatter plot analysis of the results shown in the lower panels of Fig. \ref{F-P}. As can be seen, for all line of sights it is more likely to find a relatively large scattering polarization amplitude associated with points of the field of view having relatively low continuum intensity values (i.e., {\em the theoretical scattering polarization signals are anti-correlated with the continuum intensity}). In Section 5 we show that the small-scale magnetic field of the 3D model used is compatible with the low-resolution observations of the scattering polarization signals all over the disk of the quiet Sun.

\begin{figure}[htp]
\centering 
\includegraphics[width=.5\textwidth]{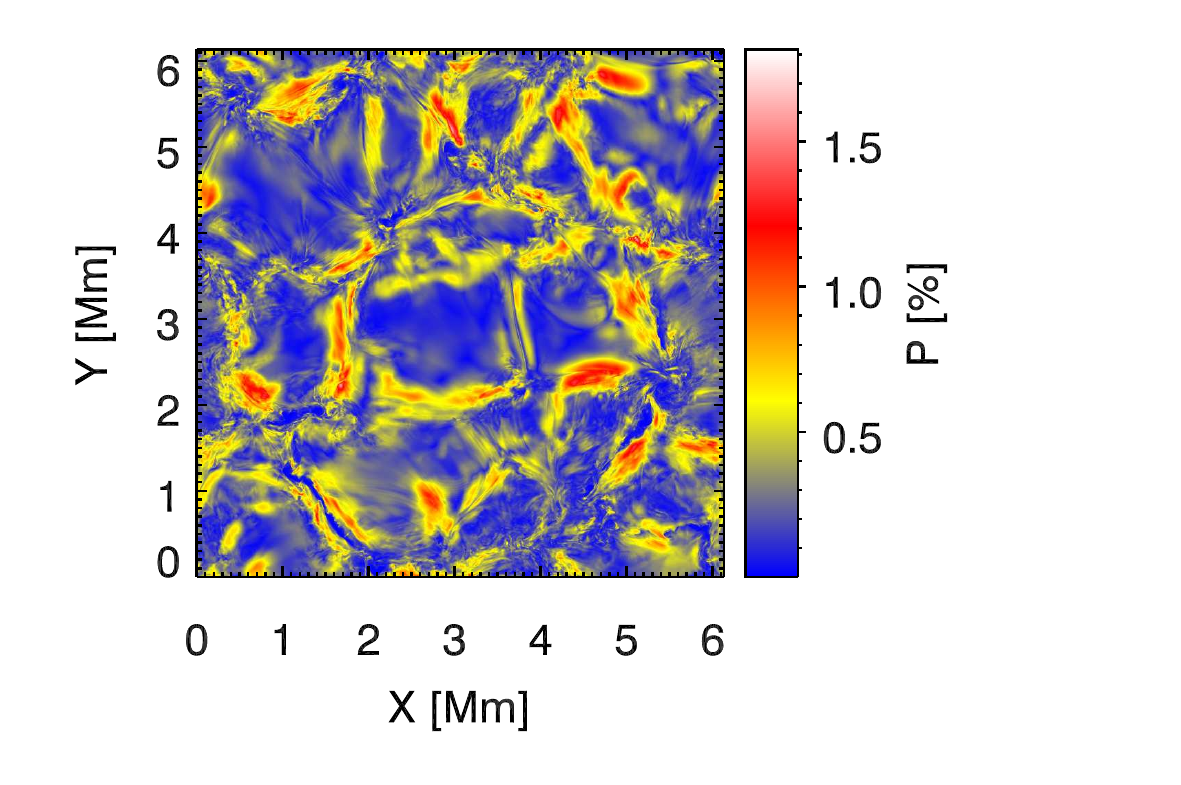}\hspace*{-6em}
\includegraphics[width=.5\textwidth]{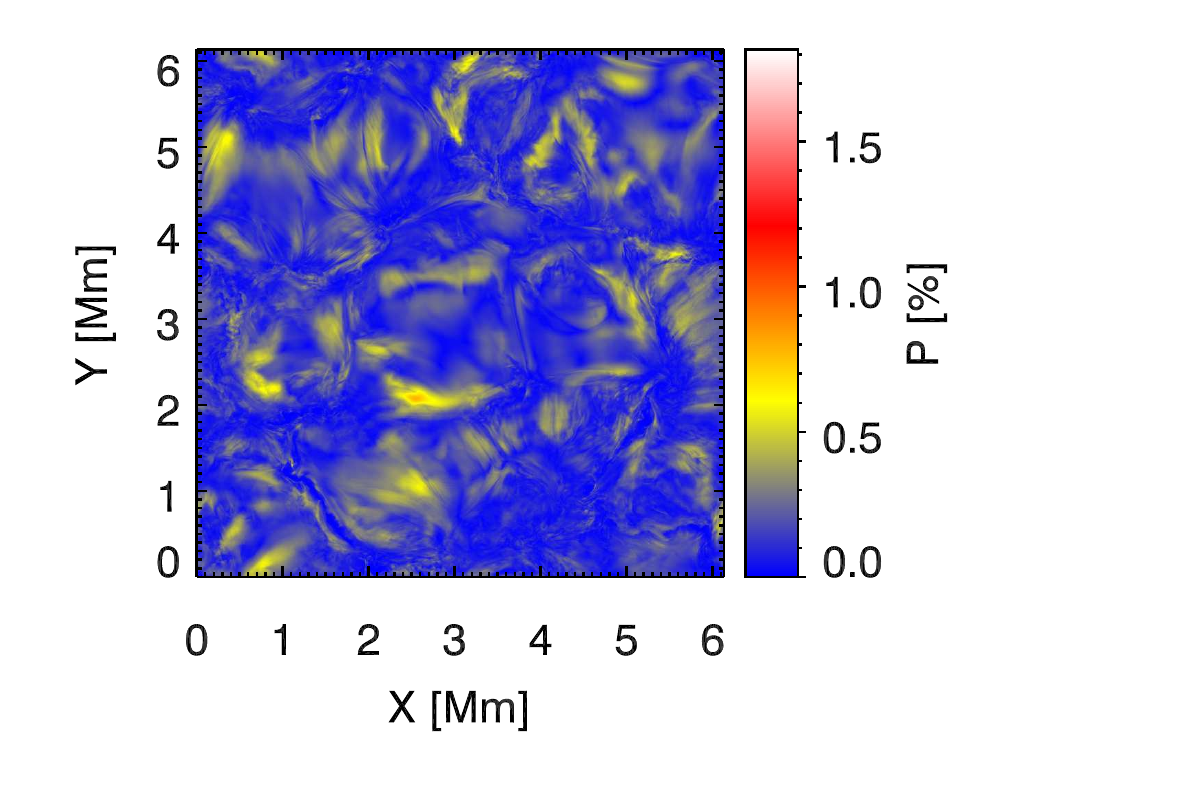}\hspace*{-9em} \\
\includegraphics[width=.5\textwidth]{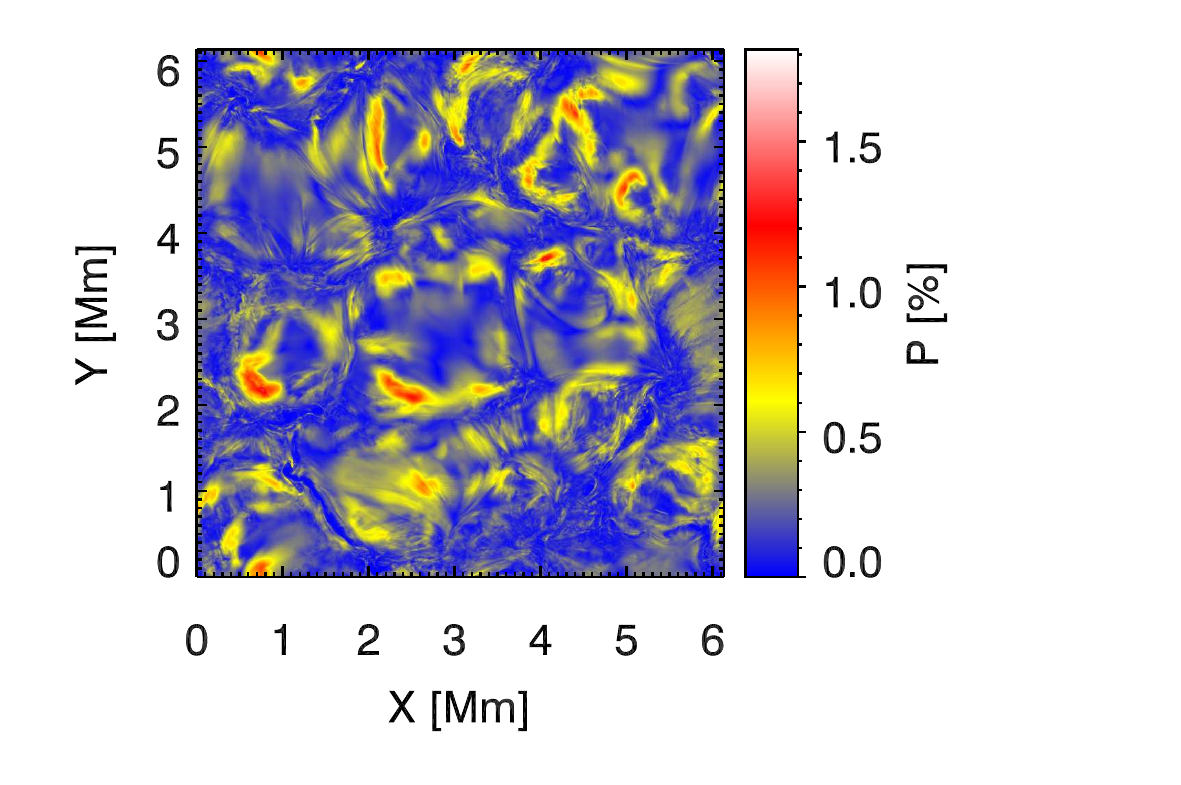}\hspace*{-6em}
\includegraphics[width=.5\textwidth]{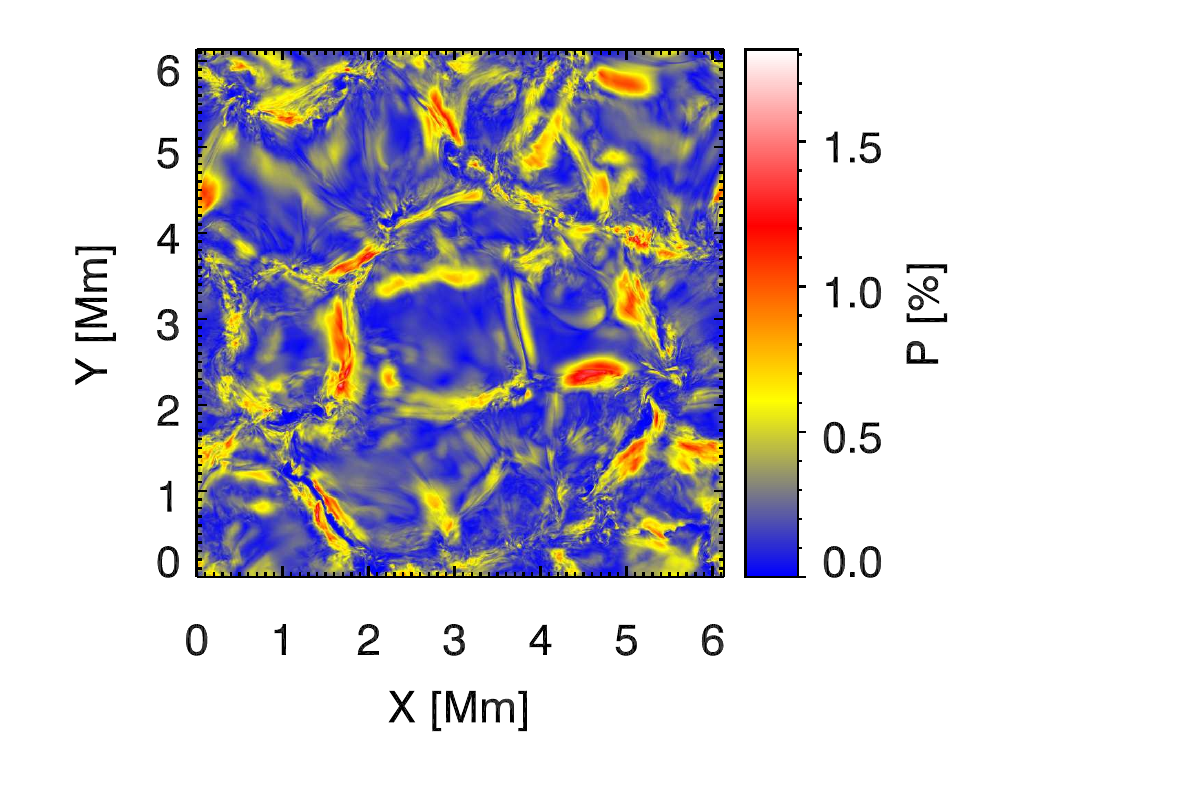}\hspace*{-9em}
\caption{Fractional total linear polarization $P$ at the wavelength where $P$ is maximum at each spatial pixel, for the disk center line of sight, taking into account (top left panel) or neglecting (top right panel) the macroscopic plasma velocities at each iterative step of the non-LTE solution. The bottom panels show the results obtained when taking into account only the vertical (bottom left panel) or only the horizontal (bottom right panel) component of the model's macroscopic velocities. We point out that in all panels we have taken into account the Doppler shifts caused by the component of the velocity along the line of sight when calculating the emergent Stokes profiles, using the previously computed self-consistent values of the atomic density matrix. The similarity between the bottom right panel and the top left panel shows the important symmetry breaking caused by the impact of the horizontal velocity components.}
\label{F-P-V} 
\end{figure}

As seen in the right panel of Fig. \ref{F-modBandV}, the dynamical state of the 3D photospheric model is significant, in the sense that the Doppler shifts corresponding to the macroscopic velocities can be a significant fraction of the local values of the line's Doppler width, especially regarding the horizontal velocity components. Figure \ref{F-IFTS} shows the important impact that such plasma motions have on the width and depth of the Stokes $I(\lambda)$ profiles. Fig. \ref{F-P-V} illustrates that such a dynamical state also has an important impact on the polarization amplitudes of the emergent spectral line radiation. The fact that also the disk center polarization signals calculated taking into account the Doppler shifts of the macroscopic velocities when iteratively solving the non-LTE problem (see the top-left panel) are much larger than those corresponding to the static case (see the top-right panel) indicates that the spatial gradients of the horizontal components of the plasma velocities produce a significant breaking of the axial symmetry of the pumping radiation. This can be clearly seen in the lower panels of the same figure, which show what happens 
with the polarization amplitudes when taking into account only the vertical component of the plasma velocity (left panel) or only the horizontal components (right panel) at each iterative step needed for the solution of the non-LTE problem.

\section{Is the model's magnetization sufficient to explain the observations\,?}\label{S-magnetization}

The 3D model we have chosen from Rempel's (\citeyear{Rempel2014}) magneto-convection simulations is characterized by the variation with height of the mean field strength $\langle B \rangle$ shown in the left panel of Fig. \ref{F-modBandV}, with $\langle B \rangle \, {\approx} \, 170 $ G at the model's visible surface and $\langle B \rangle \, {\approx} \, 70 $ G at a relative height of 300 km. The magnetic field in 3D models resulting from magneto-convection experiments with small-scale dynamo action is tangled at very small spatial scales, to the extent that many of the polarization signals produced by the Zeeman effect in photospheric lines, like those of Fe {\sc i} at 6301.5 \AA\ and 6302.5 \AA, cancel out when considering the spatial resolution achievable with today's telescopes \citep[e.g.,][]{PietarilaGrahametal2009,Danilovicetal2010}. Via the Hanle effect the model's magnetic field mainly depolarizes the linear  polarization signals caused by scattering processes in the Sr {\sc i} 4607 \AA\ line (see Fig. \ref{F-P}), and this allows us to detect the magnetic field that is hidden at subresolution scales and to investigate whether the variation with height of the model's mean field strength can be considered realistic. To this end, we have confronted observations of the center-to-limb variation (CLV) of the (line-center) scattering polarization amplitudes of the Sr {\sc i} 4607 \AA\ line with our 3D radiative transfer calculations. 

\begin{figure}[htbp]
\centering
\includegraphics[width=.7\textwidth]{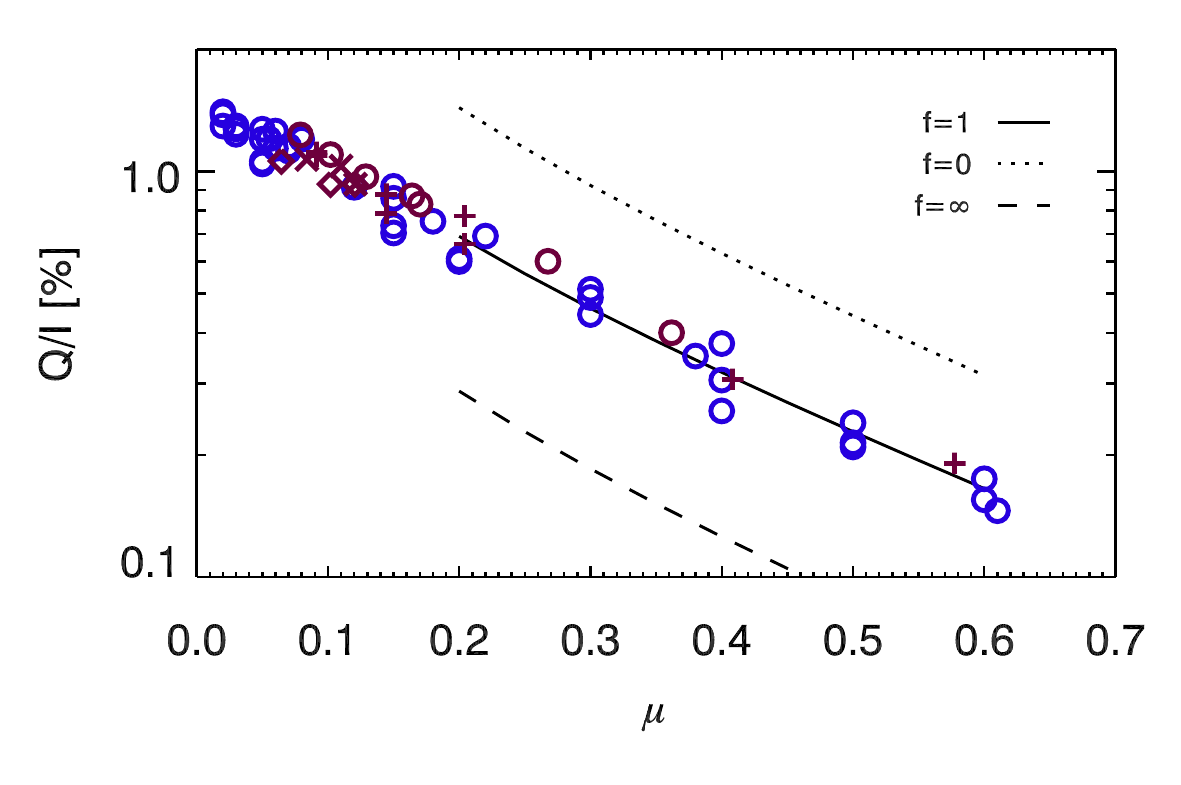}
\caption{Center-to-limb variation of the $Q/I$ scattering amplitudes of the Sr {\sc i} 4607 \AA\ line. The symbols correspond to various observations taken during a minimum and a maximum of the solar activity cycle. The dotted, solid and dashed lines show the result of scattering polarization calculations in the 3D model chosen from Rempel's (2014) magneto-convection experiments, using the elastic collisional rates given by \cite{Faurobertetal1995}. The solid line takes into account the Hanle effect produced by the model's magnetic field, while the dotted line neglects it completely. The dashed line corresponds to the Hanle saturation regime. Note that the solid curve results are in very good agreement with the observational data.}
\label{F-faurobert}
\end{figure}

The spectropolarimetric observations considered in this paper are those used by \cite{Trujilloetal2004}, which lack spatial resolution. They show $U/I {\approx} 0 $ and the CLV of the $Q/I$ line-center amplitudes indicated by the data points of Fig. \ref{F-faurobert}. Accordingly, at each LOS we have spatially averaged the calculated Stokes $I$, $Q$, and $U$ profiles, in order to obtain the ensuing $Q/I$ and $U/I$ line-center signals. We find $U/I {\approx} 0 $, while for the $Q/I$ line-center signals we obtain the CLV indicated by the dotted, solid, and dashed lines of Fig. \ref{F-faurobert}. While the dotted line corresponds to the zero-field reference case (the 3D radiative transfer calculations were performed ignoring the model's magnetic field), the dashed line shows the result for the Hanle saturation limit (the 3D radiative transfer calculations were carried out after artificially increasing the magnetic field strength at each grid point by multiplying it by a very large scaling factor $f$). The solid line shows the result that corresponds to the Hanle depolarization produced by the model's magnetic field ($f=1$). The fact that this theoretical CLV of the $Q/I$ line-center amplitudes turns out to provide an excellent fit to the observed CLV demonstrates that the strength and structure of the magnetic field of Rempel's (2014) model is compatible with the scattering polarization observations of the Sr {\sc i} 4607 \AA\ line.    

\begin{figure}[htbp]
\centering
\includegraphics[width=.7\textwidth]{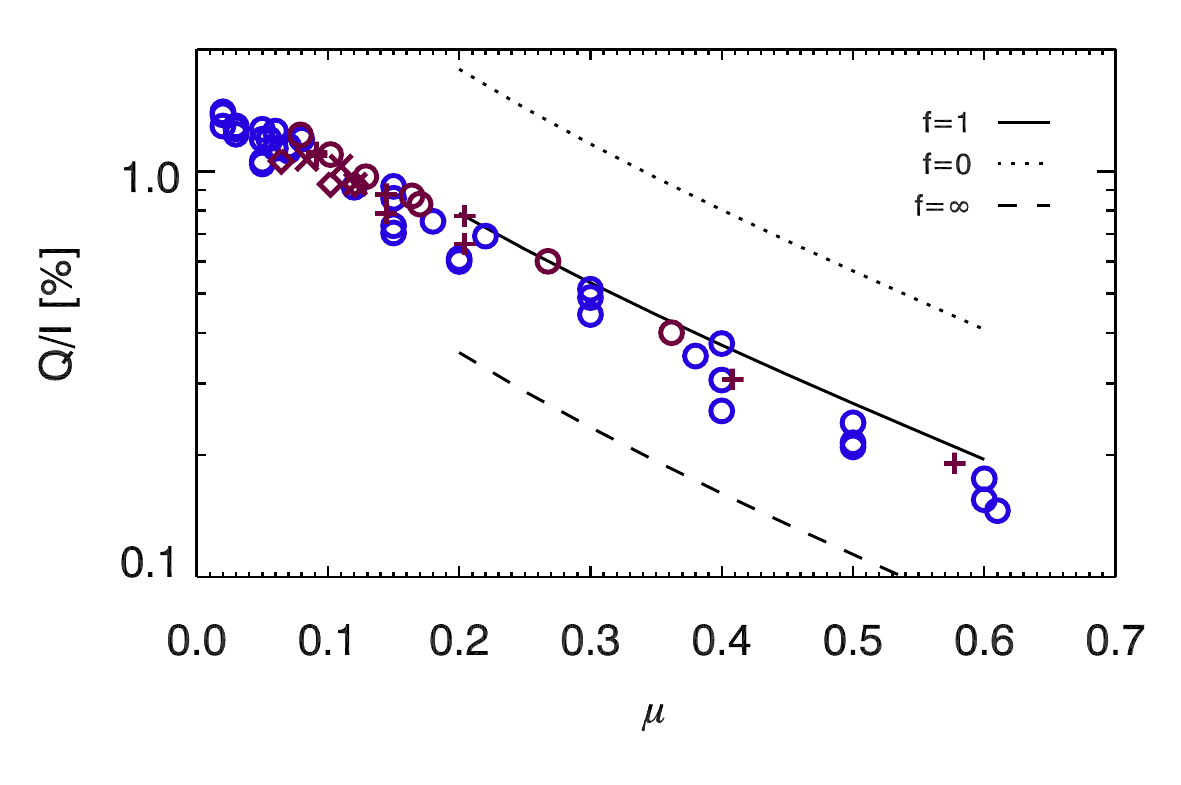}
\caption{Center-to-limb variation of the $Q/I$ scattering amplitudes of the Sr {\sc i} 4607 \AA\ line. The symbols correspond to various observations taken during a minimum and a maximum of the solar activity cycle. The dotted, solid and dashed lines show the result of scattering polarization calculations in the 3D model chosen from Rempel's (2014) magneto-convection experiments, using the elastic collisional rates given by \cite{Mansoetal2014}. The solid line takes into account the Hanle effect produced by the model's magnetic field, while the dotted line neglects it completely. The dashed line corresponds to the Hanle saturation regime. Note that the agreement between the 
solid curve results and the observational data is now not as excellent as in Fig. \ref{F-faurobert}.}
\label{F-manso}
\end{figure}

Since depolarization by elastic collisions with neutral hydrogen atoms is significant for the Sr {\sc i} 4607 \AA\ line (see Fig. \ref{F-delta}), it is important to provide information on the sensitivity of our results to the choice of collisional rates. The results of Fig. \ref{F-faurobert} were obtained using the elastic collisional rates given by \cite{Faurobertetal1995}, which are the largest among those found in the literature and agree with those that can be obtained applying the semi-classical theory of \cite{AnsteeOmara1995}. However, as mentioned in Section 1, ab-initio quantum mechanical calculations by \cite{Mansoetal2014} and \cite{Kerkeni2002} give significantly smaller elastic collisional rates (see the dashed curve of Fig. \ref{F-delta}), which turn out to be similar to those that can be obtained applying the approximate equation (7.108) of \cite{BLandiLandolfi2004}. When in our 3D radiative transfer calculations we use the elastic collisional rates of \cite{Mansoetal2014} we obtain instead the solid-line results shown in Fig. \ref{F-manso}. The agreement with the observed CLV is now not as good as with the elastic collisional rates of \cite{Faurobertetal1995}, although it is still within the observational uncertainties. In any case, it is of interest to point out that in order to achieve a similarly good fit when using the elastic collisional rates of \cite{Mansoetal2014} the magnetic field strength of Rempel's (2014) model would have to be scaled by a factor $f=3/2$ in the region of formation of the core of the Sr {\sc i} 4607 \AA\ line, which would imply a mean field strength of about 100 gauss at a height of 300 km above the model's visible surface.

\section{The impact of finite spectral resolution}\label{S-degrade-spectral}

\begin{figure}[htbp]
\centering
\includegraphics[width=.40\textwidth]{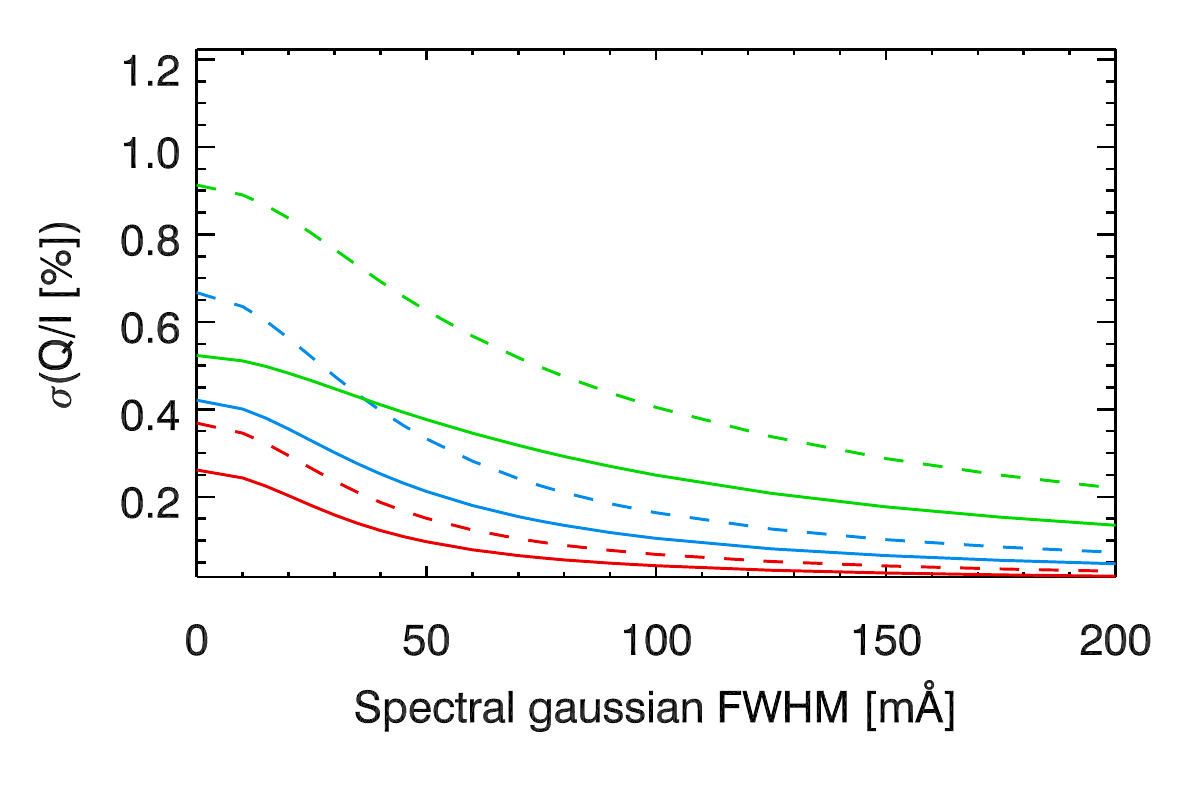}
\includegraphics[width=.40\textwidth]{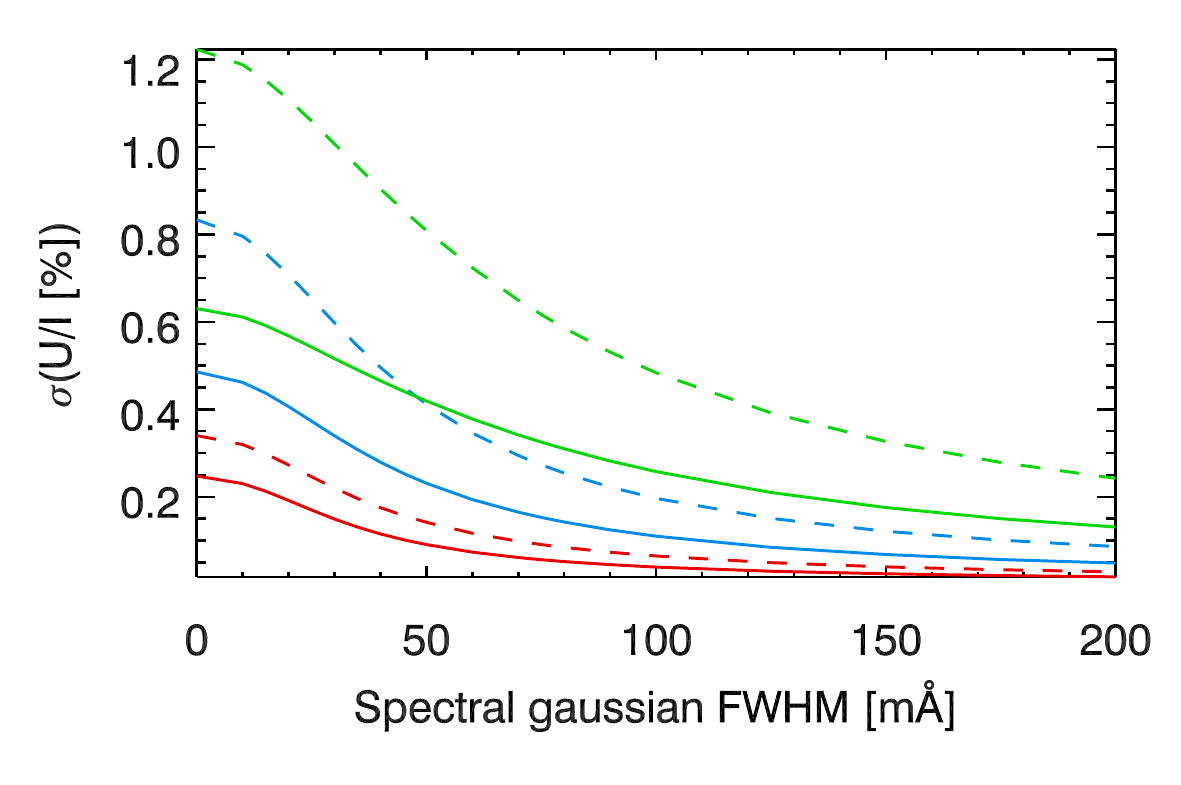}
\caption{The variation with the spectral resolution of the standard deviation of the calculated $Q/I$ (left panel) and $U/I$ (right panel) spatial variations of the scattering amplitudes of the Sr {\sc i} 4607 \AA\ line, at the wavelength where the total fractional linear polarization $P$ is maximum at the spatial pixel under consideration, for lines of sight with $\mu=0.1$ (green curves), $\mu=0.5$ (blue curves), and $\mu=1$ (red curves), taking into account (solid curves) or neglecting (dashed curves) the Hanle effect produced by the model's magnetic field. The diffraction limit of a 1.5m telescope has been taken into account.}
\label{F-sigma-QU}
\end{figure}

The spectral resolution of the instrument used has a significant impact on the scattering polarization amplitudes and on the standard deviation of their horizontal fluctuations across the field of view. We illustrate this fact in Fig. \ref{F-sigma-QU} assuming very high spatial resolution observations, corresponding to the diffraction limit of a 1.5 m telescope. In this figure the dashed curves correspond to the zero-field reference case, while the solid curves take into account the Hanle effect of the model's magnetic field. A spectral resolution of 50 m\AA\ reduces the standard deviation of the horizontal fluctuations by about a factor two. \cite{TrujilloShchukina2007} have shown that the magnetic sensitivity of the standard deviation of the line-center $Q/I$ and $U/I$ horizontal fluctuations is of great diagnostic interest. This turns out to be  very significant for the magnetic field of Rempel's (\citeyear{Rempel2014}) model (compare the dashed and solid curves of the figure for each line-of-sight $\mu$ value).

\section{The impact of seeing}\label{S-degrade-seeing}

In order to illustrate the impact of seeing on the scattering polarization signals of the Sr {\sc i} 4607 \AA\ line we assume that the ensuing degradation corresponds to the diffraction limit of a telescope of diameter $r_0$ located outside the Earth's atmosphere, $r_0$ being the so-called Fried's parameter. In order to mimic the effect of a given seeing, at each point of the field of view we have convolved the emergent Stokes $I$, $Q$ and $U$ signals with a Gaussian function having a full width at half maximum ${\rm FWHM} = 0.99 \cdot \frac{\lambda}{D}\cdot 206264.8$, in arcseconds, with $\lambda$ the spectral line wavelength and $D$ the telescope's diameter or the Fried parameter ($r_0$, \citealt{Fried1966}), as applicable. This approximate way of accounting for the seeing, which slightly underestimates the ensuing degradation effects, is sufficient for our illustrative purposes. 

\begin{figure}[htbp]
\centering
\includegraphics[width=.4\textwidth]{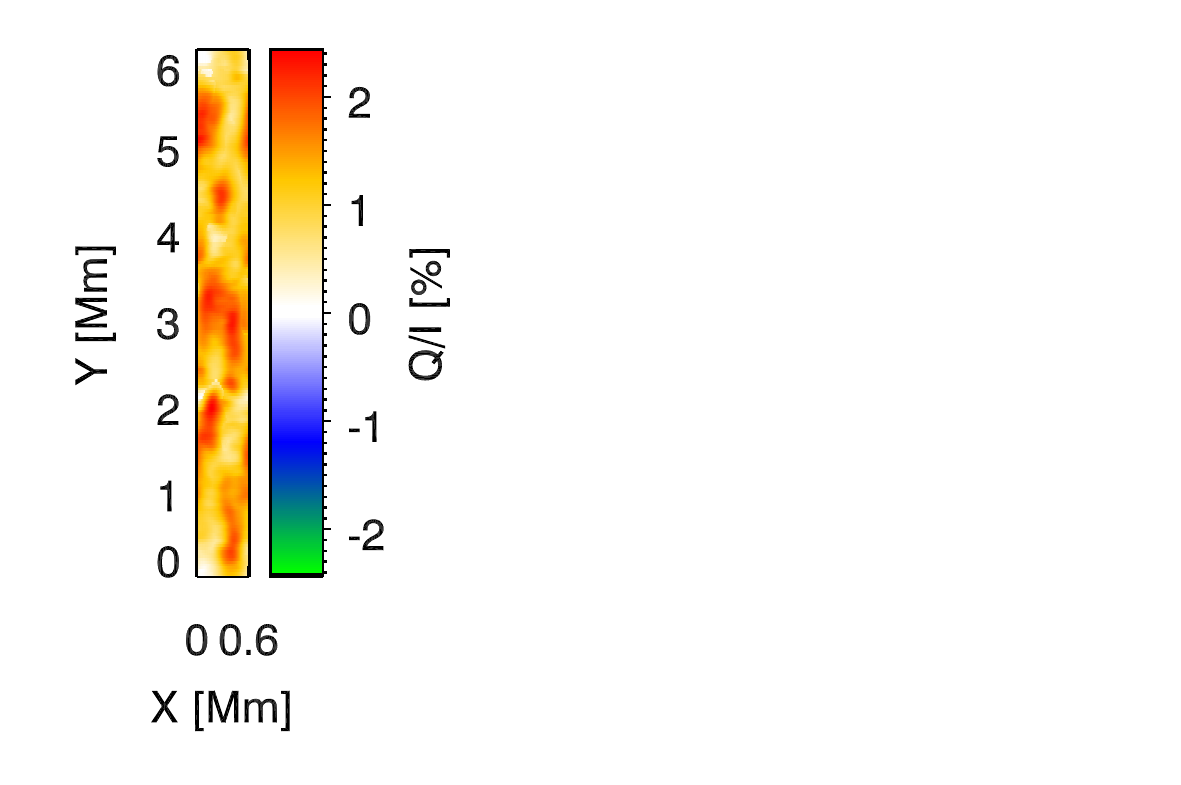}\hspace*{-14em}
\includegraphics[width=.4\textwidth]{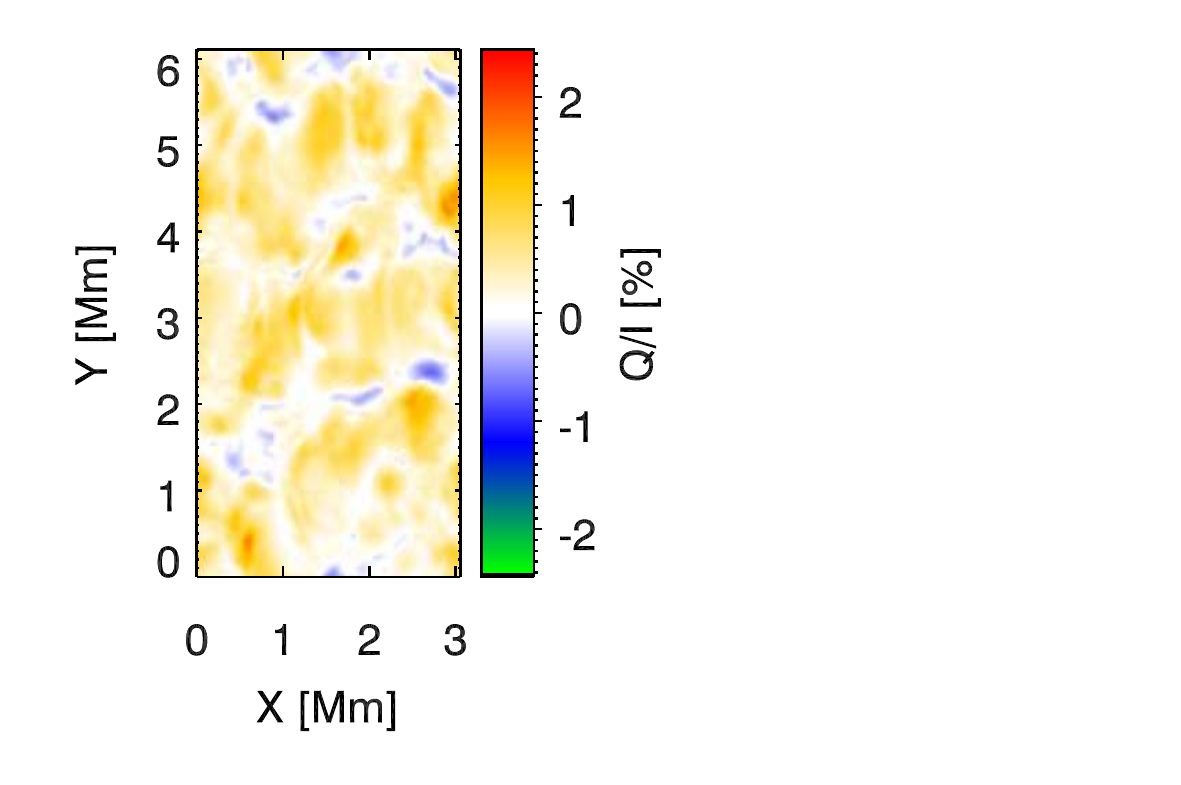}\hspace*{-10em}
\includegraphics[width=.4\textwidth]{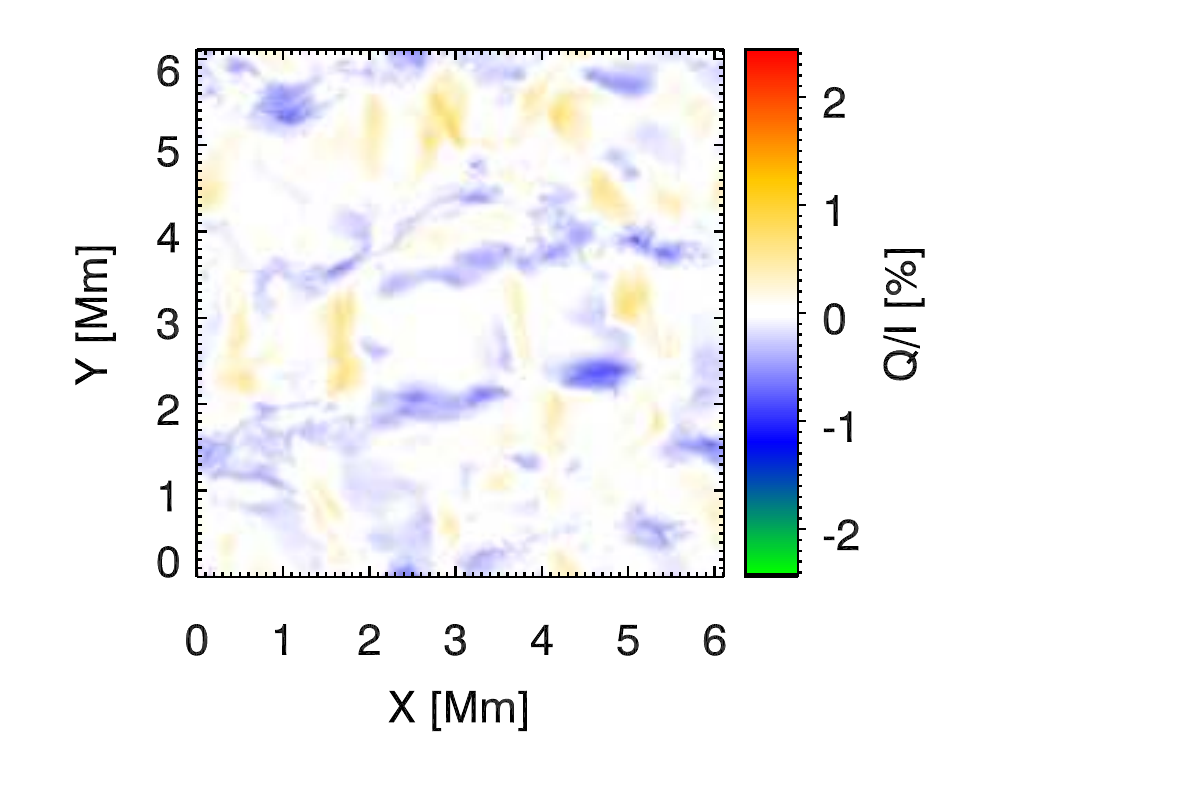}\\
\includegraphics[width=.4\textwidth]{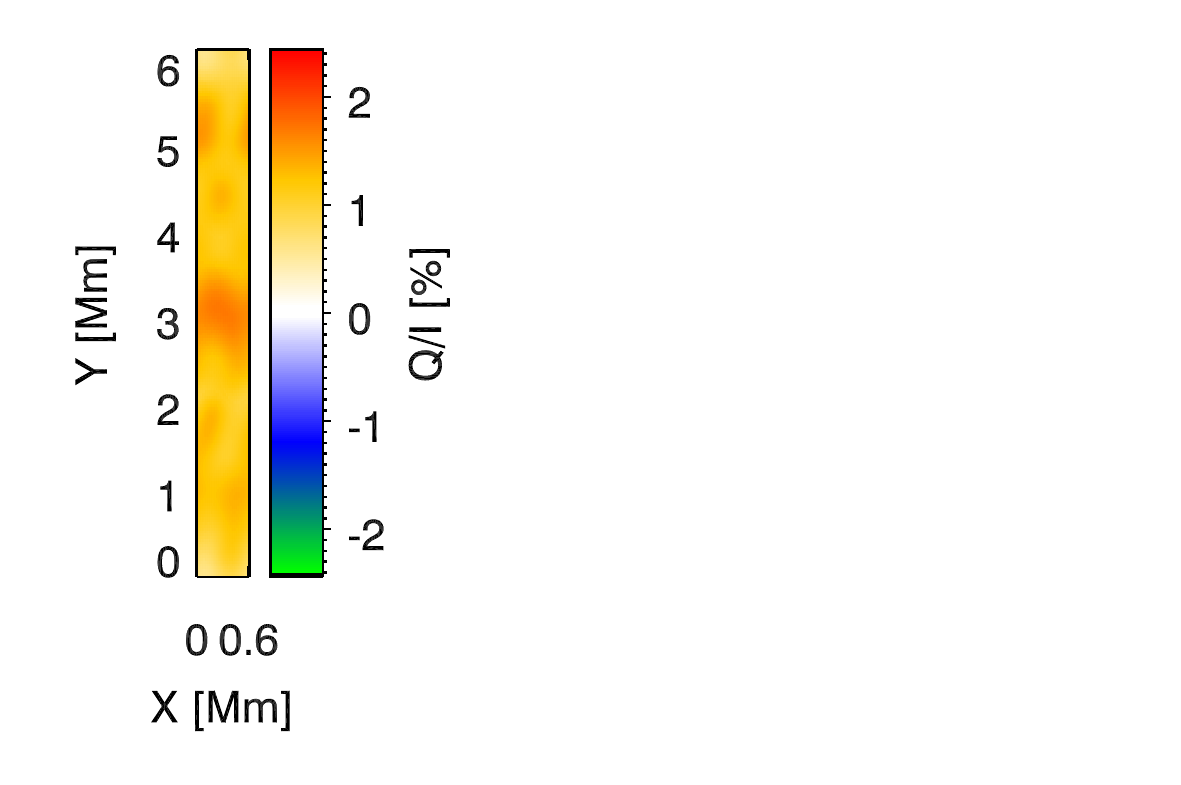}\hspace*{-14em}
\includegraphics[width=.4\textwidth]{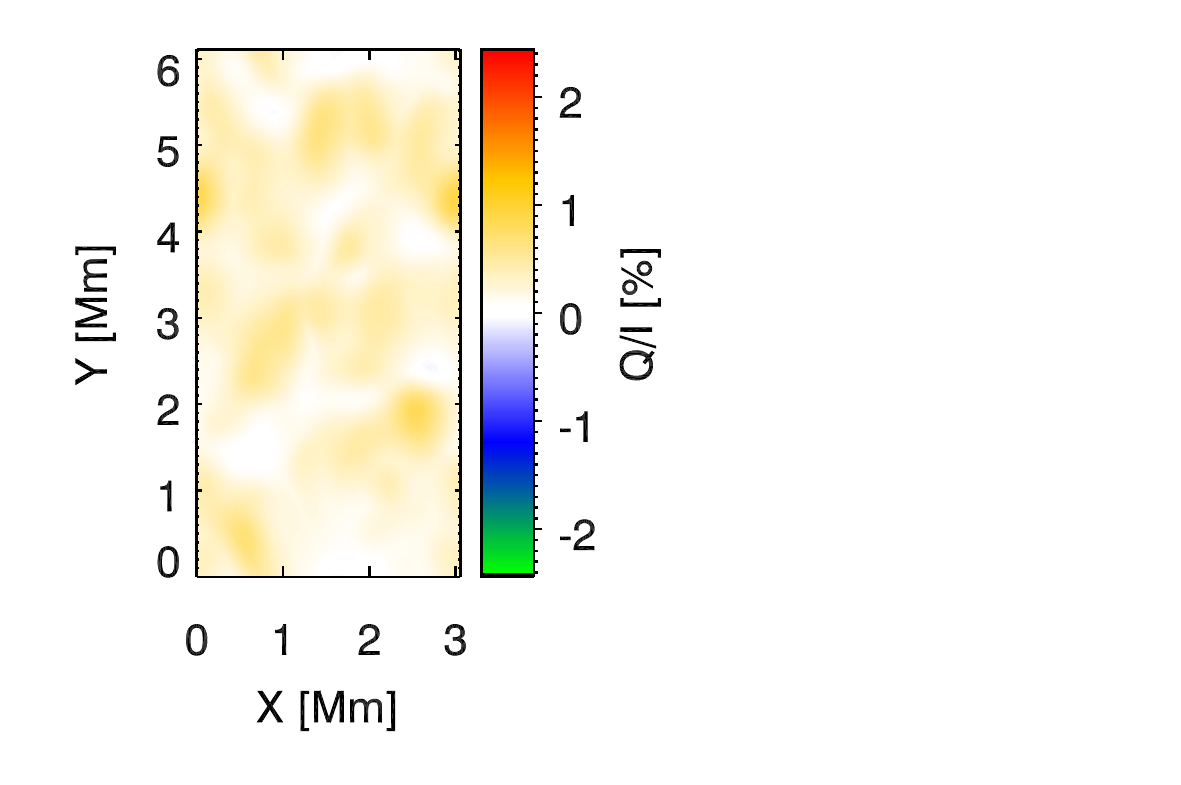}\hspace*{-10em}
\includegraphics[width=.4\textwidth]{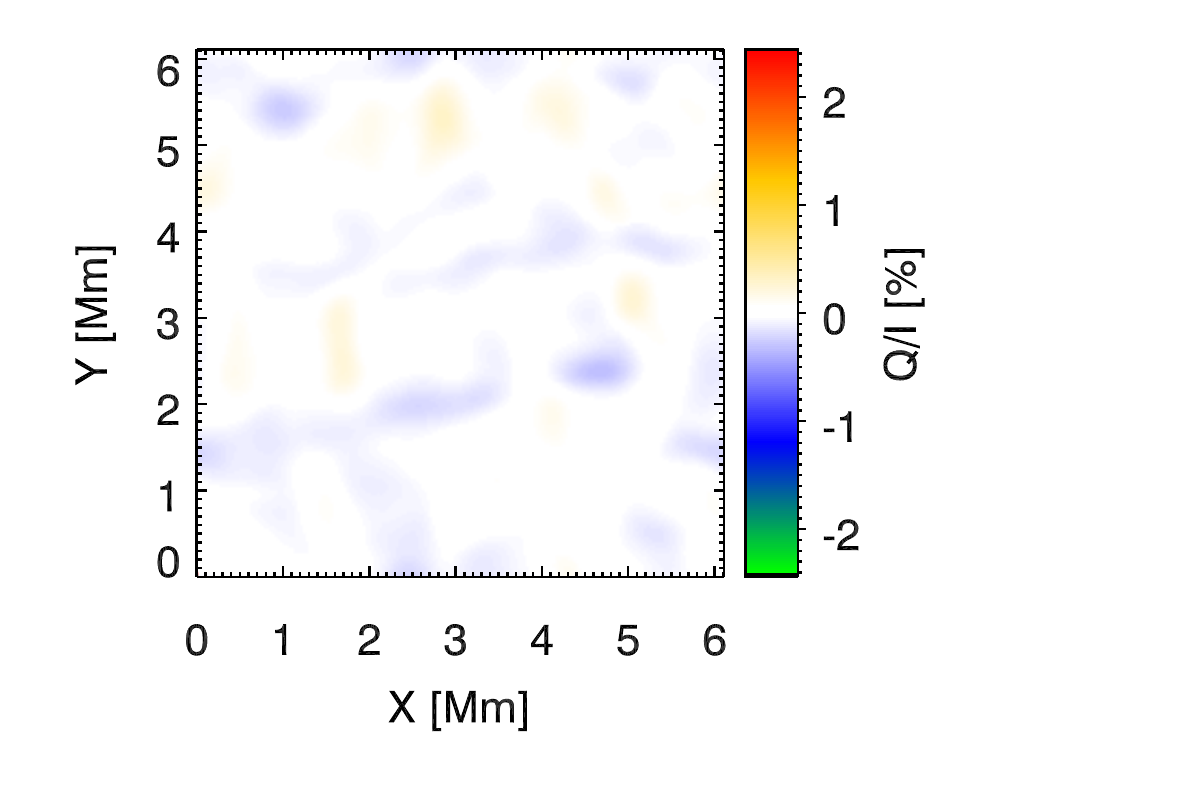}\\
\includegraphics[width=.4\textwidth]{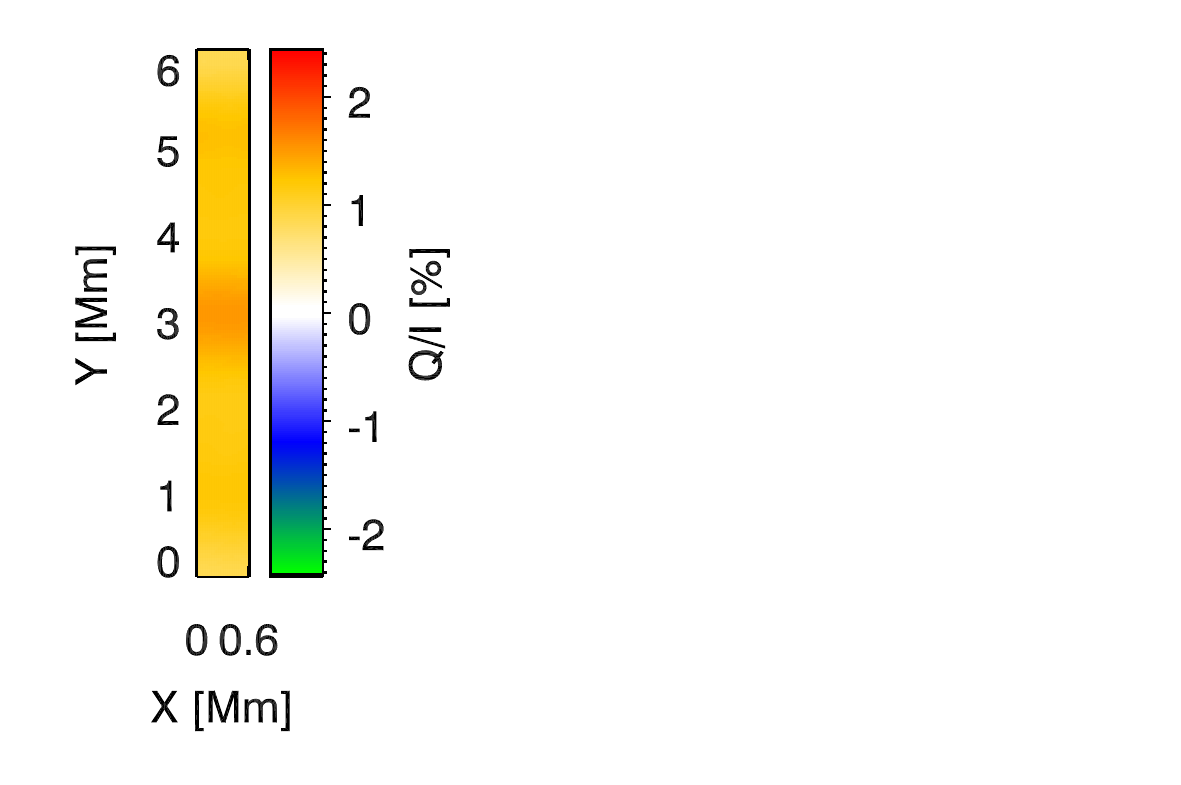}\hspace*{-14em}
\includegraphics[width=.4\textwidth]{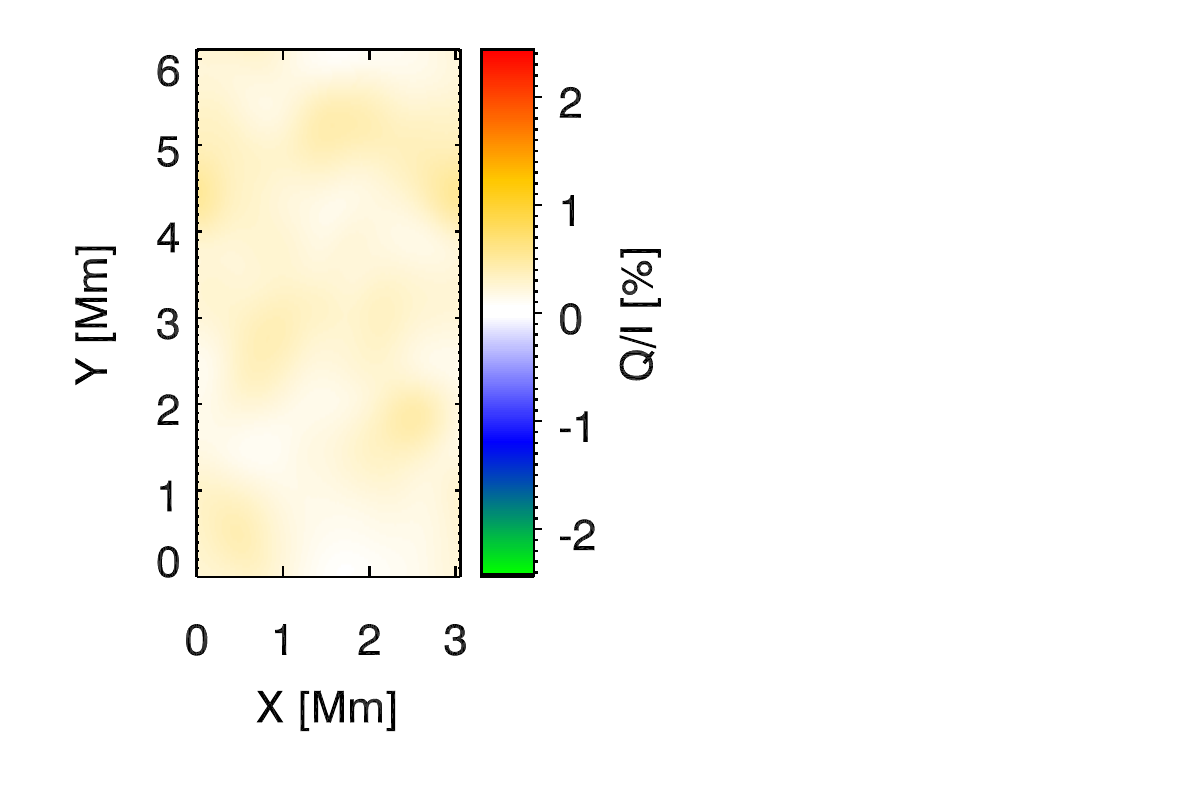}\hspace*{-10em}
\includegraphics[width=.4\textwidth]{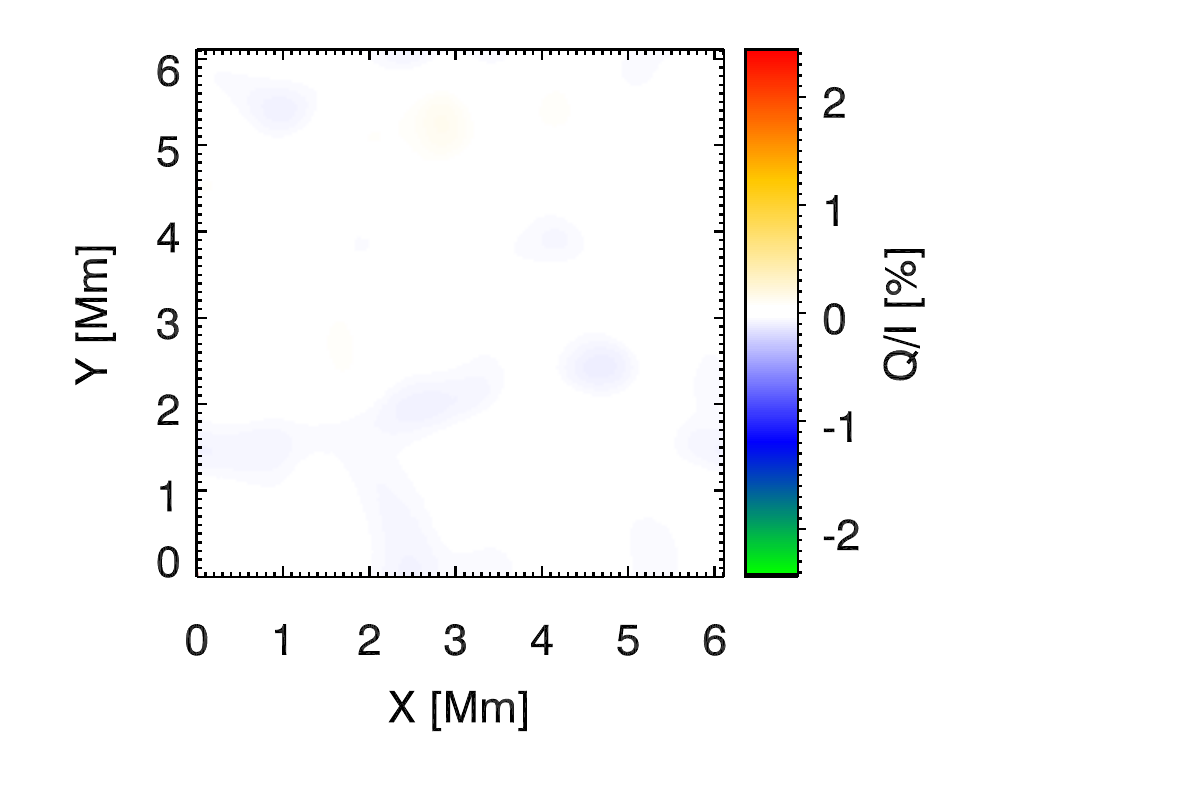}
\caption{Fractional linear polarization $Q/I$ at the wavelength where the linear polarization $P$ is maximum at the spatial pixel under consideration, for the lines of sight with $\mu=0.1$ (left column), $\mu=0.5$ (middle column), and $\mu=1$ (right column), for the cases described in the text: the diffraction limit ($r_0=150$ cm) case (top row), the case with a seeing of $0.5$ arcseconds (middle row), and the $1$ arcsecond seeing case (bottom row).} 
\label{F-seeing-Q}
\end{figure}

\begin{figure}[htbp]
\centering
\includegraphics[width=.4\textwidth]{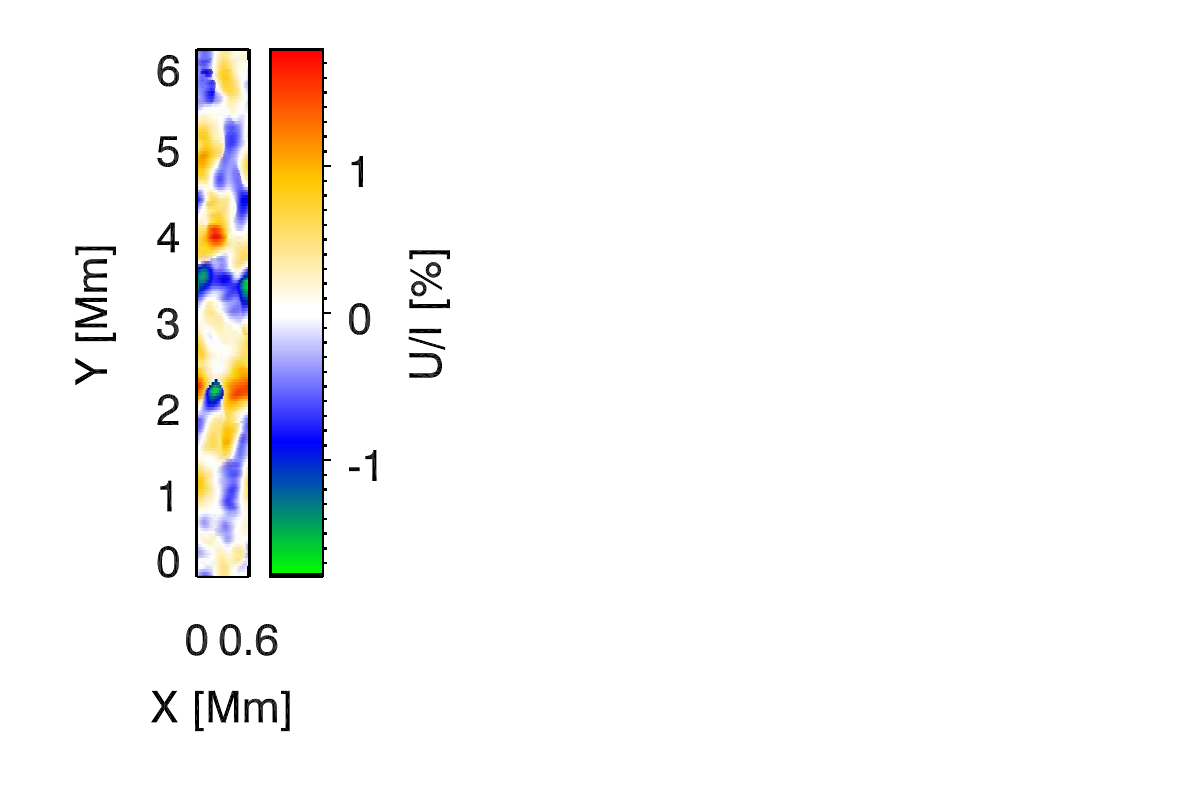}\hspace*{-14em}
\includegraphics[width=.4\textwidth]{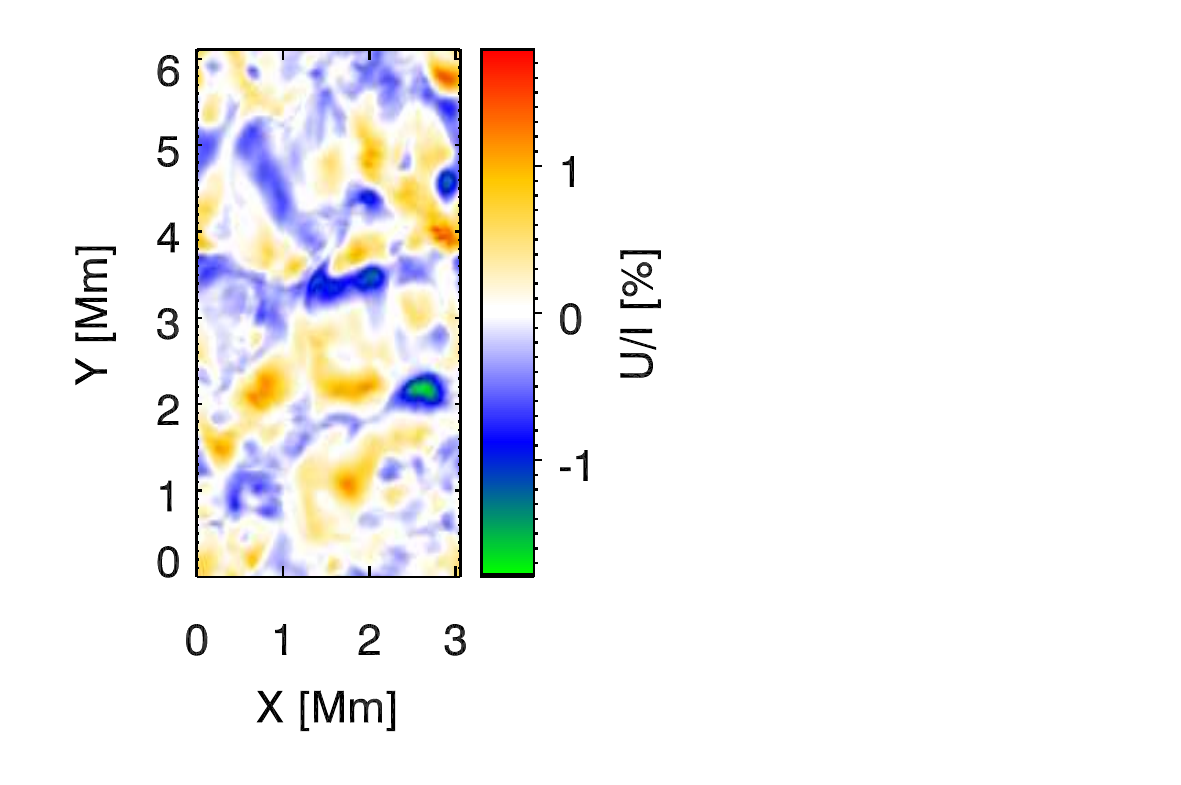}\hspace*{-10em}
\includegraphics[width=.4\textwidth]{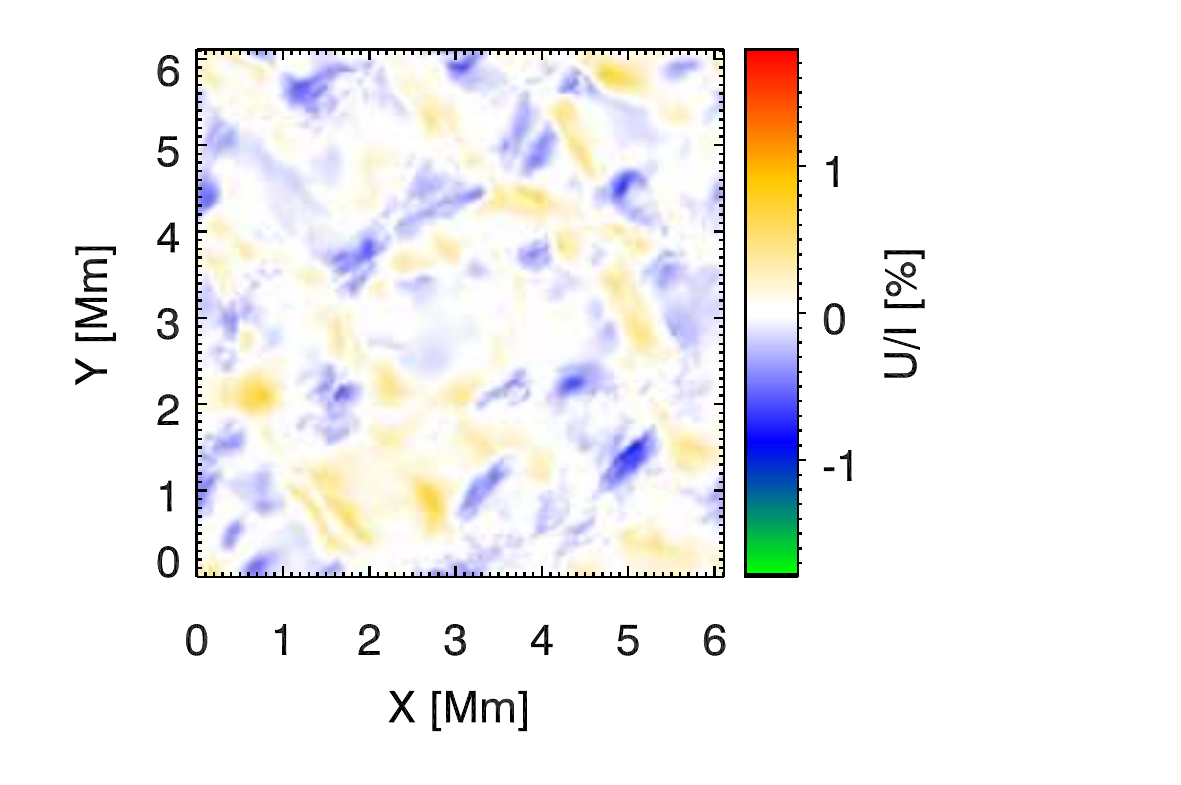}\\
\includegraphics[width=.4\textwidth]{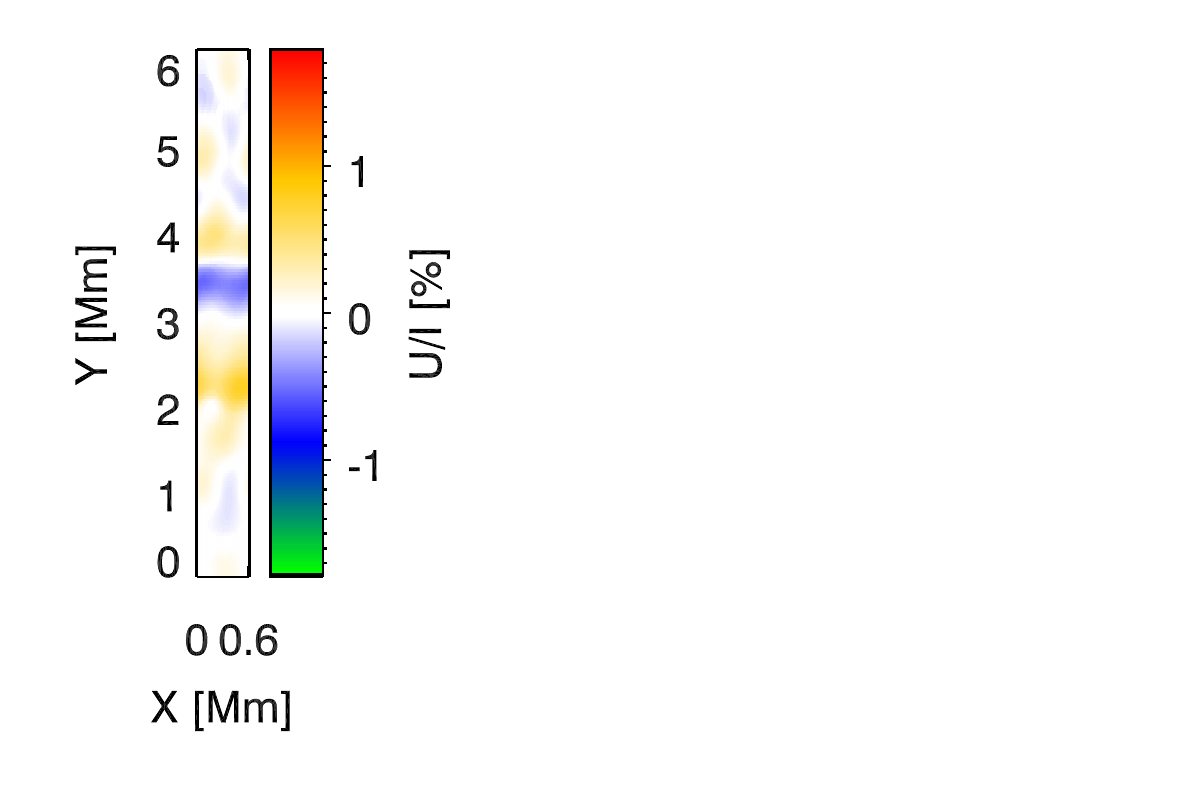}\hspace*{-14em}
\includegraphics[width=.4\textwidth]{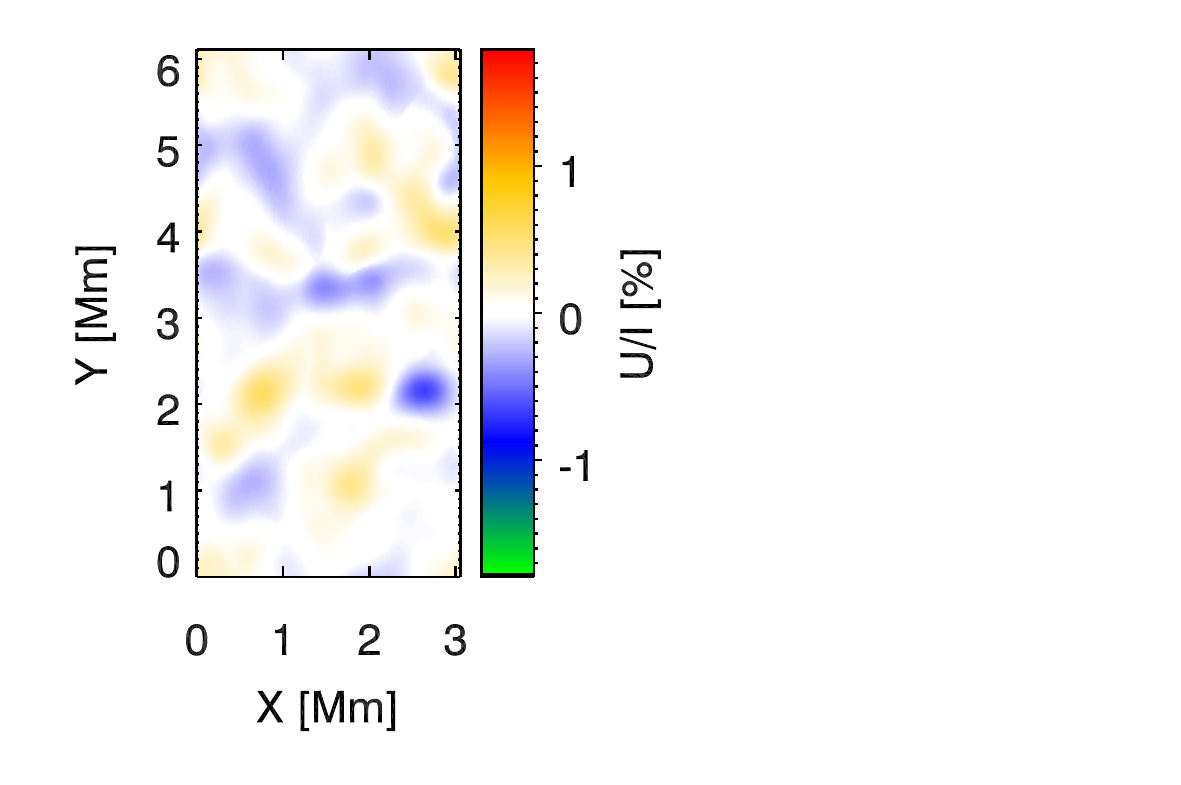}\hspace*{-10em}
\includegraphics[width=.4\textwidth]{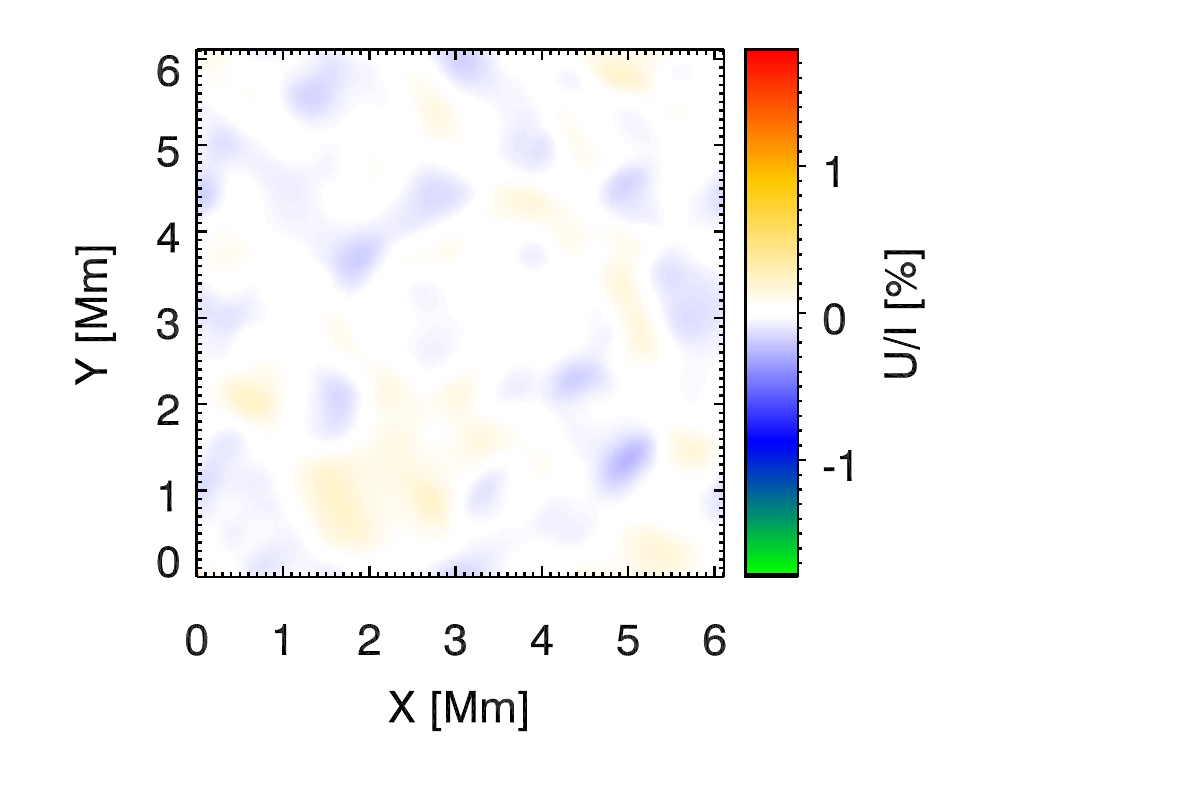}\\
\includegraphics[width=.4\textwidth]{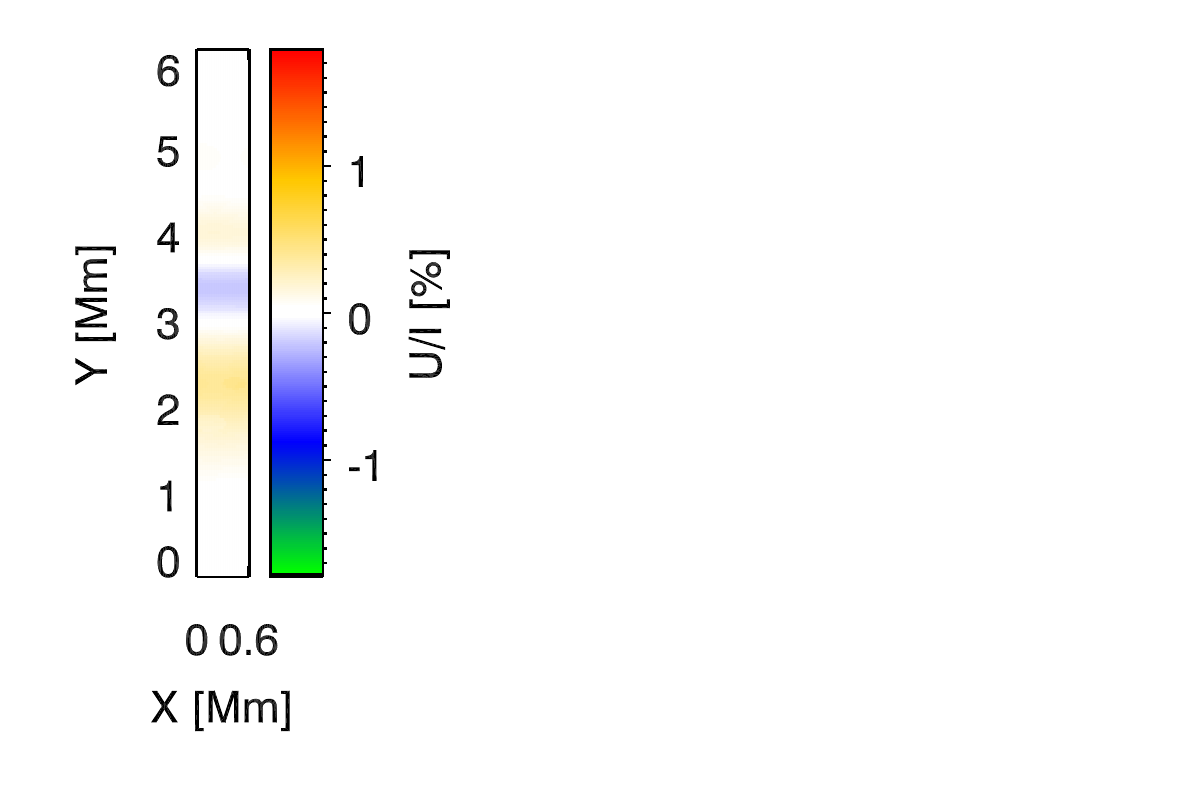}\hspace*{-14em}
\includegraphics[width=.4\textwidth]{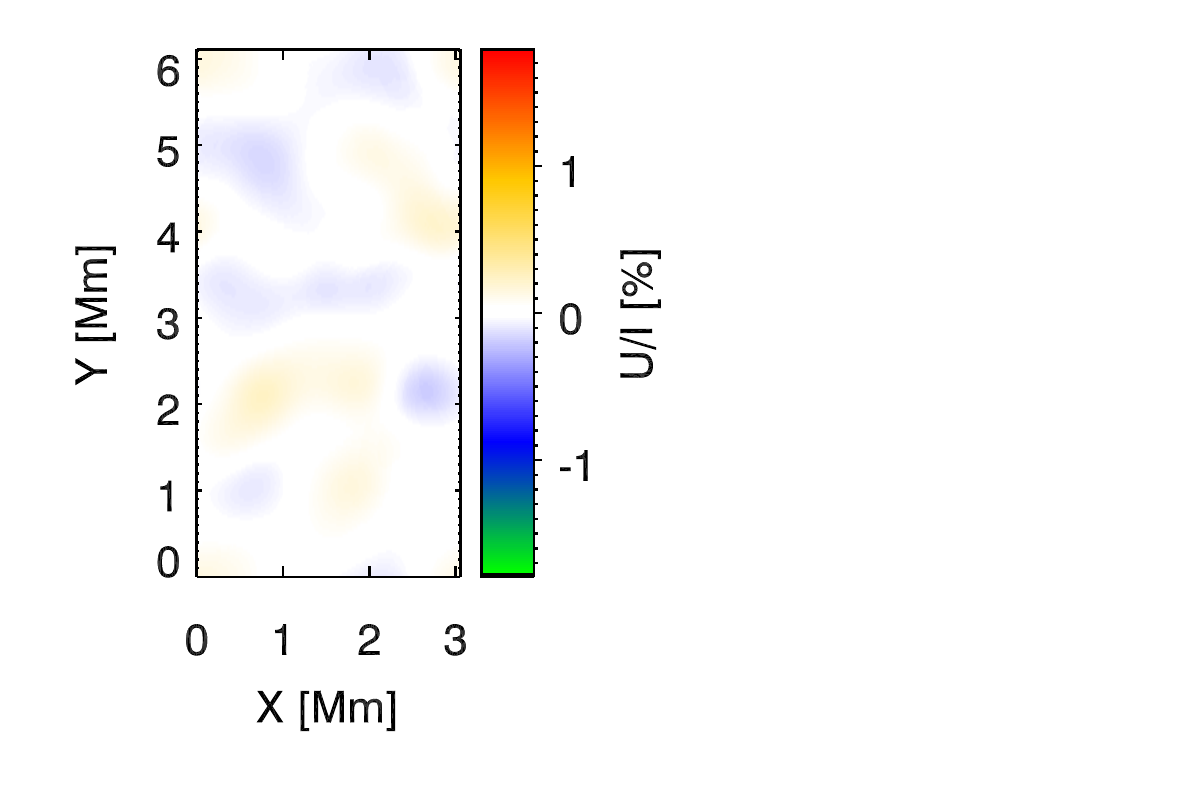}\hspace*{-10em}
\includegraphics[width=.4\textwidth]{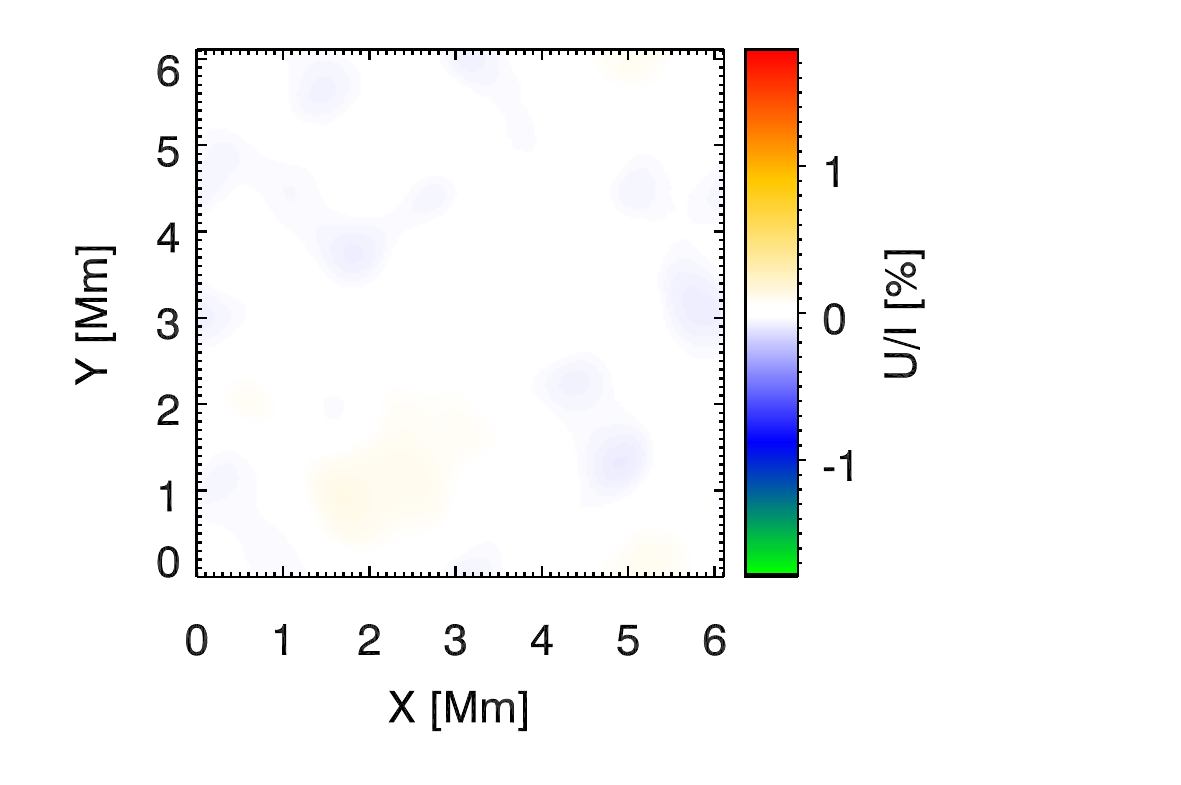}
\caption{Fractional linear polarization $U/I$ at the wavelength where the linear polarization $P$ is maximum at the spatial pixel under consideration, for the lines of sight with $\mu=0.1$ (left column), $\mu=0.5$ (middle column), and $\mu=1$ (right column), for the cases described in the text: the diffraction limit ($r_0=150$ cm) case (top row), the case with a seeing of $0.5$ arcseconds (middle row), and the $1$ arcsecond seeing case (bottom row).}
\label{F-seeing-U}
\end{figure}

\begin{figure}[htbp]
\centering
\includegraphics[width=.4\textwidth]{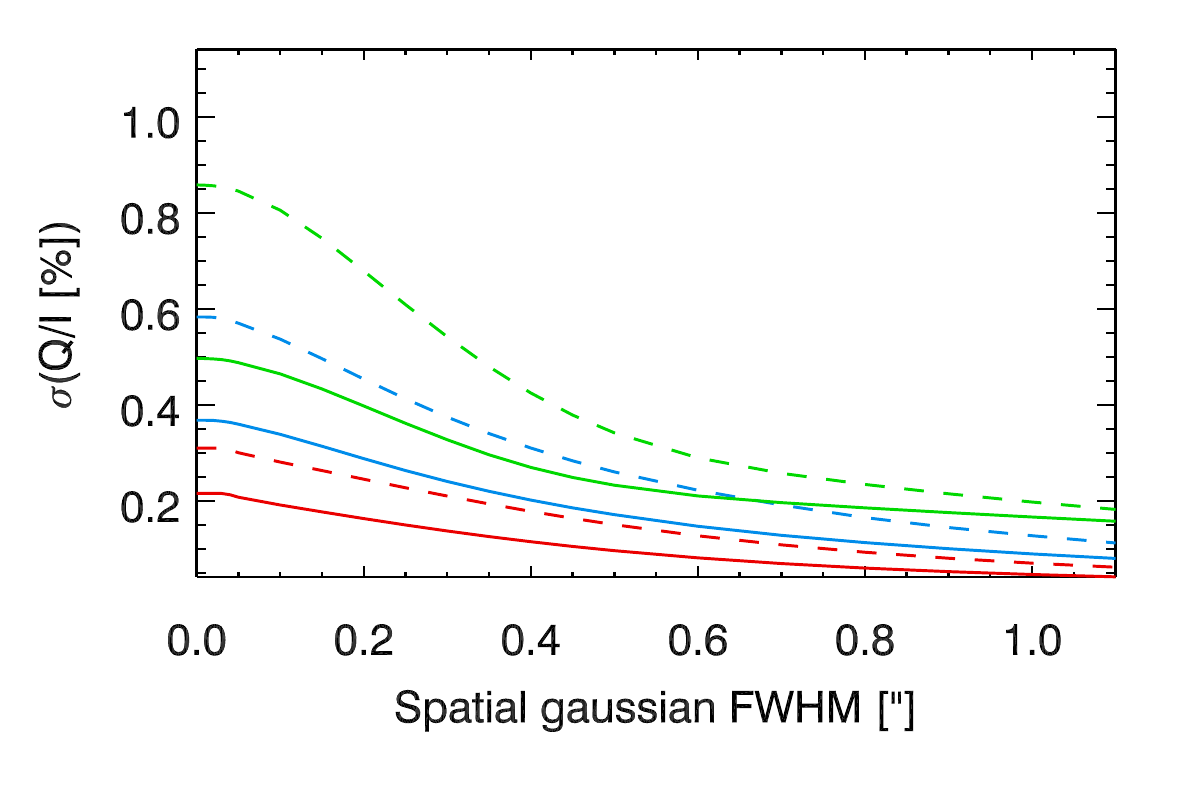}
\includegraphics[width=.4\textwidth]{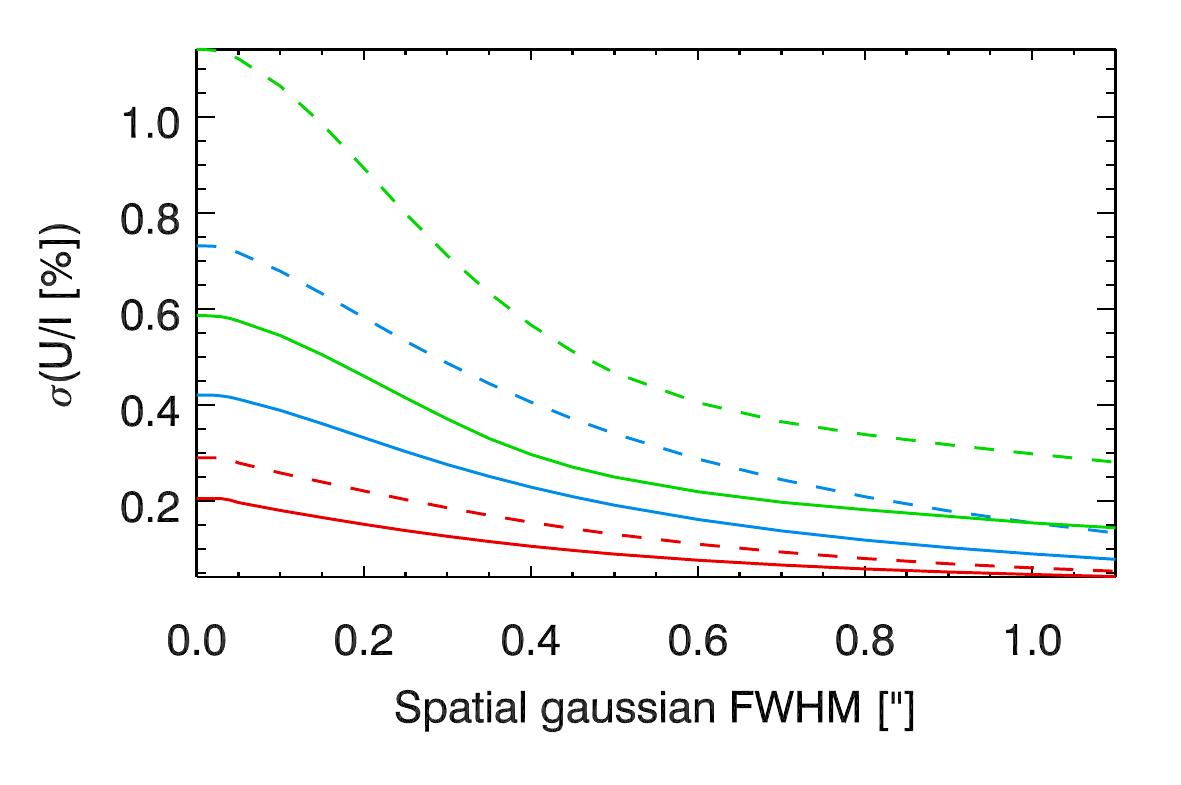}
\caption{The variation with the seeing conditions of the standard deviation of the calculated $Q/I$ (left panel) and $U/I$ (right panel) spatial variations of the scattering amplitudes of the Sr {\sc i} 4607 \AA\ line, at the wavelength where the total fractional linear polarization $P$ is maximum at the spatial pixel under consideration, for lines of sight with $\mu=0.1$ (green curves), $\mu=0.5$ (blue curves), and $\mu=1$ (red curves), taking into account (solid curves) or neglecting (dashed curves) the Hanle effect produced by the model's magnetic field. A spectral degradation of 20 m\AA\ has been taken into account.}
\label{F-sigmaspace-QU}
\end{figure}

Assuming a spectral resolution of 20 m\AA, in Figs. \ref{F-seeing-Q} and \ref{F-seeing-U} we consider three cases for illustrating the difficulty of detecting the predicted horizontal fluctuations of the linear polarization signals when the seeing conditions deteriorate: (a) $r_0=150$ cm, (b) $r_0=20$ cm and (c) $r_0=10$ cm. These cases correspond to (a) the diffraction limit of a 1.5 m telescope (top panels), (b) a seeing of 0.5 arcseconds (middle panels) and (c) a seeing of 1 arcseconds (bottom panels). The above-mentioned figures show the results for the line-center values of $Q/I$ (Fig. \ref{F-seeing-Q}) and $U/I$ (Fig. \ref{F-seeing-U}). The impact of the seeing on the spatial variations of the calculated $Q/I$ and $U/I$ signals is better quantified in Fig. \ref{F-sigmaspace-QU}, which shows how the standard deviation of the line-center amplitudes is reduced as the seeing conditions deteriorate.

The results presented in this section indicate that in order to be able to detect the predicted spatial variations of the linear polarization signals produced by scattering processes in the Sr {\sc i} 4607 \AA\ line it is crucial to observe under the best possible seeing conditions, clearly better than 0.5 arcseconds. Note that we have not accounted for other degradation effects, such as possible optical aberrations in the instrument or the blurring produced by the time variability of the solar granulation pattern during the exposure time necessary to reach the  high polarimetric sensitivity needed for proper diagnostics. Clearly, observing with high temporal resolution is also important.

\section{Simulating observations with slit-based spectropolarimeters}\label{S-slit-spectrograph}

This section presents the results of our simulations of spectropolarimetric observations with slit-based spectrographs, such as those by \cite{Malherbeetal2007} and \cite{Biandaetal2018}. Of particular interest for us are those by \cite{Biandaetal2018}, which achieved a polarimetric sensitivity of about $7.5 {\times} 10^{-4}$. Such observations have been carried out recently using the Z\"urich Imaging Polarimeter (ZIMPOL) attached to the GREGOR telescope of the Observatorio del Teide (Tenerife, Spain). Therefore, we assume a slit width of 0.3 arcseconds and CCD camera effective pixel sizes of 0.33 arcseconds along the spatial direction and of 4.4 m\AA\ along the spectral direction. The spectral resolution of the simulated observations is 20 m\AA. Although \cite{Biandaetal2018} presented their observational results for a line of sight with $\mu=0.3$, for convenience we consider here the case of $\mu=0.5$, which is qualitatively very similar, and allows us to compare the results of the instrumental effects with the scatter plots of Fig. \ref{F-scatter-P} (central panel). 

\begin{figure}[htbp]
\centering
\includegraphics[width=.36\textwidth]{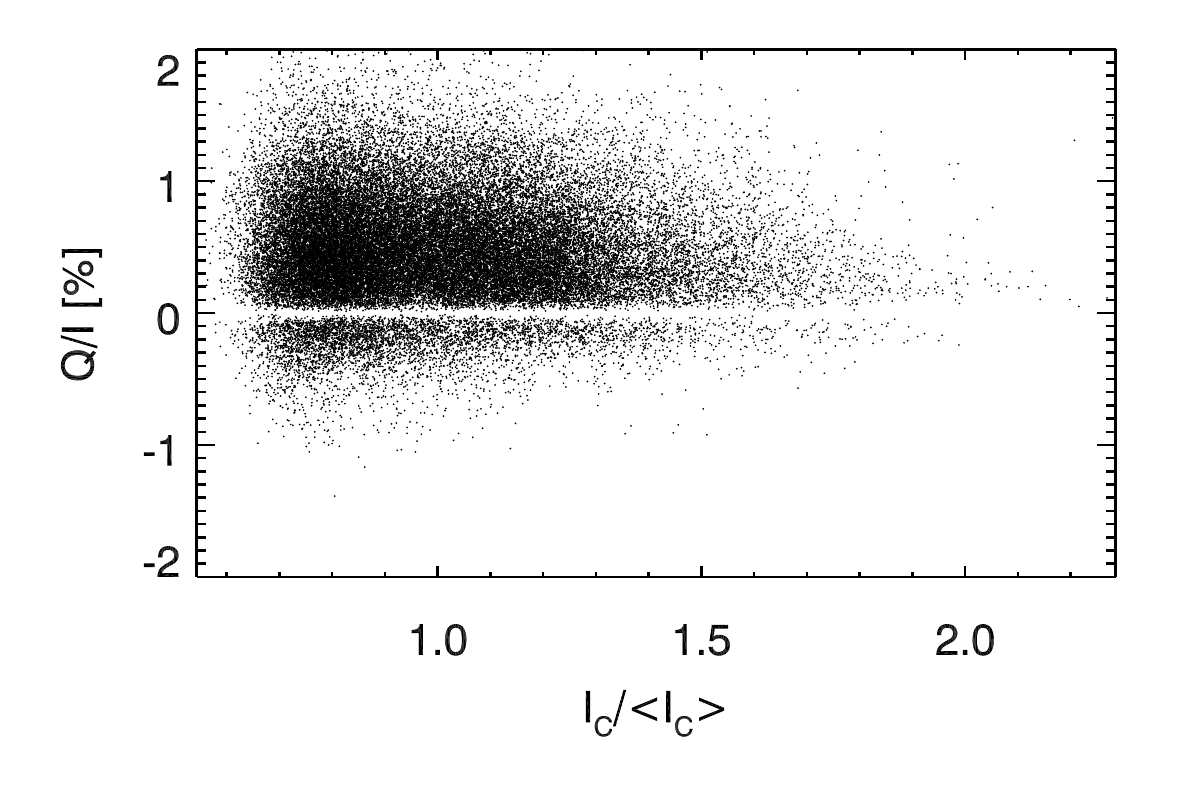}
\includegraphics[width=.36\textwidth]{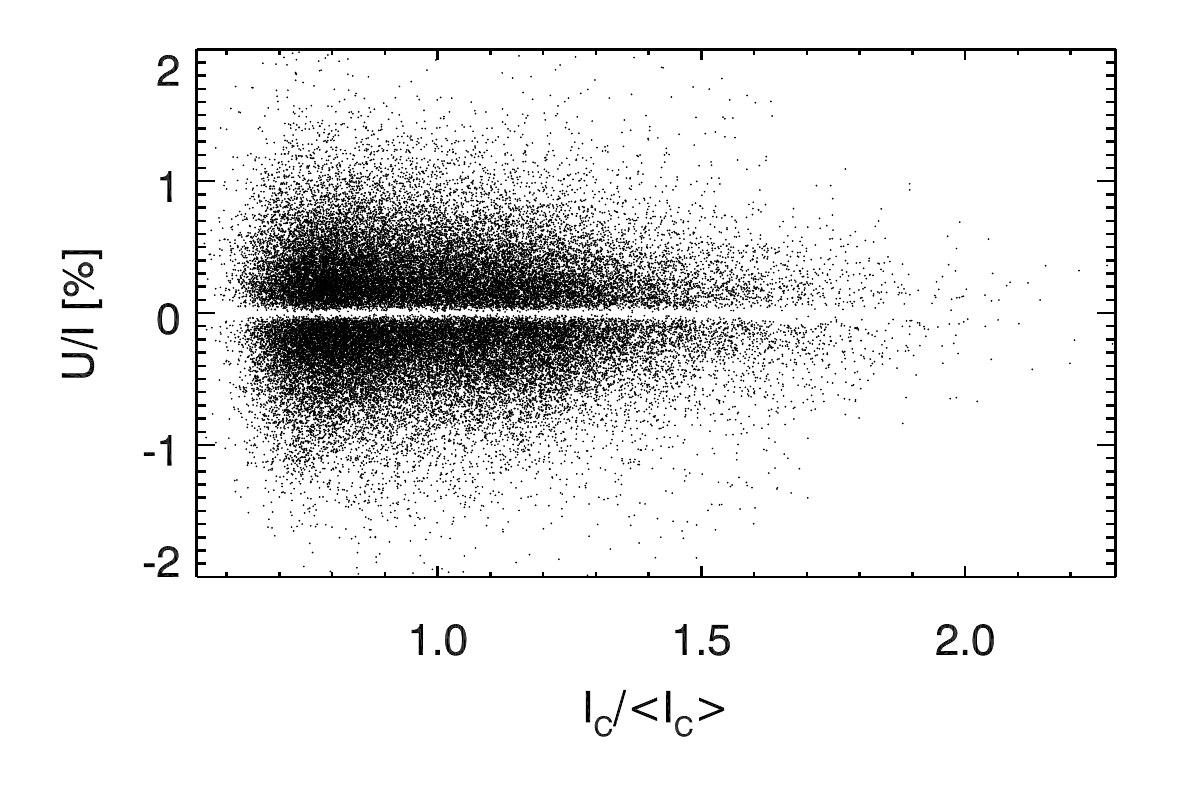} \\
\includegraphics[width=.36\textwidth]{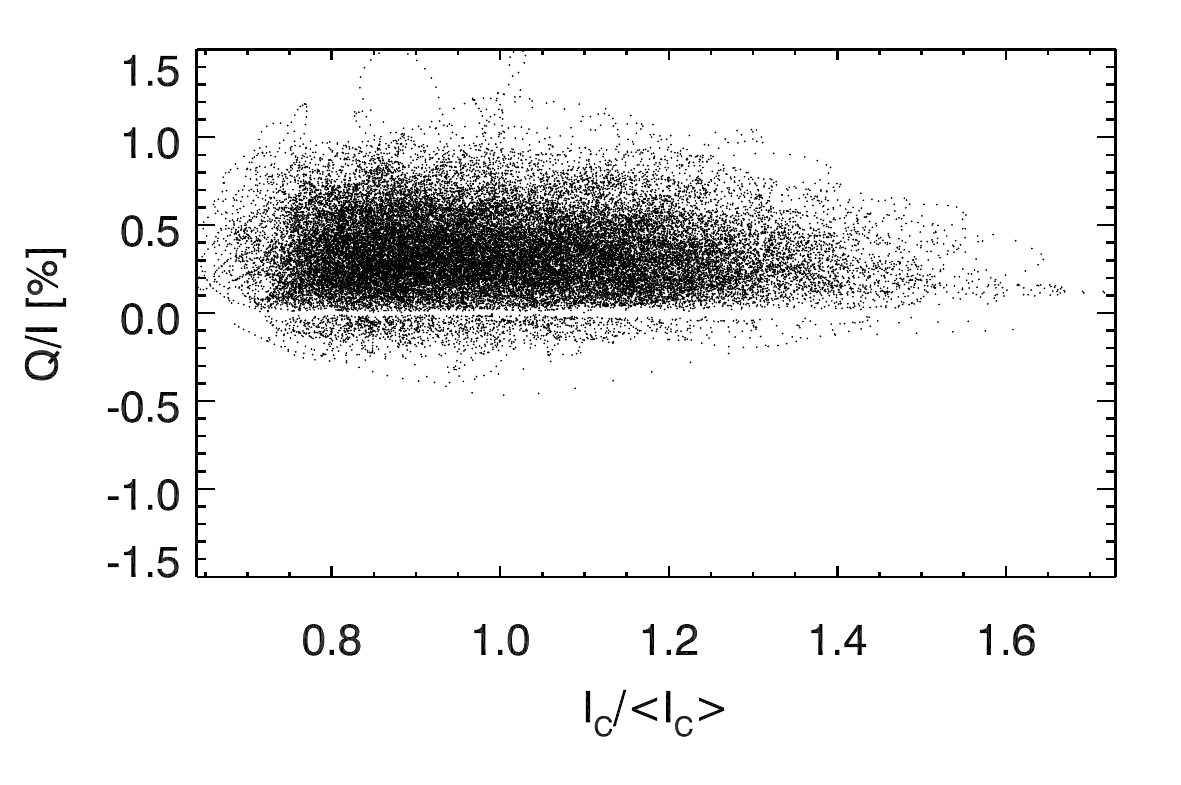}
\includegraphics[width=.36\textwidth]{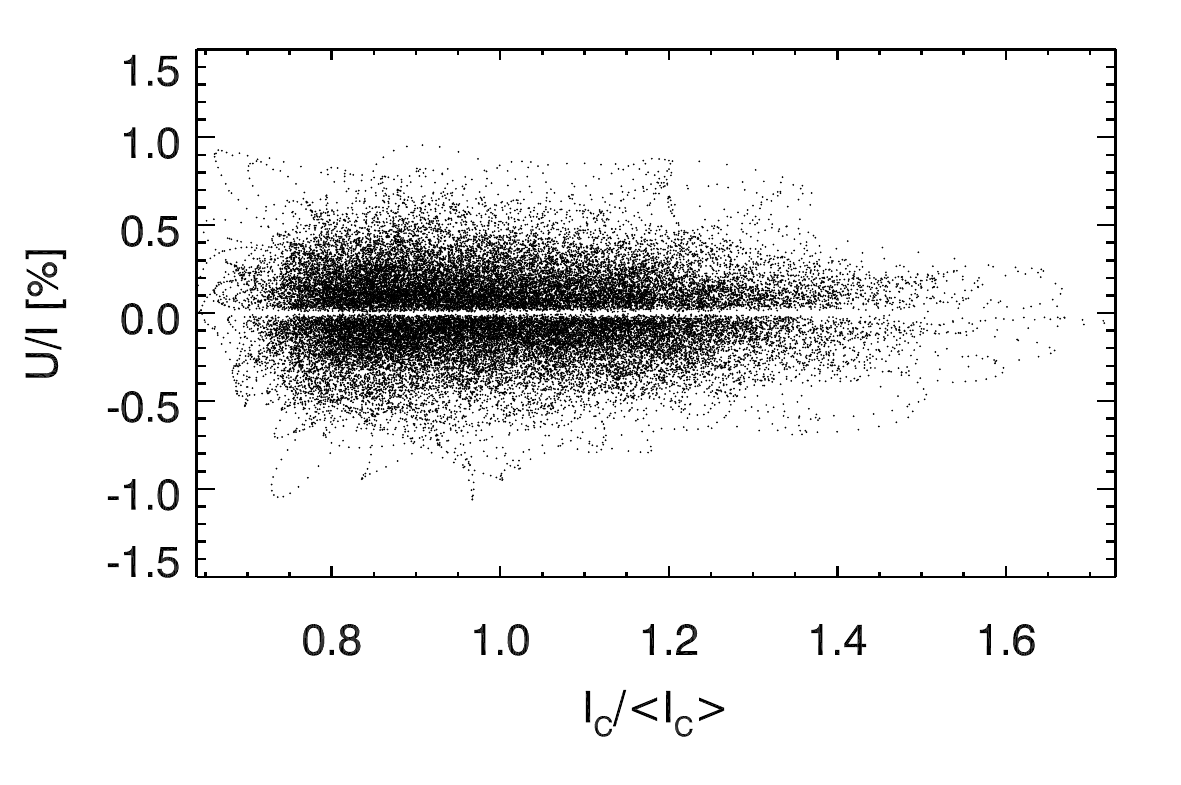} \\
\includegraphics[width=.36\textwidth]{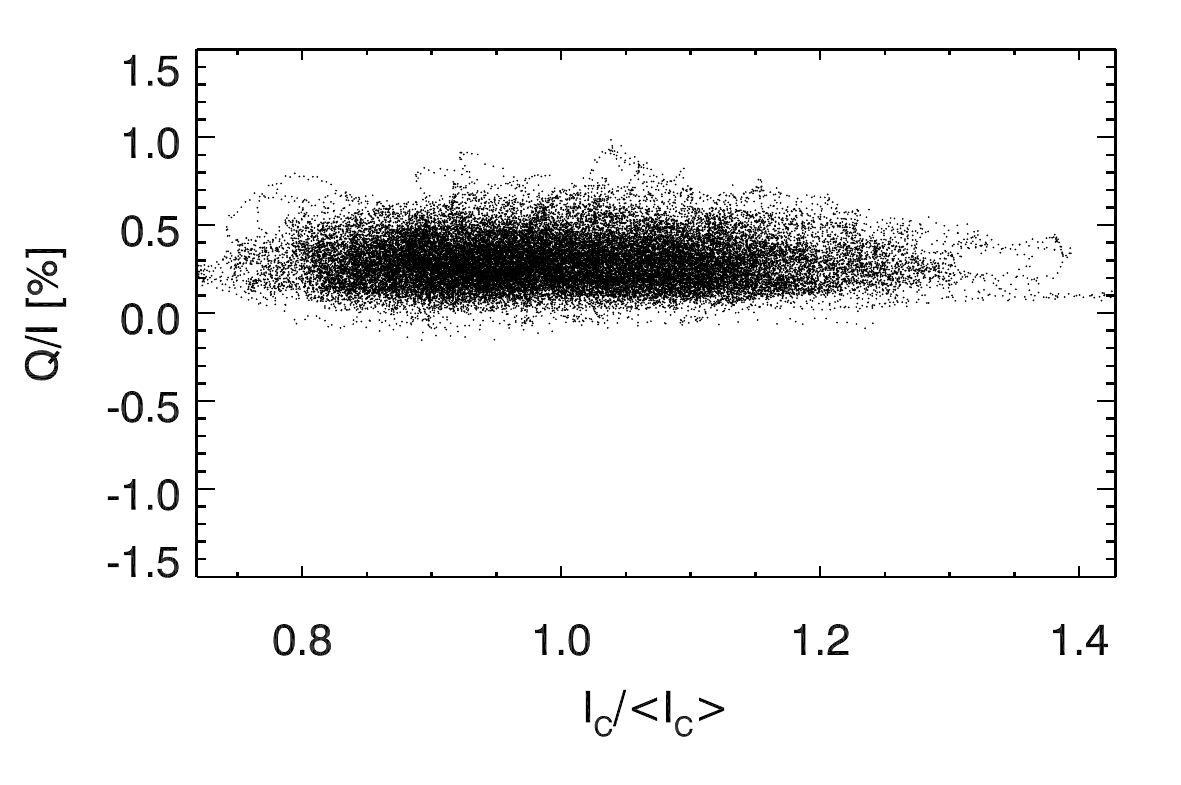} 
\includegraphics[width=.36\textwidth]{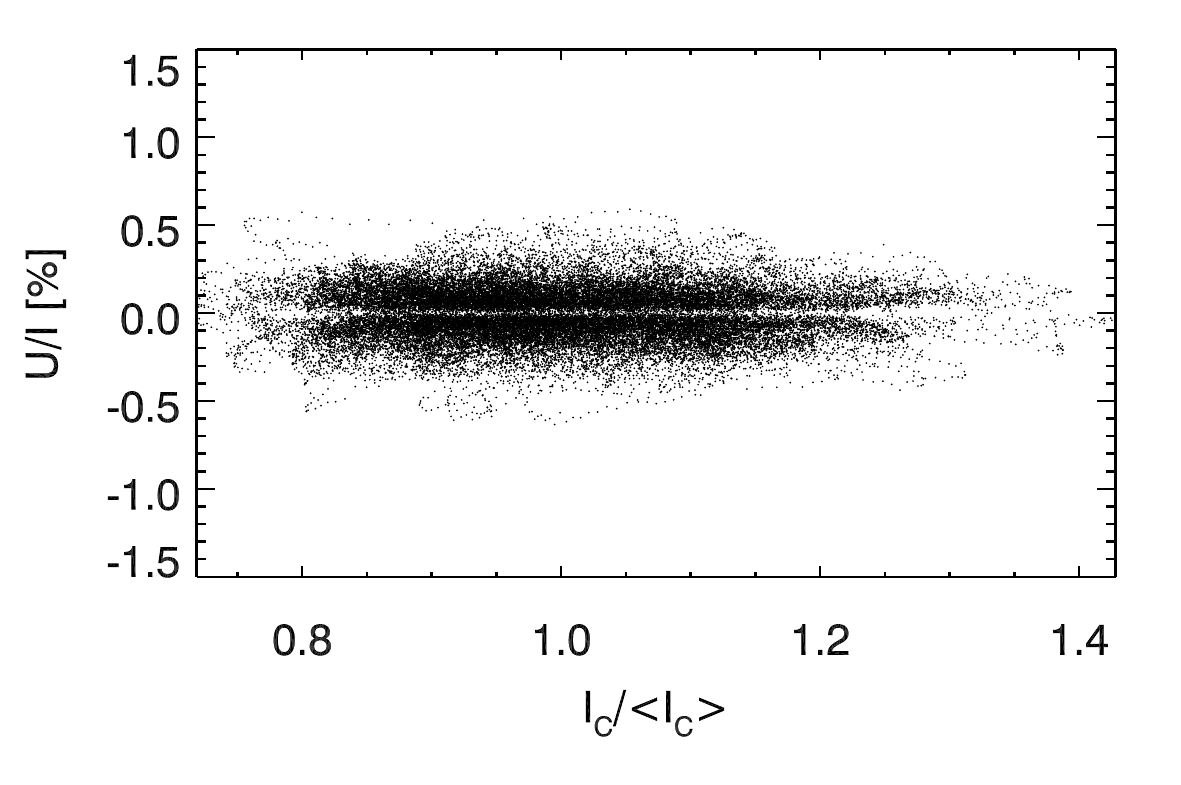} \\
\includegraphics[width=.36\textwidth]{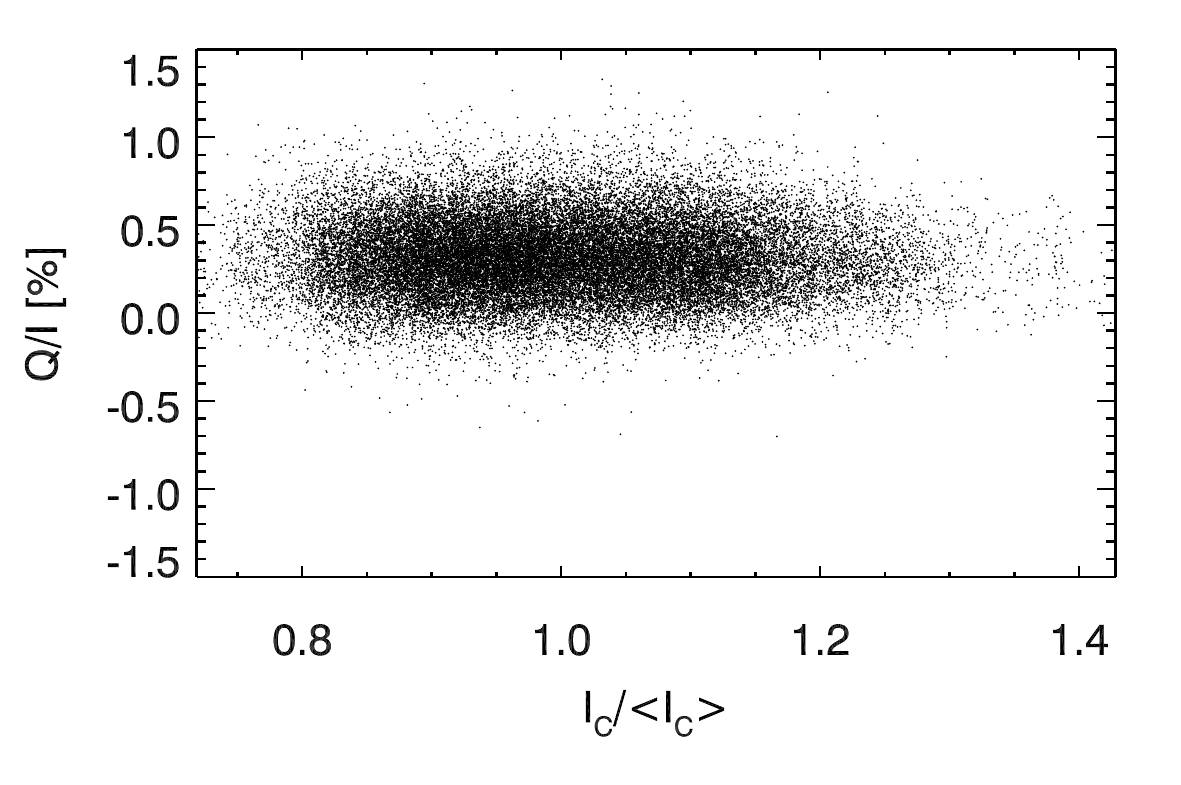}
\includegraphics[width=.36\textwidth]{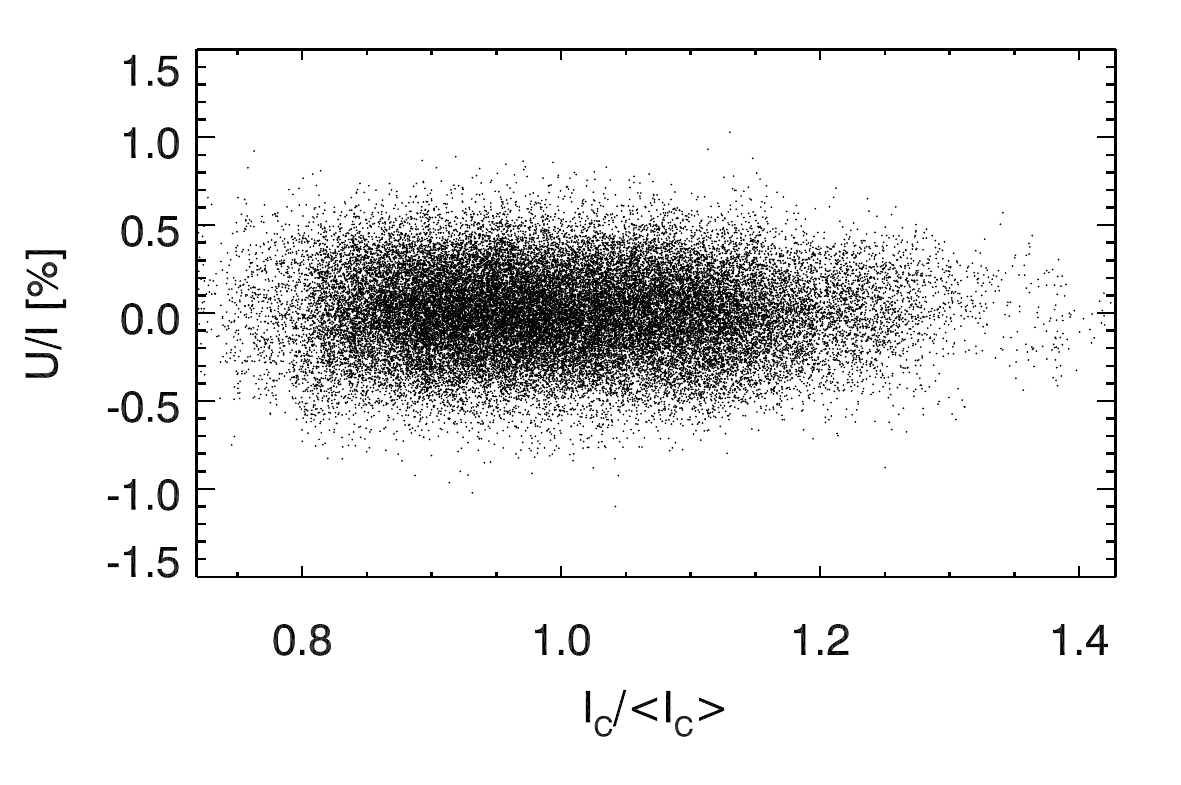} \\
\includegraphics[width=.36\textwidth]{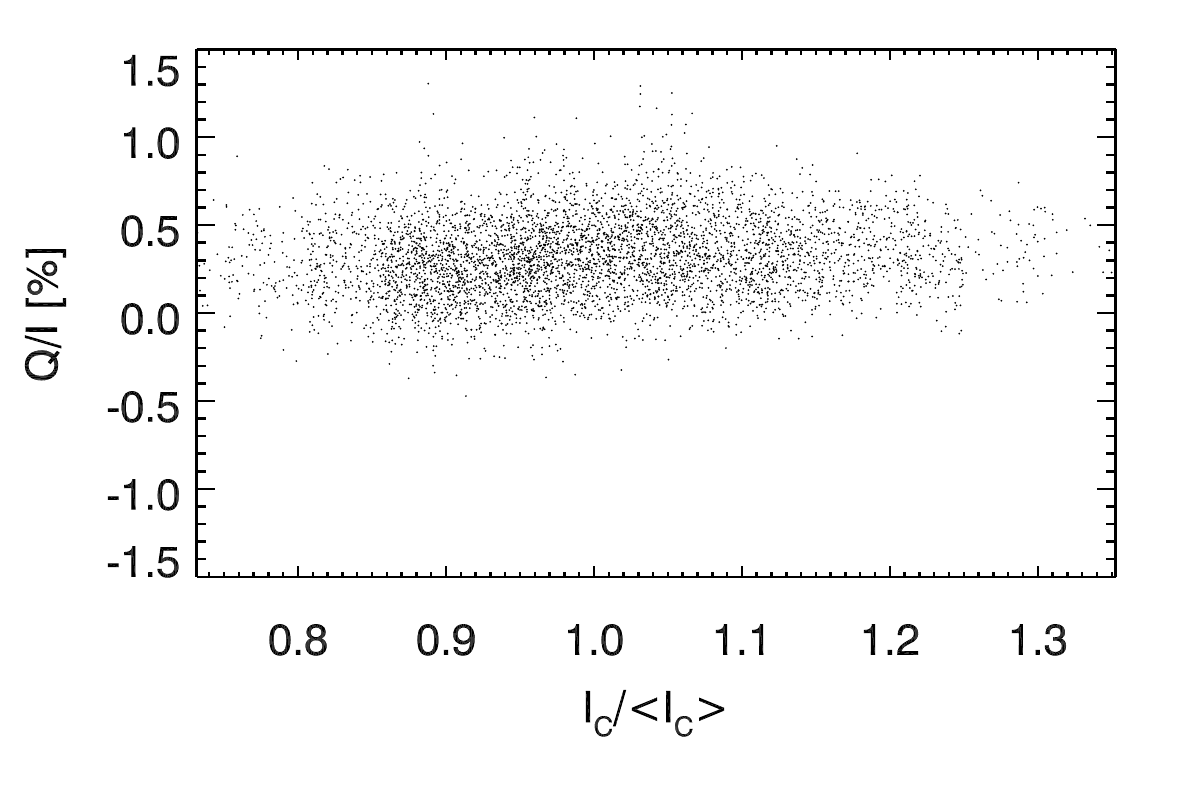} 
\includegraphics[width=.36\textwidth]{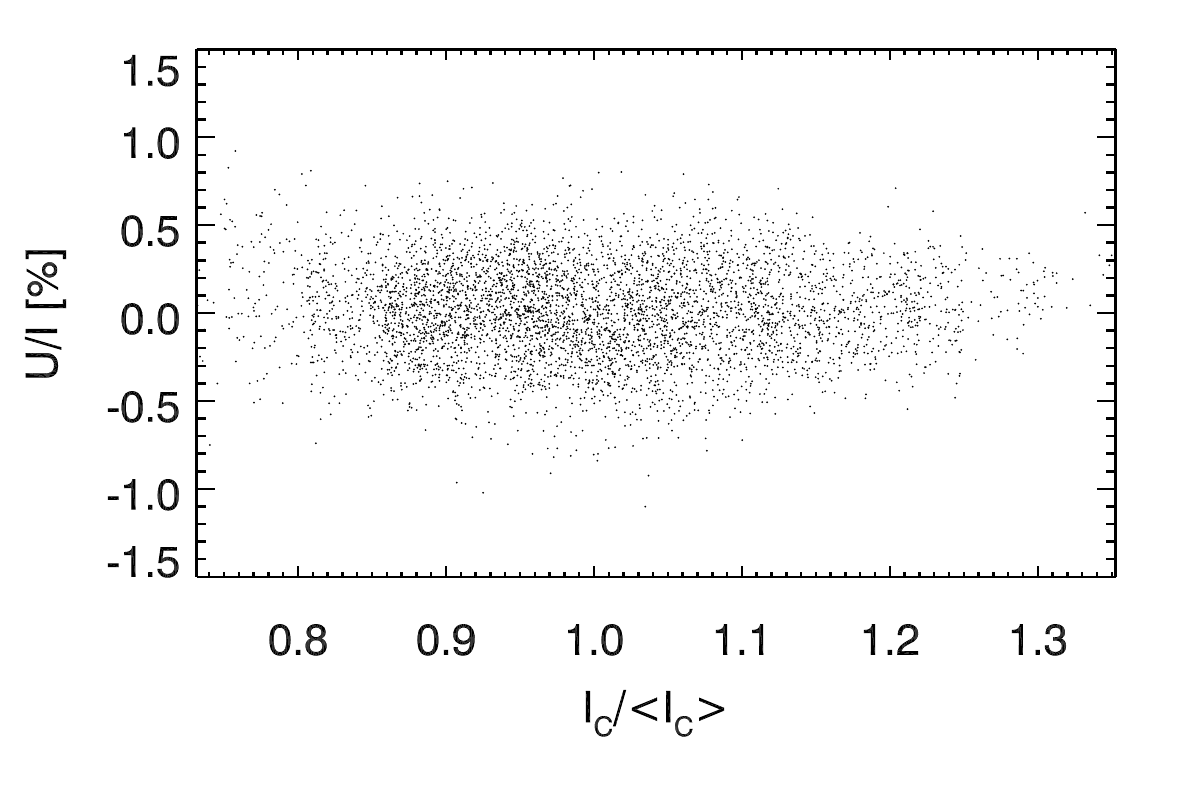}
\caption{Simulated scatter plots of the $Q/I$ (left panels) and $U/I$ (right panels) amplitudes 
against the continuum intensity, calculated all over the field of view for line of sights with inclination $\mu=0.5$. The top panels do not include any instrumental degradation (i.e., they show the theoretical signals themselves). Note that the theoretical scattering polarization amplitudes are inversely correlated with the continuum intensity (see also Fig. \ref{F-scatter-P}). The rest of the panels account for the degradation of the slit-based instrument described in the text, assuming the diffraction limit of a 1.5 m telescope (second-row panels), a seeing of 0.5 arcseconds with a polarimetric sensitivity of $7.5\times10^{-5}$ (third-row panels) and a seeing of 0.5 arcseconds with a polarimetric sensitivity of $7.5\times10^{-4}$ (the panels of the two last rows). The statistical significance is very considerable in all panels, except in the bottom ones (see the text for more explanations). Note the difficulty of detecting with the assumed observing conditions the theoretically predicted anti-correlation shown in the top panels.}
\label{F-slit-Q}
\end{figure}

Fig. \ref{F-slit-Q} shows scatter plots for the computed $Q/I$ (left panels) and $U/I$ (right panels) amplitudes against the continuum intensity, for a line of sight with $\mu=0.5$. From top to bottom we show (a) the idealized case without any noise or degradation, (b) the diffraction limit case of a 1.5 m aperture telescope without noise, (c) the case of an observation with a seeing of 0.5 arcseconds and a polarimetric sensitivity of $7.5\times10^{-5}$, (d) the case of an observation with a seeing of 0.5 arcseconds and a polarimetric sensitivity of $7.5\times10^{-4}$, and (e) the same as in the previous case, but with less statistical significance (see below). Cases (b), (c), (d) and (e) include the degradation caused by the finite spectral resolution, the width of the spectrograph's slit and the finite pixel size. We point out also that in order to mimic the procedure used by \cite{Biandaetal2018} for determining their  observed $Q/I$ amplitudes, we have used also a Gaussian function to fit the simulated fractional linear polarization profiles. In order to have a very significant statistics in panels (a), (b), (c) and (d) we used, for each point of the field of view, the calculated $Q/I$ and $U/I$ signals for lines of sights with fixed inclination $\mu=0.5$ but many azimuths $\chi$. In case (e), in order to represent better the reduced statistical significance of the observations by \cite{Malherbeetal2007} and \cite{Biandaetal2018}, we used only the polarization signals corresponding to the line of sight with $\chi=0$.  

The top panels of Fig. \ref{F-slit-Q} show clearly that, independently of the line of sight, the $Q/I$ and $U/I$ maximum scattering polarization amplitudes of the Sr {\sc i} 4607 \AA\ line are inversely correlated with the continuum intensity. In these panels, the empty horizontal strips around $Q/I=0$ and $U/I=0$ indicate that there is no point of the field of view with exactly zero polarization. The same strips are seen in the second-row panels of the figure, which show what happens if we were able to observe with ZIMPOL at GREGOR without noise at the diffraction limit of the telescope. The panels in the next two rows of the same figure illustrate what happens when observing with a seeing of 0.5 arcseconds, assuming polarimetric sensitivities of $7.5\times10^{-5}$ (third row panels) and $7.5\times10^{-4}$ (fourth row panels). Clearly, the observing conditions of the slit-based spectropolarimetric measurements of \cite{Malherbeetal2007} and \cite{Biandaetal2018} were not suitable for finding the anti-correlation with the continuum intensity we have pointed out in Fig. \ref{F-scatter-P} and in the top panels Fig. \ref{F-slit-Q}. Instead, both authors tentatively reported on a very small correlation of their observed $Q/I$ amplitudes with the continuum intensity, which we believe is an artificial correlation resulting from the relatively poor statistical significance of their slit-based observations. Interestingly enough, when we reduce the statistical significance of our simulation of observations with a seeing of 0.5 arcseconds and a polarimetric sensitivity of $7.5{\times}10^{-4}$ \citep[i.e., as in the observations of][]{Biandaetal2018} we also end up with a similarly small but positive correlation (see the bottom left panel of Fig. \ref{F-slit-Q}).  

Concerning the $U/I$ signals, we note that the ones measured by \cite{Biandaetal2018} were just above the noise level and could not be analyzed. As shown in the right panels of Fig. \ref{F-slit-Q}, the theoretical $U/I$ signals fluctuate in sign across the field of view, which make them much more difficult to detect.

Clearly, a better instrument for investigating the scattering polarization signals of the Sr {\sc i} 4607 \AA\ line would be a two-dimensional spectropolarimeter capable of measuring simultaneously the four Stokes profiles at each point of the field of view with a spectral resolution not worse than 20 m\AA, a spatial resolution ${\sim}0.1$ arcseconds and a polarimetric sensitivity better than $10^{-4}$. Such a goal could perhaps be achieved by combining a fast imaging polarimeter \citep[e.g.,][]{Iglesiasetal2016} with DKIST and/or EST.

\section{Simulating observations with a filter polarimeter}\label{S-FILTER-polarimetry}

\begin{figure}[htbp]
\centering
\includegraphics[width=.4\textwidth]{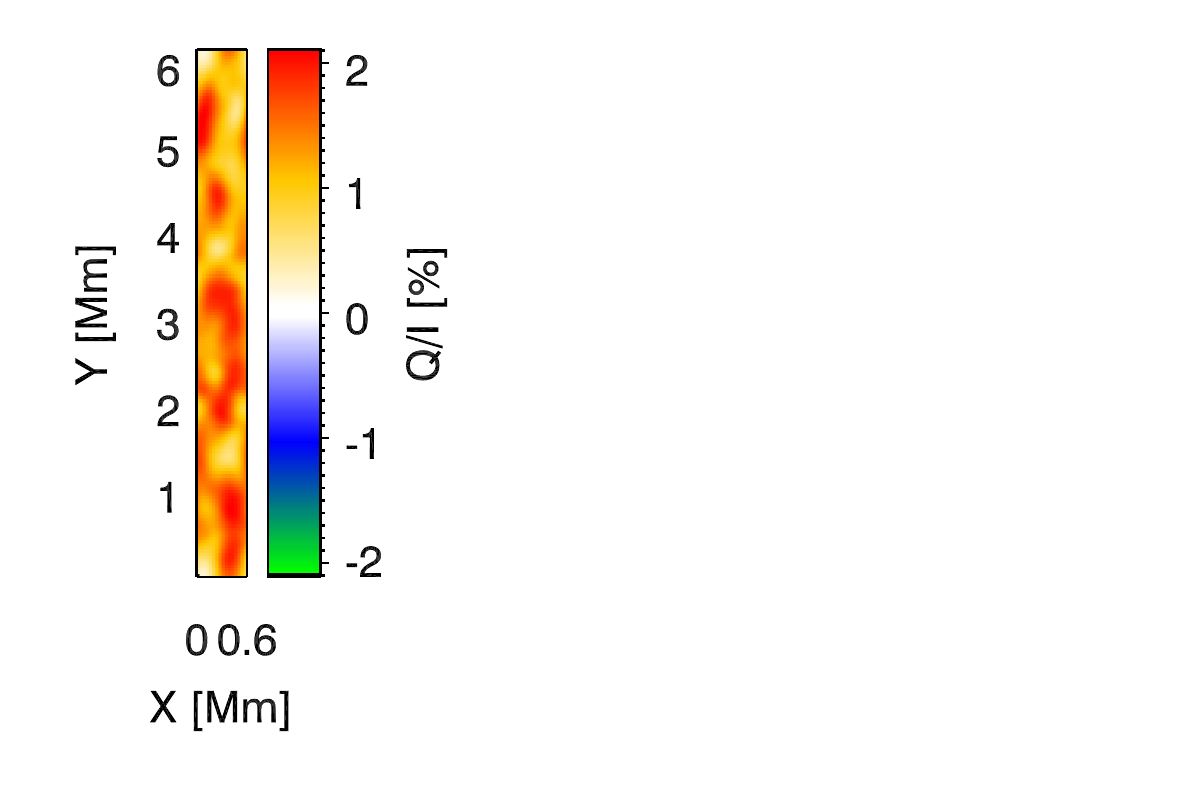}\hspace*{-14em}
\includegraphics[width=.4\textwidth]{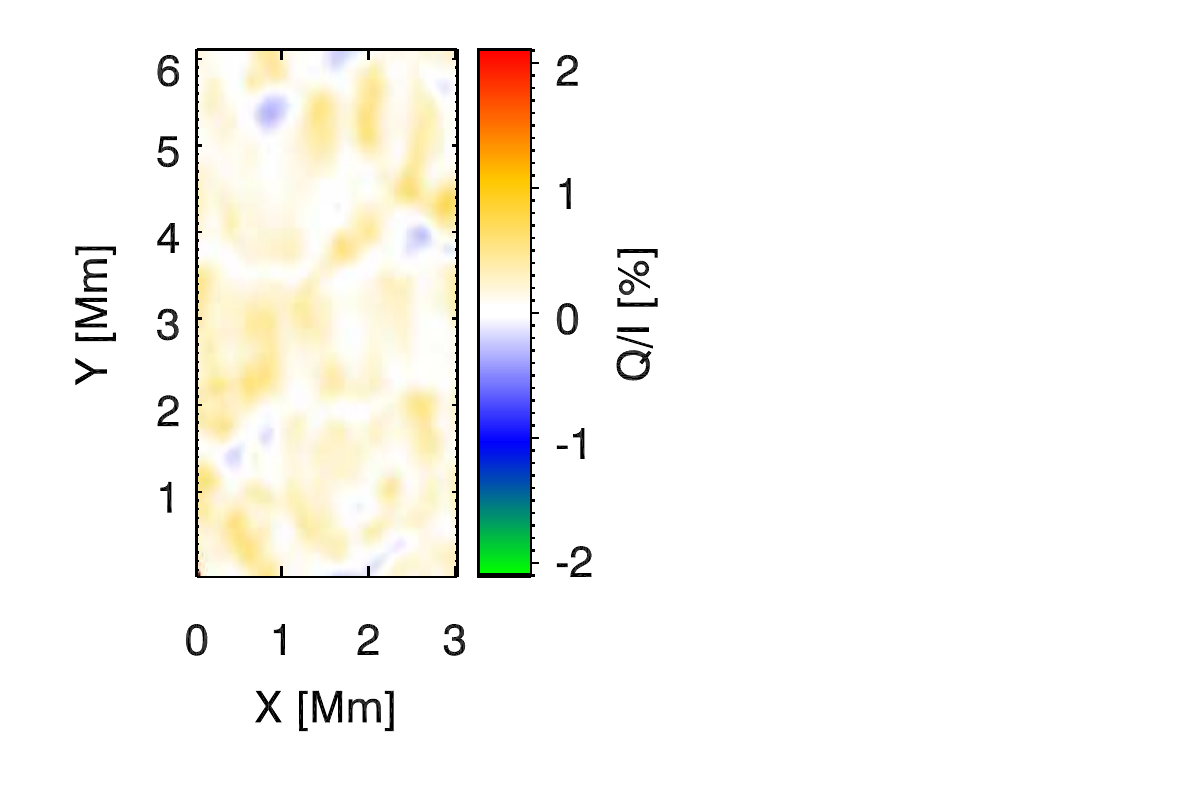}\hspace*{-10em}
\includegraphics[width=.4\textwidth]{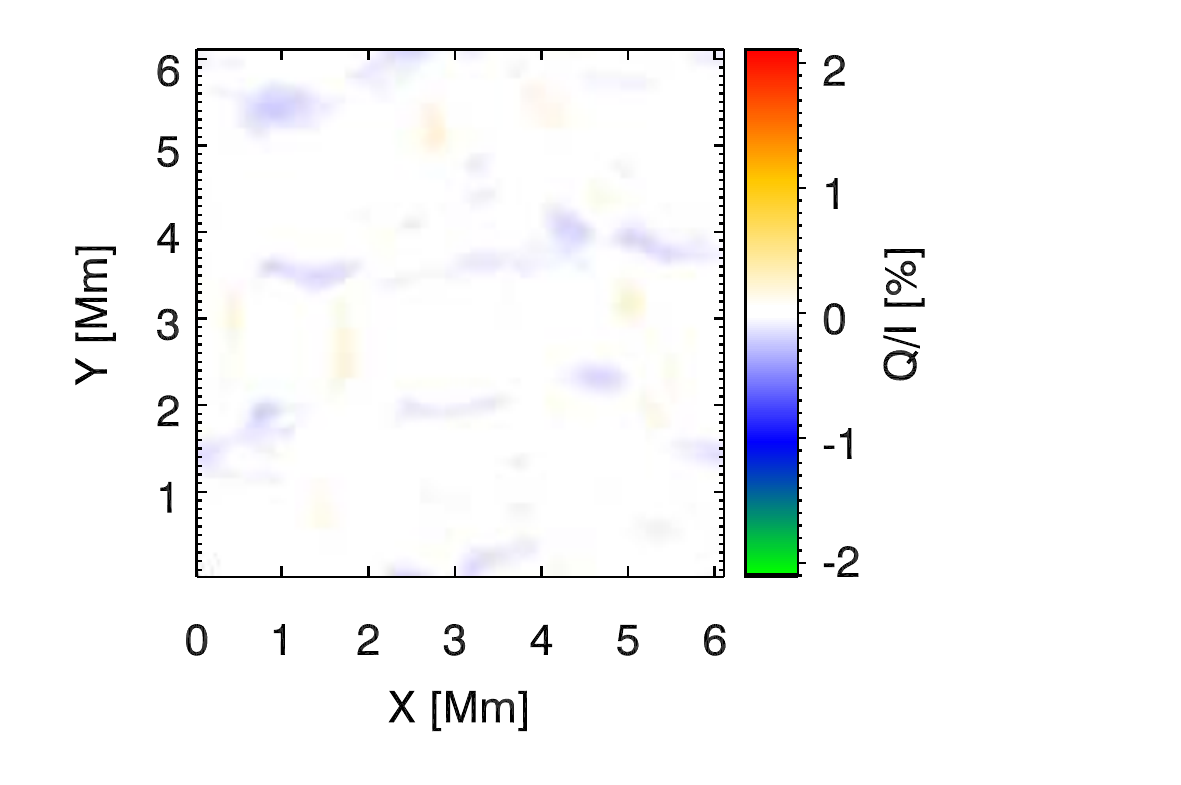}\\
\includegraphics[width=.4\textwidth]{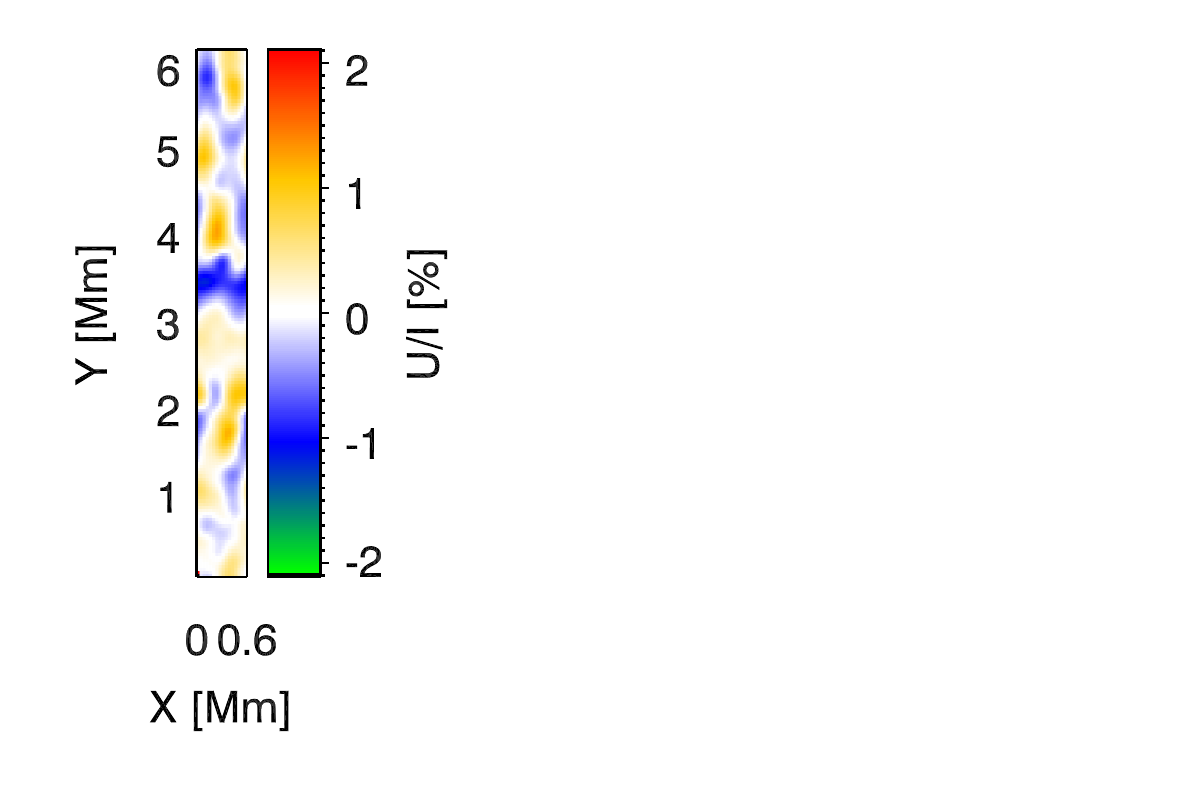}\hspace*{-14em}
\includegraphics[width=.4\textwidth]{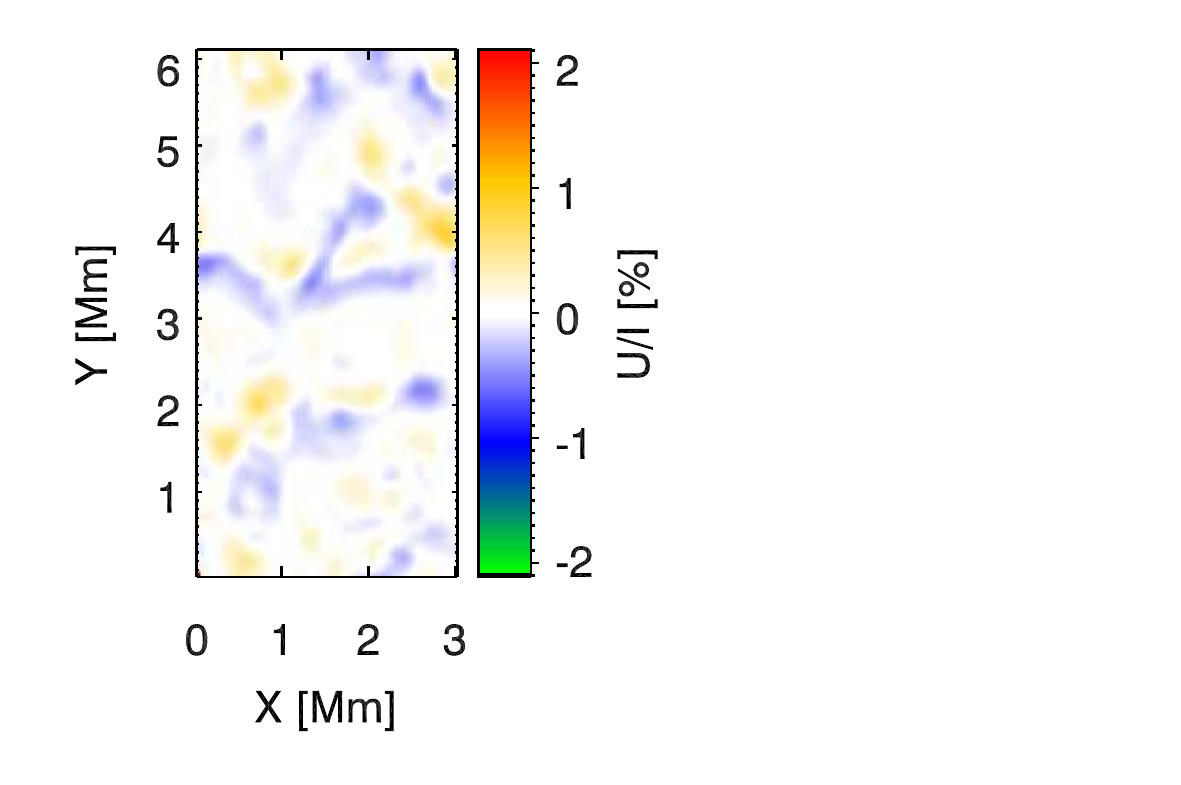}\hspace*{-10em}
\includegraphics[width=.4\textwidth]{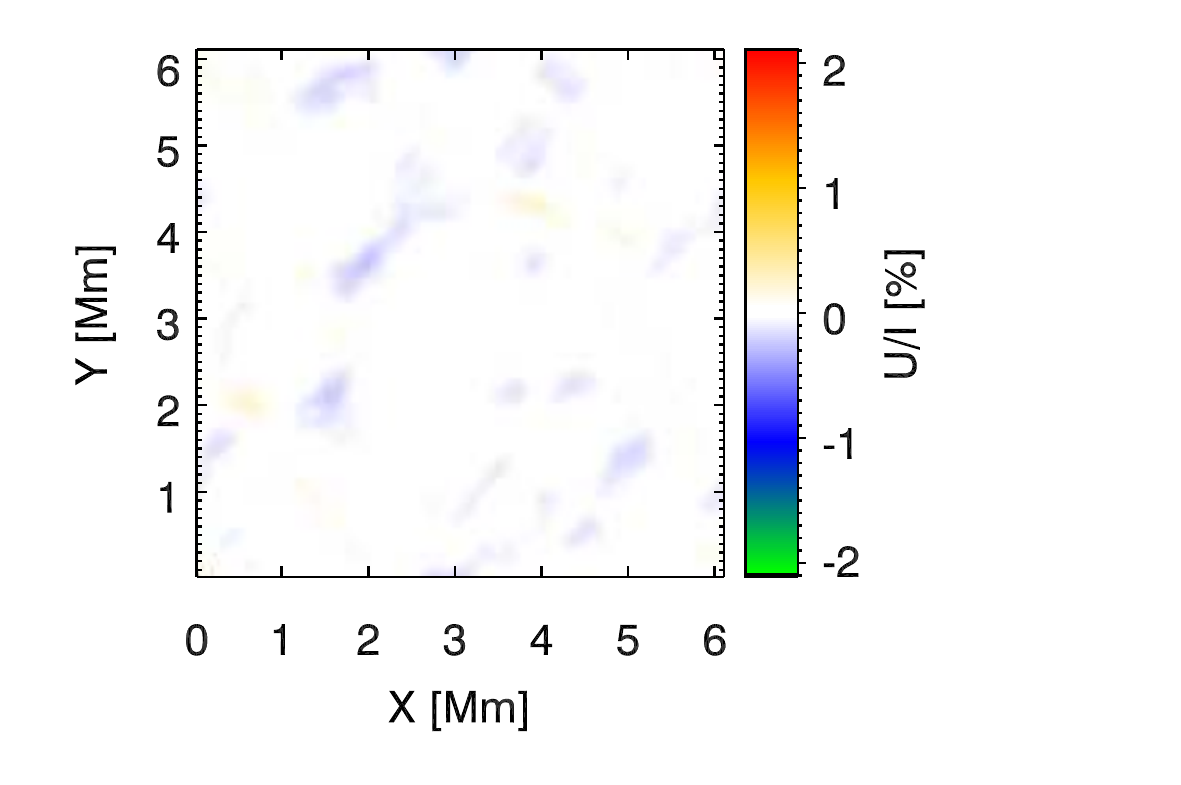}\\
\includegraphics[width=.4\textwidth]{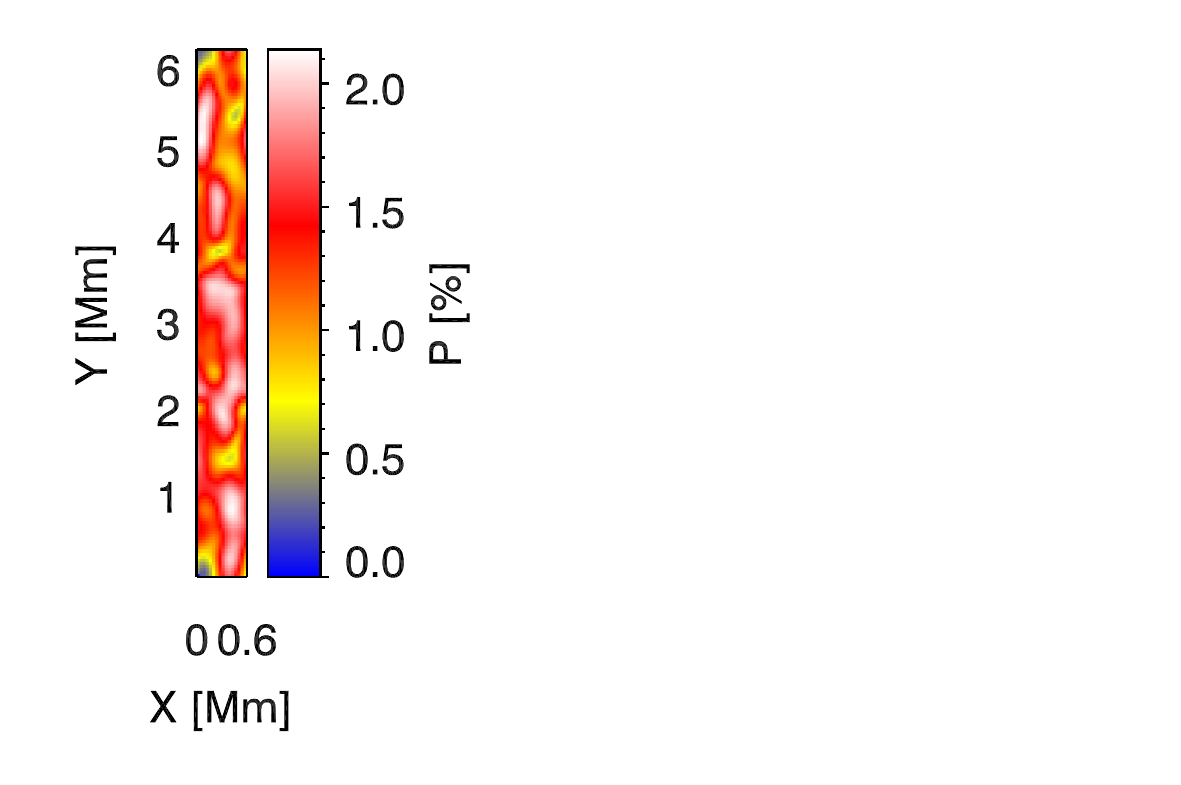}\hspace*{-14em}
\includegraphics[width=.4\textwidth]{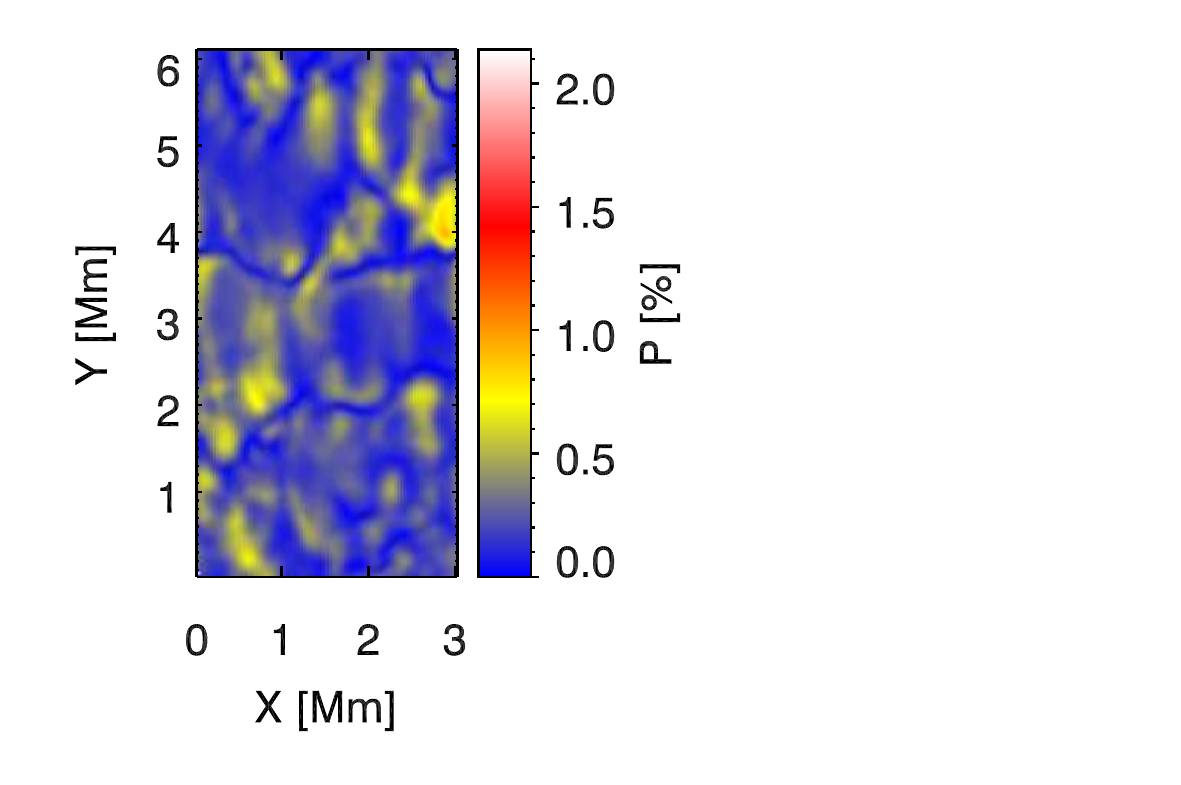}\hspace*{-10em}
\includegraphics[width=.4\textwidth]{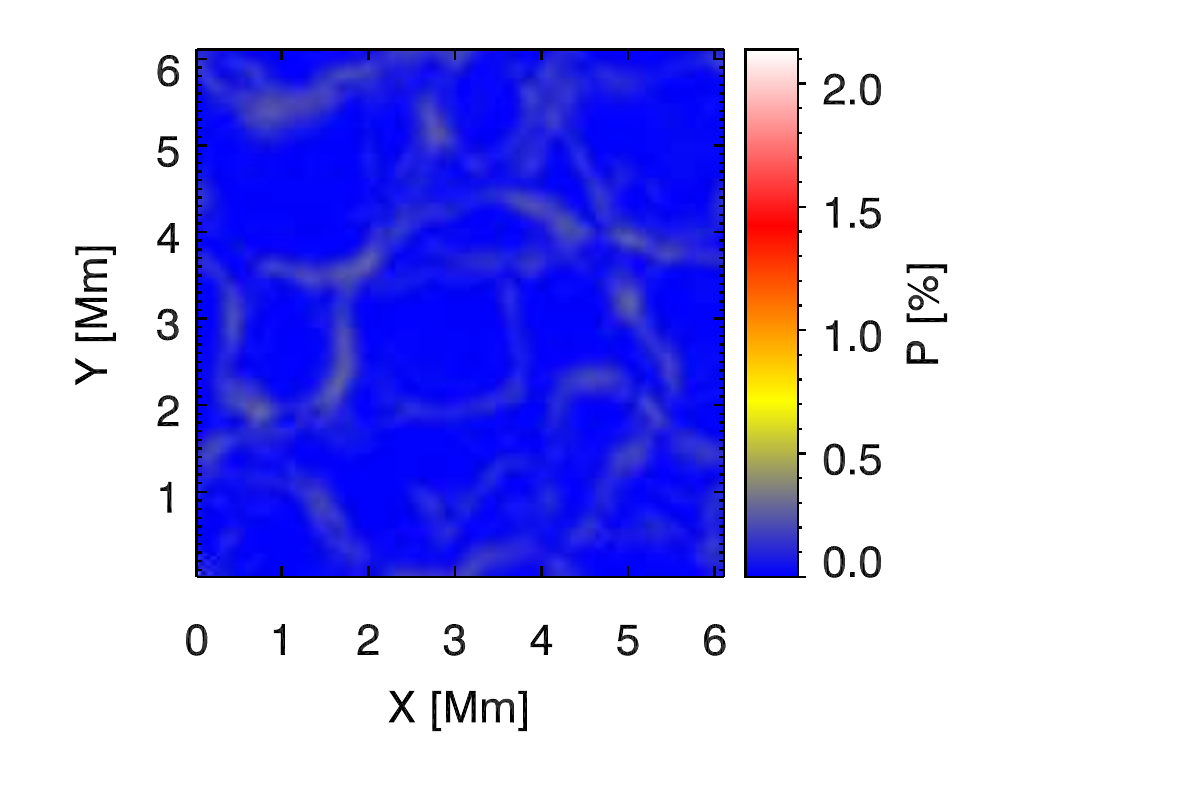}\\
\caption{The spatial variation of the calculated $Q/I$ (top panels), $U/I$ (central panels) and $P$ (bottom panels) assuming the filter polarimeter described in the text.}
\label{F-filter-polarimetry}
\end{figure}

Another instrument worthwhile to consider is a filter-polarimeter, such as that being pursued by M. Bianda et al. (2018; private communication). Such option allows to obtain simultaneous information across a two-dimensional field of view by measuring the wavelength-integrated Stokes signals over a given spectral bandwidth around the center of the spectral line. Here, we assume a Gaussian filter having a ${\rm FWHM}=100$ m\AA, a CCD camera with square pixels of 0.05 arcseconds, and a telescope with a difraction limit of 0.1 arcseconds. The results for the $Q/I$, $U/I$ and $P$ signals are shown in Fig. \ref{F-filter-polarimetry}. With such an instrument spatial fluctuations could perhaps be measured, especially off the disk center, provided that the FWHM of the filter is not larger than 100 m\AA\ and that a spatial resolution not worse than 0.1 arcseconds is reached with a polarimetric sensitivity better than $10^{-4}$.

\section{Concluding comments}\label{S-conclusions}

We have carried out a detailed radiative transfer investigation of the linear polarization of the Sr {\sc i} 4607 \AA\ line produced by anisotropic radiation pumping and the Hanle effect in a 3D model of the quiet solar photosphere resulting from Rempel's (\citeyear{Rempel2014}) numerical experiments of magneto-convection. The 3D model used is the most magnetized one of his simulations, which is characterized by a complex small-scale magnetic field (see Fig. \ref{F-Blines}) with a mean field strength ${\langle B \rangle}$ that varies with height as indicated in the left panel of Fig. \ref{F-modBandV} (i.e., it has ${\langle B \rangle}{\approx}170$ G at the model's visible surface and ${\langle B \rangle}{\approx}70$ G at a height of 300 km). It is important to note that the small-scale magnetic activity of this 3D model is not solely due to small-scale dynamo acting at the granular scales, as it also accounts for a significant amount of (small-scale) horizontal magnetic flux advected in through the bottom boundary in order to model the strong coupling between the photosphere and the deeper convection zone, where ``small-scale" dynamo action takes place as well. We point out also that the model's {\em net} magnetic flux is zero, so that it can in principle be considered representative of the inter-network regions of the quiet Sun. 

Our 3D radiative transfer calculations with the PORTA code of \cite{StepanTrujillo2013} have been carried out preserving the very high spatial resolution of the original 3D model, which is 8 km along any of the three spatial directions. The first important conclusion we want to highlight is that when the calculated Stokes profiles of the Sr {\sc i} 4607 \AA\ line corresponding to each $\mu$ value are spatially averaged, we find $U/I{\approx}0$ and a center-to-limb variation for the line-center $Q/I$ signals that is compatible with low resolution observations of the scattering polarization in the Sr {\sc i} 4607 \AA\ line. As seen in Fig. \ref{F-faurobert}, the agreement is excellent when the radiative transfer calculations are performed using the elastic collisional rates of \cite{Faurobertetal1995}, while Fig. \ref{F-manso} shows that the theoretical $Q/I$ amplitudes turn out to be slightly larger than the observed ones when the calculations are done using instead the elastic collisional rates given by \cite{Mansoetal2014}.\footnote{Fig. \ref{F-delta} informs on the quantitative difference between both rates of elastic collisions with neutral hydrogen atoms.} In order to end up with an equally good fit to the observations when using the elastic collisional rates of \cite{Mansoetal2014}, we would have to scale the model's magnetic strength by a factor $f=3/2$ in the region of formation of the core of the Sr {\sc i} 4607 \AA\ line, which would imply ${\langle B \rangle}{\approx}100$ G instead of 70 G at a height of 300 km. This illustrates why it is of crucial importance to clarify which elastic collisional rates are the most accurate ones. In any case, we can conclude that the small-scale magnetic field of Rempel's (2014) most magnetized model, which has a subsurface rms field strength increasing with depth at the same rate as the equipartition field strength, produces a Hanle depolarization of the scattering polarization of the Sr {\sc i} 4607 \AA\ line that is compatible with the (low resolution) spectropolarimetric observations. 

The scattering polarization of the Sr {\sc i} 4607 \AA\ line is due to the atomic polarization of its upper level whose $J_u=1$ (i.e., by the population imbalances and quantum coherence among its three magnetic sublevels). As shown by Eqs. \eqref{E-S00}-\eqref{E-SUline}, such upper-level polarization is caused by a ``transfer of order" from the radiation field to the atomic system (see the first term on the right hand side of Eq. \eqref{E-SKQ}) and by the Hanle effect (see the second term on the right hand side of Eq. \eqref{E-SKQ}). The ``order" of the line's radiation field is quantified by the radiation field tensors given in the Appendix, whose quantitative values at the corrugated surface of line-center optical depth unity along the $\mu=1$ line of sight are shown in the left panels of Fig. \ref{F-JKQrhoKQ}. The $\bar{J}^2_1$ and $\bar{J}^2_2$ panels indicate that a very significant breaking of the axial symmetry of the pumping radiation field must be expected, especially at the granular-intergranular boundary. The $\bar{J}^2_0$ panel indicates that, while the anisotropy tensor is larger inside the upflowing granular plasma, its largest values are found close to such  boundary.

As indicated by Eqs. \eqref{E-SQline} and \eqref{E-SUline}, for the forward scattering case of a disk center observation (LOS with $\mu=1$) the linear polarization of the emergent spectral line radiation is solely caused by the $S^2_2$ quantum coherence between the $M=1$ and $M=-1$ states of the upper-level, which is controlled by the $\bar{J}^2_2$ radiation field tensor and the Hanle effect. The symmetry breaking produced by the horizontal thermal and density inhomogeneities, as well as by the dynamical state of the photospheric plasma, produces very significant forward scattering signals without the need of any magnetic field, with zero-field total polarization amplitudes of about 2\% around the intergranular lanes (see the top right panel of Fig. \ref{F-P}). Interestingly, in these regions of the solar granulation pattern the Hanle effect of the model's magnetic field produces mainly depolarization (around a factor two) of the forward scattering signals corresponding to the unmagnetized case (compare with the bottom right panel of Fig. \ref{F-P}).

For increasingly inclined ($\mu{<1}$) line of sights, the population imbalances quantified by the $S^2_0$ density-matrix element, which is significantly influenced by the $\bar{J}^2_0$ radiation field tensor, increasingly contribute to the $Q/I$ signals of the emergent spectral line radiation (see Eq. \eqref{E-SQline}). The smaller the $\mu$ value of the line of sight the larger the number of points of the field of view with positive $Q/I$ signals (see the top panels of Fig. \ref{F-QU}), while the $U/I$ signals always fluctuate in sign across the field of view (see bottom panels of Fig. \ref{F-QU}). Therefore, ${\langle U \rangle}/{\langle I \rangle}{\approx}0$ (with the symbol ${\langle \dots \rangle}$ indicating spatial averaging), while ${\langle Q \rangle}/{\langle I \rangle}$ steadily increases from zero at $\mu=1$ to its largest positive values for close to the limb line of sights (Figs. \ref{F-faurobert} and \ref{F-manso}). Clearly, also for the line of sights with $\mu{<}1$ the Hanle effect produced by the model's magnetic field causes mainly depolarization (see Fig. \ref{F-P}).

Another point worth noting is that the Doppler shifts produced by the macroscopic velocities of the 3D model have an important impact on the symmetry properties of the pumping radiation field (quantified by the radiation field tensors) and, therefore, on the amplitudes of the fractional polarization signals (compare the two upper panels of Fig. \ref{F-P-V}). This physical ingredient was taken into account in previous 3D radiative transfer investigations of the scattering polarization in the Sr {\sc i} 4607 \AA\ line \citep[e.g.,][]{TrujilloShchukina2007}, but it was not illustrated. Even more interesting is the very significant symmetry breaking produced by the spatial gradients of the horizontal components of the macroscopic velocity, as it can be deduced by comparing the top left panel (taking into account all the velocity components) and the bottom right panel (ignoring the vertical velocity component) of the figure.

The standard deviation $\sigma$ of the spatial variations of the calculated $Q/I$ and $U/I$ signals is magnetically sensitive; for instance, the Hanle effect produced by the model's magnetic field reduces $\sigma$ by about a factor 3/2 (compare the solid and dashed lines of Fig. \ref{F-sigma-QU}). This must be taken into account for interpreting high spatio-temporal resolution observations of the scattering polarization in the Sr {\sc i} 4607 \AA\ line \citep[cf.,][]{TrujilloShchukina2007}. It is also important to achieve high spectral resolution, as $\sigma$ is strongly affected by it (see Fig. \ref{F-sigma-QU}). Obviously, a proper quantification of the spatial variations of the scattering polarization in the Sr {\sc i} 4607 \AA\ line requires sufficiently high spatial resolution, significantly better than 0.5 arcseconds (see Fig. \ref{F-sigmaspace-QU}).

As seen in Fig. \ref{F-scatter-P}, which shows scatter plots of the linear polarization amplitude $P$ against the continuum intensity, without accounting for instrumental degradation, the largest polarization signals are preferentially associated with regions with the lowest continuum intensity values, independently of the line of sight. In other words, the scattering polarization amplitudes of the Sr {\sc i} 4607 \AA\ line are inversely correlated with the continuum intensity (see also the top panels of Fig. \ref{F-slit-Q}).

We have investigated also the case of slit-based spectropolarimeters, which provide simultaneous information only along one spatial direction, finding that with a spatial resolution of 0.5 arcseconds or worse they are not suitable to properly quantify the spatial fluctuations of the scattering polarization signals of the Sr {\sc i} 4607 \AA\ line. One option to obtain simultaneous information across a two-dimensional field of view is filter-polarimetry to measure the wavelength-integrated Stokes signals over a given bandwidth. With this motivation, we have modeled the scattering polarization signals of the Sr {\sc i} 4607 \AA\ line that would be observed by an instrument based on a filter characterized by a ${\rm FWHM}=100$ m\AA\ around the central wavelength of the Sr {\sc i} 4607 \AA\ line, assuming a spatial resolution of 0.1 arcseconds (see Fig. \ref{F-filter-polarimetry}). 

An optimal instrument setup would consist of a two-dimenional spectro-polarimeter attached to a large-aperture telescope to reach a spatial resolution of at least $0.1$ arcseconds and a polarimetric sensitivity better than $10^{-4}$. Ideally, such instrument should measure the Stokes profiles with a spectral resolution not worse than 20 m\AA. This kind of instruments is what the next generation of solar telescopes (e.g., DKIST and EST) will need in order to reveal new aspects of the Sun's hidden magnetism.
 
\acknowledgements

We are grateful to Matthias Rempel (HAO) for having kindly provided the 3D magneto-convection model used in this investigation, and for helpful scientific conversations.
Thanks are also due to the referee for helping with the presentation of the paper. We acknowledge the funding received from the European Research Council (ERC) under the European Union's Horizon 2020 research and innovation programme (ERC Advanced Grant agreement No 742265). This research was also supported by the projects \mbox{AYA2014-60476-P} and \mbox{AYA2014-55078-P} of the Spanish Ministry of Economy and Competitiveness, as well as by the grant \mbox{16--16861S} of the Grant Agency of the Czech Republic and the project \mbox{RVO:67985815} of the Czech Academy of Sciences. The 3D radiative transfer simulations were carried out with the MareNostrum supercomputer of the Barcelona Supercomputing Center (National Supercomputing Center, Barcelona, Spain), and we gratefully acknowledge the resources and assistance provided.


\bibliographystyle{apj}
\bibliography{apj-jour,biblio}

\appendix

The spherical components of the radiation field tensor read:
\begin{subequations}\label{E-JKQ}
\begin{align}
  \bar{J^0_0}&=  \int {\rm d}{\nu}\oint {\phi\bigg(\nu\Big[1-{\vec{\rm v}.\vec{\Omega}\over{c}}\Big]\bigg)}
  \frac{{\rm d}
    \vec{\Omega}}{4\pi}\,I_{\nu \vec{\Omega}},                          
                                                              \label{E-J00}
\displaybreak[0]\\
  \bar{J}^2_0&= \frac{1}{2\sqrt{2}} \int {\rm d}{\nu}\oint {\phi\bigg(\nu\Big[1-{\vec{\rm v}.\vec{\Omega}\over{c}}\Big]\bigg)}
  \frac{{\rm d} \vec{\Omega}}{4\pi}
  \big[(3\mu^2-1)I_{\nu \vec{\Omega}}+3(\mu^2-1)Q_{\nu
  \vec{\Omega}}\big],
                                                              \label{E-J20}
\displaybreak[0]\\
  {\rm Re}[{\bar J}^2_1]&= \frac{\sqrt{3}}{2} \int {\rm d}{\nu} \oint {\phi\bigg(\nu\Big[1-{\vec{\rm v}.\vec{\Omega}\over{c}}\Big]\bigg)}
  \frac{{\rm d} \vec{\Omega}}{4\pi}
  \sqrt{1-\mu^2}\big[-\mu\cos\chi(I_{\nu \vec{\Omega}}+Q_{\nu
  \vec{\Omega}})+\sin\chi U_{\nu \vec{\Omega}}\big],
                                                              \label{E-RJ21}
\displaybreak[0]\\
  {\rm Im}[{\bar J}^2_1] &= \frac{\sqrt{3}}{2} \int {\rm d}{\nu} \oint {\phi\bigg(\nu\Big[1-{\vec{\rm v}.\vec{\Omega}\over{c}}\Big]\bigg)}
  \frac{{\rm d} \vec{\Omega}}{4\pi}
  \sqrt{1-\mu^2}\big[-\mu\sin\chi(I_{\nu \vec{\Omega}}+Q_{\nu
  \vec{\Omega}})-\cos\chi U_{\nu \vec{\Omega}}\big],
                                                              \label{E-IJ21}
\displaybreak[0]\\
  {\rm Re}[{\bar J}^2_2] &= \frac{\sqrt{3}}{4}\int {\rm d}{\nu} \oint {\phi\bigg(\nu\Big[1-{\vec{\rm v}.\vec{\Omega}\over{c}}\Big]\bigg)}
  \frac{{\rm d} \vec{\Omega}}{4\pi}\,
  \Big[\cos(2\chi)\big[(1-\mu^2)I_{\nu \vec{\Omega}}-(1+\mu^2)Q_{\nu
  \vec{\Omega}}\big]+2\sin(2\chi)\mu U_{\nu \vec{\Omega}}\Big],
                                                              \label{E-RJ22}
\displaybreak[0]\\
  {\rm Im}[{\bar J}^2_2] &= \frac{\sqrt{3}}{4}\int {\rm d}{\nu} \oint {\phi\bigg(\nu\Big[1-{\vec{\rm v}.\vec{\Omega}\over{c}}\Big]\bigg)}
  \frac{{\rm d} \vec{\Omega}}{4\pi}\,
  \Big[\sin(2\chi)\big[(1-\mu^2)I_{\nu \vec{\Omega}}-(1+\mu^2)Q_{\nu
  \vec{\Omega}}\big]-2\cos(2\chi)\mu U_{\nu \vec{\Omega}}\Big].
                                                              \label{E-IJ22}
\end{align}
\end{subequations}
In these equations $\phi$ is the Voigt absorption profile, with $\nu$ the frequency and $\vec{\Omega}$ the propagation direction of the radiation. This direction is characterized by $\mu={\rm cos} \theta$ (with $\theta$ the inclination of the ray with respect to the local vertical) and by the azimuth $\chi$. The results presented in this paper were obtained using 25 rays per octant for the angular integration, with Gaussian quadrature for the 5 inclinations per octant and the trapezoidal rule for the 5 azimuths per octant. For the frequency integration we used the trapezoidal rule, with 101 frequency points, equidistant in the line-core, covering ${\pm}$ 0.45 \AA\ around the 4607 \AA\ central wavelength of the Sr {\sc i} resonance line. Obviously, $I_{\nu\vec{\Omega}}$, $Q_{\nu\vec{\Omega}}$ and $U_{\nu\vec{\Omega}}$ are the Stokes parameters of each ray considered, with the reference direction for the quantification of $Q_{\nu\vec{\Omega}}$ and $U_{\nu\vec{\Omega}}$ in the plane formed by $\vec{\Omega}$ and the local vertical.

The non-zero coefficients $M_{ij}$ of the magnetic kernel of Eq. \eqref{E-SKQ} are:
\begin{subequations}\label{E-Mij}
\begin{align}
  M_{12} &= \sqrt{6}\sin{\theta_B}\sin{\chi_B}  ,
                                                              \label{E-M12}
\displaybreak[0]\\
  M_{13} &= \sqrt{6}\sin{\theta_B}\cos{\chi_B}  ,
                                                              \label{E-M13}
\displaybreak[0]\\
  M_{21} &= -\sqrt{\frac{3}{2}}\sin{\theta_B}\sin{\chi_B}  ,
                                                              \label{E-M21}
\displaybreak[0]\\
  M_{23} &= -M_{32} = \cos{\theta_B}  ,
                                                              \label{E-M23}
\displaybreak[0]\\
  M_{24} &= M_{35} = -M_{42} = -M_{53} = \sin{\theta_B}\sin{\chi_B}  ,
                                                              \label{E-M24}
\displaybreak[0]\\
  M_{25} &= M_{43} = -M_{34} = -M_{52} = \sin{\theta_B}\cos{\chi_B}  ,
                                                              \label{E-M25}
\displaybreak[0]\\
  M_{31} &= -\sqrt{\frac{3}{2}}\sin{\theta_B}\cos{\chi_B}  ,
                                                              \label{E-M31}
\displaybreak[0]\\
  M_{45} &= -M_{54} = 2\cos{\theta_B}  ,
                                                              \label{E-M45}
\end{align}
\end{subequations}

Note that $\theta_B$ is the inclination of the magnetic field vector with respect to the local vertical and $\chi_B$ is the magnetic field azimuth, measured anticlockwise from the X axis. In all the figures of this paper the reference direction for Stokes $Q>0$ is along the Y axis.

\end{document}